\documentclass[letter,twoside,titlepage,11pt]{article}
\usepackage{amssymb}
\def\clock{{\count0=\time
           \divide\count0 60
           \ifnum\count0<10 0\fi\the\count0
           \multiply\count0 -60 \advance\count0 \time
           :\ifnum\count0<10 0\fi \the\count0
         }}
\newcommand{\timestamp}{{\small\vbox{\hbox{\tt\jobname.tex}
\hbox{\the\day/\the\month/\the\year, \clock}}}}

\usepackage{epsfig}

\makeatletter \@addtoreset{equation}{section} \makeatother

\newcommand{\ads}{\mbox{AdS}}

\newcommand{\eqref}[1]{(\ref{#1})}

\newcommand{\id}{\hbox{1\kern-.27em l}}
\newcommand{\sid}{\hbox{\scriptsize1\kern-.27em l}}

\newcommand{\we}{\kern-.1em\wedge\kern-.1em}
\newcommand{\scal}{\kern-.13em\cdot\kern-.13em}

\newcommand{\II}{I\kern-.09em I}

\newcommand{\Z}{\mathbb{Z}}
\newcommand{\R}{\mathbb{R}}

\newcommand{\smallsec}[1]{\vskip 0.2cm \noindent {\bf \underline{#1}} \nopagebreak }

\begin{document}

\begin{titlepage}

\rightline{HUTP-02/A006}
\rightline{NBI-HE-02-06}
\rightline{hep-th/0204047}
\rightline{April, 2002}
\vskip 1.4cm

\centerline{\Large \bf Black Holes on Cylinders}

\vskip 1.3cm
\centerline{{\bf Troels Harmark}\footnote{e-mail: harmark@bose.harvard.edu} }
\vskip 0.2cm
\centerline{\sl Jefferson Physical Laboratory}
\centerline{\sl Harvard University}
\centerline{\sl Cambridge, MA 02138, USA}

\vskip 0.7cm

\centerline{{\bf Niels A. Obers}\footnote{e-mail: obers@nbi.dk} }
\vskip 0.2cm
\centerline{\sl The Niels Bohr Institute}
\centerline{\sl Blegdamsvej 17, DK-2100 Copenhagen \O, Denmark}

\vskip 1.3cm
\centerline{\bf Abstract}
\vskip 0.2cm
\noindent
We take steps toward constructing explicit solutions that describe
non-extremal charged dilatonic branes of string/M-theory with
a transverse circle.
Using a new coordinate system
we find an ansatz for the solutions with only one unknown function.
We show that this function is independent of the charge and our
ansatz can therefore also be used to construct neutral black holes
on cylinders and near-extremal charged dilatonic branes with
a transverse circle.
For sufficiently large mass $M > M_c$ these solutions have a horizon that
connects across the cylinder but they are not translationally
invariant along the circle direction.
We argue that the neutral solution has larger entropy than
the neutral black string for any given mass.
This means that for
$M > M_c$ the neutral black string can gain entropy by
redistributing its mass to a solution that breaks
translational invariance along the circle, despite the fact that it is
classically stable.
We furthermore explain how our construction can be used to study
the thermodynamics of Little String Theory.


\end{titlepage}


\newcommand{\nn}{\nonumber}
\newcommand{\spa}{\ \ ,\ \ \ \ }
\newcommand{\str}{\mathop{{\rm Str}}}
\newcommand{\tr}{\mathop{{\rm Tr}}}
\newcommand{\sn}{\mathop{{\rm sn}}}

\newcommand{\gym}{g_{\mathrm{YM}}}
\newcommand{\geff}{g_{\mathrm{eff}}}
\newcommand{\gseff}{g_s^{\mathrm{eff}}}
\newcommand{\Ord}{{\cal{O}}}
\newcommand{\tlst}{T_{\rm LST}}
\newcommand{\los}{l_{\rm os}}

\newcommand{\vecto}[2]{\left( \begin{array}{c} #1 \\ #2 \end{array} \right) }
\newcommand{\matrto}[4]{\left( \begin{array}{cc} #1 & #2 \\ #3 & #4 \end{array} \right) }

\newcommand{\calF}{\mathcal{F}}
\newcommand{\calL}{\mathcal{L}}
\newcommand{\calH}{\mathcal{H}}

\newcommand{\eps}{\epsilon}


\setcounter{page}{1}

{\footnotesize

\contentsline {section}{\numberline {1}Introduction}{\pageref{secintro}}
\contentsline {section}{\numberline {2}A new coordinate system}{\pageref{secnewcoord}}
\contentsline {subsection}{\numberline {2.1}Defining the new coordinates}{\pageref{secdefnewc}}
\contentsline {subsection}{\numberline {2.2}Critical curve}{\pageref{seccritcurv}}
\contentsline {subsection}{\numberline {2.3}The flat metric}{\pageref{secflatmet}}
\contentsline {section}{\numberline {3}Ansatz}{\pageref{secbigansatz}}
\contentsline {subsection}{\numberline {3.1}Conditions on a solution}{\pageref{secconditions}}
\contentsline {subsection}{\numberline {3.2}The ansatz}{\pageref{secansatz}}
\contentsline {section}{\numberline {4}Map of solution to neutral and near-extremal solutions}{\pageref{secmapsol}}
\contentsline {subsection}{\numberline {4.1}Neutral black holes on a cylinder}{\pageref{neutralmap}}
\contentsline {subsection}{\numberline {4.2}Near-extremal branes on transverse circle}{\pageref{secnemap}}
\contentsline {section}{\numberline {5}Newton limit of small black holes on cylinders}{\pageref{secsmallBH}}
\contentsline {subsection}{\numberline {5.1}The Newton limit of Einsteins equation}{\pageref{secnewteins}}
\contentsline {subsection}{\numberline {5.2}The Newton potential and the $g_{00}$ component}{\pageref{secg00}}
\contentsline {subsection}{\numberline {5.3}The metric in the Newton limit}{\pageref{secnewton}}
\contentsline {subsection}{\numberline {5.4}Consequences for small black hole on cylinder}{\pageref{secconsBH}}
\contentsline {subsection}{\numberline {5.5}Consequences for measurements of mass}{\pageref{secconmass}}
\contentsline {section}{\numberline {6}Finding solutions}{\pageref{secfindsol}}
\contentsline {subsection}{\numberline {6.1}Equations of motion}{\pageref{seceoms}}
\contentsline {subsection}{\numberline {6.2}General considerations}{\pageref{secgencons}}
\contentsline {subsection}{\numberline {6.3}Analysis of equations of motion}{\pageref{secaneom}}
\contentsline {section}{\numberline {7}Thermodynamics}{\pageref{secthermodynamics}}
\contentsline {subsection}{\numberline {7.1}Properties of the horizon}{\pageref{secprophor}}
\contentsline {subsection}{\numberline {7.2}Thermodynamics}{\pageref{secthermo}}
\contentsline {subsection}{\numberline {7.3}Physical subspace of solutions}{\pageref{secsolspace}}
\contentsline {subsection}{\numberline {7.4}Further study of $\chi (R_0)$ and $\gamma (R_0)$}{\pageref{secgamchi}}
\contentsline {subsection}{\numberline {7.5}Neutral solution has larger entropy than black string}{\pageref{secpunch}}
\contentsline {section}{\numberline {8}Black holes on cylinders and thermal Little String Theory}{\pageref{secLST}}
\contentsline {subsection}{\numberline {8.1}Review of supersymmetric Little String Theory from supergravity}{\pageref{secsusyLST}}
\contentsline {subsection}{\numberline {8.2}Review of thermal Little String Theory from supergravity}{\pageref{secthLST}}
\contentsline {subsection}{\numberline {8.3}The new solutions and thermal Little String Theory}{\pageref{secnewsolLST}}
\contentsline {section}{\numberline {9}Discussion and
Conclusions}{\pageref{secconc}} \ \newline
{\noindent {\bf Appendices}}

\contentsline {subsection}{\numberline {A}Charged dilatonic $p$-brane solutions}{\pageref{appextrsol}}
\contentsline {subsection}{\numberline {B}The functions $F_{2s}$ and $G_{2s}$}{\pageref{appfctF}}
\contentsline {subsection}{\numberline {C}Coordinate change in large $R$ limit}{\pageref{appcoord}}
\contentsline {subsection}{\numberline {D}Details on the expansion of the equations of motion}{\pageref{appdet}}
\contentsline {subsection}{\numberline {E}M5 and NS5-branes}{\pageref{appM5NS5}}
\contentsline {section}{References}{\pageref{refs}}

}

\newpage

\section{Introduction}
\label{secintro}

Apart from being an interesting subject in its own right,
black holes on cylinders $\R^{d-1} \times S^1$
show up in various contexts in String Theory
as branes of String/M-theory on a transverse circle.
Examples include S-duality between M2 and M5-branes in M-theory
and D2 and NS5-branes in Type IIA String Theory
and T-duality between D$p$-branes in Type IIA/B String Theory.
Moreover, the near-extremal limits of these branes with transverse
circles are important
in relation to the AdS/CFT correspondence
\cite{Maldacena:1997re,Itzhaki:1998dd,Witten:1998qj,Gubser:1998bc}
and Matrix Theory \cite{Banks:1997vh,Sen:1997we,Seiberg:1997ad}.

Black holes on cylinders $\R^{d-1} \times S^1$ have a more interesting
dynamics and richer phase structure than black holes on flat space $\R^d$.
In flat space the static neutral black hole of a certain
mass $M$ is uniquely described by the Schwarzschild solution with mass $M$.
Black holes on cylinders can have a richer phase structure since the
radius of the circle provides a macroscopic scale in the system.
Moreover, $\R^{d-1} \times S^1$ has a non-trivial topology
in the sense that it is non-contractible
with a non-trivial fundamental group.
Because of this there exist black strings, which are black
objects with an event horizon that wraps the circle.
Also, if a black hole gets sufficiently heavy, its event horizon
can ``meet itself'' across the cylinder.
Thus, on the cylinder $\R^{d-1} \times S^1$ we have
black objects both with event horizons of topology $S^{d-1}$
and $S^{d-2} \times S^1$.

Gregory and Laflamme \cite{Gregory:1993vy}
found that a neutral black string wrapping a cylinder is
classically unstable if its mass is sufficiently small.
Since the entropy of a black hole
with the same mass is larger it was conjectured that
the black string decays to a black hole.
This obviously involves a transition from a black object of
topology $S^{d-2} \times S^1$ to one of topology $S^{d-1}$.
However, recently Horowitz and Maeda \cite{Horowitz:2001cz}
argued that an event horizon cannot have a collapsing
circle in a classical evolution.
So it is not possible for the black string to change
the topology of the event horizon.
This lead them to conjecture that there
exist new classical solutions with
event horizon of topology $S^{d-2} \times S^1$
that are non-translationally invariant along the circle
whenever the black string is classically unstable.

While the Gregory-Laflamme and Horowitz-Maeda papers concerned
themselves with stability of black strings of small masses,
we shall in this paper mainly concern ourselves with
stability of black strings with large masses in which case
the black strings are in fact classically stable.

The basic claim of this paper is that
when a black hole on a cylinder grows sufficiently large
so that its event horizon shifts its
topology to $S^{d-2} \times S^1$ the solution corresponding
to this is {\sl not} the black string solution.
To be more precise,
let $M_c$ be the critical mass of a black hole on a cylinder
so that the topology of the horizon of the black hole is
$S^{d-1}$ for $M < M_c$ and $S^{d-2} \times S^1$ for $M > M_c$.
Then we claim that solutions with mass $M > M_c$ are
non-translationally invariant along the circle and are thus
not black strings.

To support this claim, we take in this paper steps toward
constructing explicit solutions for black holes on cylinders
$\R^{d-1} \times S^1$ for $d \geq 4$ %
\footnote{Explicit solutions have been constructed for black
holes on $\R^2 \times S^1$
\cite{Myers:1987rx,Bogojevic:1991hv,Lu:1997kg,Lavrinenko:1998rc}.
However, the methods used
there are highly particular to that case and cannot be generalized
to $\R^{d-1} \times S^1$ for $d \geq 4$.}.
We study in this paper three classes of black holes on cylinders.
The non-extremal charged dilatonic branes of
String/M-theory with a transverse circle%
\footnote{We can consider the $p$-brane as a black hole
if we ignore the $p$ directions on the world-volume of
the $p$-brane. Concretely, one can compactify the $p$-branes
of String/M-theory on $T^p$. This is reviewed in
Appendix \ref{appextrsol}.
In this paper we will in this spirit loosely refer to $p$-branes with
$d$ transverse directions as black holes in a $(d+1)$-dimensional
space-time.},
the near-extremal limit of these branes and finally
neutral black holes on cylinders.
One of the main results of this paper is that the same
construction applies to all three classes of black holes.

As part of the construction we find a new coordinate system in which we
are able to conjecture a general ansatz for non-extremal charged dilatonic
branes with a transverse circle.
We show that the EOMs imply that the ansatz in fact is fully described
by only one function.
Moreover, we find that this function is independent of the charge
which means that
we can map the ansatz for non-extremal charged dilatonic
branes with a transverse circle to an ansatz for the near-extremal
limit of those and, moreover,
to an ansatz for neutral black holes on cylinders.

The existence of the new solutions has the consequence
that for $M > M_c$ we have two different black objects
with the same horizon topology:
The black strings, which are translationally
invariant along the circle, and our new solutions, which are
not translationally invariant along the circle.
The natural question is therefore which of these solutions
has the highest entropy for a given mass.
We argue via our construction that
the new solutions have larger entropy than the black strings.
This means that for $M > M_c$ the black string can gain entropy
by spontaneously breaking the translational invariance and
redistributing its mass according to our new solution.
We have thus found a new instability of the black string for
large mass $M > M_c$ which is not classical in nature.

Generalizing the argument for the neutral case,
we show that our solution for
non-extremal charged dilatonic $p$-branes on a transverse circle
has higher entropy than that of
non-extremal charged dilatonic $p$-branes smeared on the
transverse circle for given mass and charge.
Moreover, the near-extremal
charged dilatonic $p$-branes on a transverse circle
has higher entropy than that of
near-extremal charged dilatonic $p$-branes smeared on the
transverse circle for given mass.
So, similarly to what happens for the neutral black string
the smeared non-extremal and near-extremal
branes will gain entropy by breaking the
translational invariance along the circle.

Finally, as an example of an application of our construction
to String Theory,
we show how to use it to study thermodynamics of Little String Theory
\cite{Seiberg:1997zk,Berkooz:1997cq,Dijkgraaf:1997ku}.
We argue that one in principle can use this to study the phase
transition between $(2,0)$ Super Conformal Field Theory
\cite{Strominger:1996ac,Witten:1995zh}
and $(2,0)$ Little String Theory.
We find two possible scenarios for the thermodynamics both of which
have important consequences for the understanding of thermodynamics
of Little String Theory and the near-extremal NS5-brane.

This paper is organized as follows.
In Section \ref{secnewcoord} we define our new coordinate system
on cylinders $\R^{d-1} \times S^1$ and discuss its properties.
In Section \ref{secbigansatz} we discuss the conditions that we
impose on our construction
and we put forward an ansatz for it in
the case of non-extremal charged dilatonic $p$-branes
with a transverse circle.
In Section \ref{secmapsol} we show that one can map our ansatz
to ans\"atze for near-extremal $p$-branes and neutral
black holes on cylinders.
In Section \ref{secsmallBH} we discuss small black holes on cylinders
and check that our ansatz is consistent in that limit.
In Section \ref{secfindsol} we consider the EOMs for the ansatz,
discuss their general structure and their consistency.
In Section \ref{secthermodynamics} we discuss thermodynamics of
our new solutions. We begin by discussing the
killing horizon and the surface gravity. Then we derive the general
expressions for the thermodynamics
and discuss in detail what we can say about the thermodynamics
of our new solutions.
Finally, in Section \ref{secLST} we apply our results and
methods to the supergravity dual of thermal Little String Theory.
In Section \ref{secconc} we discuss our results and draw conclusions.

A number of appendices has been included, supplying the discussion
of the text with further details. In Appendix \ref{appextrsol}
we recall the extremal and non-extremal
charged dilatonic $p$-brane solutions and some related results, such
as compactification of the solution and the thermodynamics. Appendix
\ref{appfctF} gives the mathematical details of two functions
that play an essential role in the new coordinate system.
In Appendix \ref{appcoord} the first few terms in a large radius
expansion in the map
from cylindrical coordinates to the new coordinates is worked out
and used to obtain the corresponding expansion of two functions
that enter the flat metric in the new coordinates. Then, Appendix
\ref{appdet} gives some of the details that are relevant to the
solution of black holes on cylinders in a large radius expansion.
Finally, Appendix \ref{appM5NS5} summarizes in our notation
the M5 and NS5-brane backgrounds that are needed for the application
of our development to Little String Theory.

\newpage

\section{A new coordinate system}
\label{secnewcoord}

Our goal in this section and in Section \ref{secbigansatz}
is to find an ansatz for non-extremal charged dilatonic $p$-branes with
transverse space $\R^{d-1} \times S^1$, or, equivalently%
\footnote{See Appendix \ref{appextrsol}
for a discussion of the dimensional reduction of
the charged dilatonic branes to black holes.},
for charged dilatonic black holes on $\R^{d-1} \times S^1$.

Finding solutions of black holes on the $d$-dimensional
cylinder $\R^{d-1} \times S^1$ involves solving
highly complicated nonlinear equations.
To see this, we can consider the covering space $\R^d$ of the cylinder.
On the covering space $\R^d$ a black
hole in $\R^{d-1} \times S^1$ is really a one-dimensional
array of black holes. Since the interactions between black holes
are in general non-linear, the geometry is very complicated once
the back-reaction is included.
Another way to see the complication is to note that neither
the spherical symmetry nor the cylindrical symmetry applies
in general for such a black hole solution. Clearly we are
forced to consider a solution with functions that depend on
two coordinates rather than one, contrary to the spherically
and cylindrically symmetric solutions.

As will be discussed
in Section \ref{secbigansatz}, an essential ingredient
in finding such an ansatz is the requirement that
the solution should interpolate between the usual black brane
with transverse space $\R^d$, which is a good description at
small mass, and the black brane smeared on the transverse circle,
which is a good description at large mass.
We furthermore demand that the solution should reduce to
the extremal charged dilatonic $p$-branes with
transverse space $\R^{d-1} \times S^1$ for zero temperature.

In order to capture these features in an ansatz
we must therefore find an appropriate coordinate system that can be used
in both the small and large mass limits and
also for the extremal solution.
Finding such a coordinate system is
the goal of this section.
Here and in the following we denote the radius of the $S^1$ as $R_T$.

\subsection{Defining the new coordinates}
\label{secdefnewc}


\smallsec{Spherical and cylindrical coordinates}

We first review the coordinate systems used for the limiting cases.
The spherical coordinates on $\R^{d}$ have the metric
\begin{equation}
\label{sphemet}
ds_d^2 = d \rho^2 + \rho^2 d\Omega_{d-1}^2
= d \rho^2 + \rho^2 d\theta^2 + \rho^2 \sin^2 \theta d\Omega_{d-2}^2
\end{equation}
where $0 \leq \theta \leq \pi$.
These are the coordinates used when the mass of the black hole is small, i.e.
with a Schwarzschild radius much smaller than $R_T$.
They can obviously only be used as coordinates on $\R^{d-1} \times
S^1$ when $\rho \ll R_T$.
The cylindrical coordinates on $\R^{d-1} \times S^1$ have the metric
\begin{equation}
ds_d^2 = dr^2 + dz^2 + r^2 d\Omega_{d-2}^2
\end{equation}
where $0 \leq z \leq 2 \pi R_T$.
These are the coordinates used when the mass of the black hole is large,
i.e. with a Schwarzschild radius much larger than $R_T$.

We note that the coordinate transformation between
$(\rho,\theta)$ and $(r,z)$ is
\begin{equation}
r = \rho \sin \theta
\spa
z = \rho \cos \theta
\end{equation}
\begin{equation}
\rho = \sqrt{r^2 + z^2}
\spa
\theta = \arctan \left( \frac{r}{z} \right) \ .
\end{equation}
For completeness, we also specify that the $S^{d-2}$
has angles $\phi_1,...,\phi_{d-2}$ with the spherical metric
\begin{equation}
d\Omega_{d-2}^2 = d\phi_1^2 + \sin^2 \phi_1 d\phi_2^2 +
\cdots + \sin^2 \phi_1 \sin^2 \phi_2 \cdots \sin^2 \phi_{d-1} d\phi_{d-2}^2 \ .
\end{equation}

\smallsec{Defining the coordinate $R$}

Our aim is to find a coordinate system that
in a convenient way interpolates between the spherical coordinates
$(\rho,\theta)$ and the cylindrical coordinates $(r,z)$.

To find this coordinate system
we first consider the extremal dilatonic $p$-brane
solutions of Appendix \ref{appextrsol}
with transverse space $\R^{d-1} \times S^1$.
The metric is
\begin{equation}
\label{ordextrmet}
ds_{D}^2 = H^{-\frac{d-2}{D-2}}
\left[ - dt^2 + \sum_{i=1}^p (dx^i)^2
+ H \Big( dr^2 + dz^2 + r^2 d\Omega_{d-2}^2 \Big) \right]
\end{equation}
while the dilaton $\phi$
and  one-form potential $A_{p+1}$ are given by
\begin{equation}
e^{2\phi} = H^{a}
\spa
\label{potA}
A_{01 \cdots p} = 1 - H^{-1}  \ .
\end{equation}
Here, the harmonic function is
\begin{equation}
\label{theHarm}
H = 1
+ \sum_{n \in \Z} \frac{L^{d-2}}{(r^2 + (z+2\pi n R_T)^2)^{\frac{d-2}{2}}}
\end{equation}
which can be written as
\begin{equation}
H = 1 + \frac{L^{d-2}}{R_T^{d-2}}
F_{d-2} \Big( \frac{r}{R_T} , \frac{z}{R_T} \Big)
\end{equation}
where $F_{2s} (a,b)$ is defined in Appendix \ref{appfctF}.

In Appendix \ref{appfctF} we derive various
properties of $F_{2s} (a,b)$, for example
it follows from \eqref{F2sexp} that
\begin{equation}
\label{theFF}
F_{d-2}\Big( \frac{r}{R_T} , \frac{z}{R_T} \Big) = k_d
\left( \frac{r}{R_T} \right)^{-(d-3)} \left[
1 + \sum_{n=1}^\infty f_d \Big( n \frac{r}{R_T} \Big)
\cos \Big( n \frac{z}{R_T} \Big) \right]
\end{equation}
with the definitions%
\footnote{The relation with $\hat f_s$ and $\hat k_s$ in \eqref{fsdef}
is $f_d \equiv \hat f_{s=(d-2)/2}$ and $k_d \equiv \hat k_{s=(d-2)/2}$.}
\begin{equation}
\label{fddef}
 f_d (y) \equiv \frac{\sqrt{2}}{2^{(d-6)/2}} \frac{1}
{\Gamma( \frac{d-3}{2}) } y^{(d-3)/2}
K_{(d-3)/2} ( y)
\end{equation}
\begin{equation}
k_d \equiv \frac{1}{2\sqrt{\pi}}
\frac{\Gamma \left( \frac{d-3}{2}\right)}{\Gamma \left( \frac{d-2}{2}\right)}
= \frac{1}{2\pi} \frac{(d-2) \Omega_{d-1}}{(d-3)\Omega_{d-2}}
\end{equation}
where $K_{s}$ is the modified Bessel function of the second kind
and $\Omega_n$ is the volume of the unit $n$-sphere.
In particular, it follows from \eqref{theFF} that for  $r/R_T \gg 1$ one finds the
leading behavior
\begin{equation}
F_{d-2} \Big( \frac{r}{R_T} , \frac{z}{R_T} \Big)
\simeq k_d
\left( \frac{r}{R_T} \right)^{-(d-3)}
\end{equation}
while for $\sqrt{r^2+z^2}/R_T \ll 1$ we have
\begin{equation}
F_{d-2} \Big( \frac{r}{R_T} , \frac{z}{R_T} \Big)
\simeq \left[ \frac{\sqrt{r^2+z^2}}{R_T} \right]^{-(d-2)}
= \left( \frac{\rho}{R_T} \right)^{-(d-2)}  \ .
\end{equation}
{} From these two results we see that if we define a new coordinate
as a function of $F_{d-2} (r/R_T,z/R_T)$ then this coordinate
will interpolate between being a function of $r$ for $r/R_T \gg 1$ and
being a function of $\rho$ for $\sqrt{r^2+z^2}/R_T \ll 1$.

Thus, we define a new coordinate $R(r,z)$ by
\begin{equation}
\label{Rdef}
R^{d-3}
= \frac{k_d}{F_{d-2} \Big( \frac{r}{R_T} , \frac{z}{R_T} \Big)} \ .
 \end{equation}
We then see that for $r/R_T \gg 1$ we have
\begin{equation}
\label{Rbigr}
R \simeq \frac{r}{R_T}
\end{equation}
while for $\sqrt{r^2+z^2}/R_T \ll 1$ we have
\begin{equation}
\label{Randrho}
R \simeq k_d^{\frac{1}{d-3}}
\left( \frac{\rho}{R_T} \right)^{\frac{d-2}{d-3}}  \ .
\end{equation}
So, as desired, the new coordinate $R$ interpolates between being a
function of  $r$ for $r/R_T \gg 1$ and
a function of $\rho$ for $\sqrt{r^2+z^2}/R_T \ll 1$.

Moreover, we see that the
harmonic function $H$ in \eqref{theHarm} is solely a function
of $R$. It thus follows from \eqref{potA} that the
surfaces defined by constant $R$ are precisely the equipotential surfaces.
This makes it a natural coordinate to use for the solution.

\smallsec{Defining the coordinate $v$}

To complete the new coordinate system we need a coordinate
that interpolates between $z$ and $\theta$ in the same way as $R$ interpolates
between $r$ and $\rho$. We now find this coordinate by imposing
that the metric should be diagonal in the new coordinate system,
and by demanding that the coordinate becomes $z/R_T$ for $r/R_T \gg 1$.
Denoting the new coordinate by $v$, we thus demand
\begin{equation}
\label{diagcond1}
A_{R} dR^2 + A_{v} dv^2 = dr^2 + dz^2
\end{equation}
for some functions $A_{R}$ and $A_{v}$.
Since it follows from  \eqref{Rdef} that $dR/R$ is proportional to
$dF_{d-2}/F_{d-2}$ we find that \eqref{diagcond1} is equivalent to
\begin{equation}
\label{diagcond2}
A_{F} (dF_{d-2})^2 + A_{v} dv^2 = dr^2 + dz^2 \ .
\end{equation}
Clearly, this is only possible provided
\begin{eqnarray}
\label{threeconds1}
1 & = & A_{F} \left( \frac{\partial F_{d-2}}{\partial r} \right)^2
+ A_{v} \left( \frac{\partial v}{\partial r} \right)^2
\\
1 & = & A_{F} \left( \frac{\partial F_{d-2}}{\partial z} \right)^2
+ A_{v} \left( \frac{\partial v}{\partial z} \right)^2
\\
\label{threeconds3}
0 &=& A_{F} \frac{\partial F_{d-2}}{\partial r} \frac{\partial F_{d-2}}{\partial z}
+ A_{v} \frac{\partial v}{\partial r} \frac{\partial v}{\partial z} \ .
\end{eqnarray}
If we now suppose that
\begin{equation}
\label{supv}
\frac{\partial v}{\partial r} = m \frac{\partial F_{d-2}}{\partial z}
\spa
\frac{\partial v}{\partial z} = - m \frac{\partial F_{d-2}}{\partial r}
\end{equation}
with $m$ being an undetermined function, then
the three equations \eqref{threeconds1}-\eqref{threeconds3}
are satisfied, provided $A_{F} = m^2 A_{v}$.
If $v(r,z)$ is to be a well-defined function we need
that
\begin{equation}
\frac{\partial^2 v}{\partial r \partial z}
= \frac{\partial^2 v}{\partial z \partial r}
\end{equation}
so combining this with \eqref{supv} we obtain the condition
\begin{equation}
\label{diagcond3}
\frac{\partial}{\partial r} \left( m \frac{\partial F_{d-2}}{\partial r} \right)
+ \frac{\partial}{\partial z} \left( m \frac{\partial F_{d-2}}{\partial z} \right)
= 0 \ .
\end{equation}
Clearly, this is only possible to satisfy if $m$ is proportional to
$r^{d-2}$ since then \eqref{diagcond3} is equivalent to
the harmonic equation $\nabla^2 F_{d-2} = 0$ (where $\nabla^2 =
\partial_r^2 + \frac{d-2}{r} \partial_r  + \partial_z^2 $ in cylindrical
coordinates).

Therefore, we define the new coordinate $v(r,z)$
by the integrable system
\begin{equation}
\label{pavr}
\frac{\partial v}{\partial r}
= \frac{1}{(d-3)k_d}  \left( \frac{r}{R_T} \right)^{d-2}
\frac{\partial F_{d-2}}{\partial z}
\end{equation}
\begin{equation}
\label{pavz}
\frac{\partial v}{\partial z}
= - \frac{1}{(d-3)k_d} \left( \frac{r}{R_T} \right)^{d-2}
\frac{\partial F_{d-2}}{\partial r} \ .
\end{equation}
We can in fact write an explicit expression for
$v(r,z)$ using \eqref{theFF} (see Appendix \ref{appfctF}, where we consider
the function $G_{d-2}(\frac{r}{R_T},\frac{z}{R_T})$
which is proportional to $v(r,z)$). We find
\begin{equation}
\label{vexpress}
v = \frac{z}{R_T} + \sum_{n=1}^\infty \sin\left( n \frac{z}{R_T} \right)
\left[ \frac{1}{n} f_d \left(n\frac{r}{R_T} \right) - \frac{1}{d-3}
\frac{r}{R_T} f_d'\left(n \frac{r}{R_T} \right)
  \right]
\end{equation}
where the function $f_d$ is defined in \eqref{fddef}.
We see that $v(r,z + 2\pi R_T) = 2\pi + v(r,z)$.

We observe from \eqref{vexpress} that for $R \gg 1$ we can write
\begin{equation}
\label{vbigR}
v \simeq \frac{z}{R_T}
\end{equation}
while for $R \ll 1$ it follows from integrating \eqref{pavr},\eqref{pavz} that
\begin{equation}
\label{vsmallR}
v \simeq \pi - \frac{d-2}{d-3} k_d^{-1} \int_{\theta'=0}^\theta d\theta'
 (\sin \theta')^{d-2} \ .
\end{equation}
We note that $\theta=0$ corresponds to $v=\pi$ and $\theta=\pi$ corresponds
to $v=-\pi$, so the interval $\theta \in [0,\pi]$ is mapped one-to-one to
$v \in [-\pi,\pi]$. Thus, $v$ is not a periodic coordinate. However,
we see from \eqref{sphemet}
that the metric can be thought of as being
periodic in $v$ with periodic $2\pi$.

As promised, we see from \eqref{vbigR} and \eqref{vsmallR} that
$v$ interpolates between a function of $z$ for $R \gg 1$
and a function of $\theta$ for $R \ll 1$.

\subsection{Critical curve}
\label{seccritcurv}

\begin{figure}[ht]
\begin{center}
\includegraphics{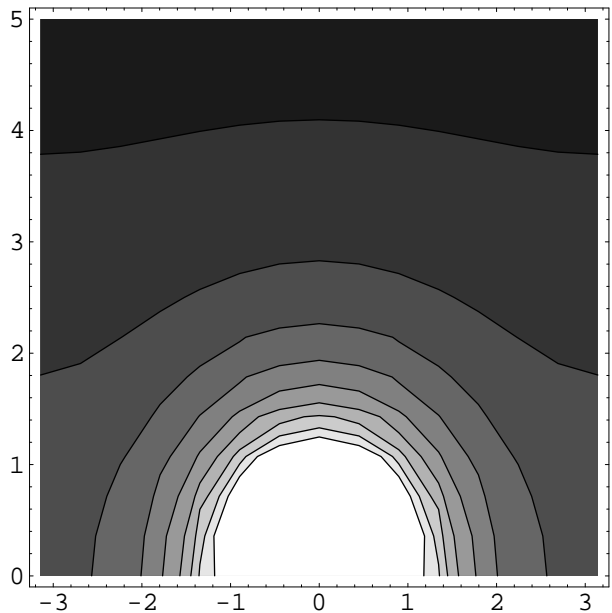}
\caption{\small Equipotential lines for $F_2(a,b)$ corresponding to $d=4$.
Horizontal axis is $b$ and vertical axis is $a$.}
\label{fig1}
\end{center}
\end{figure}

As an illustration of the $(R,v)$ coordinate system, we have depicted
the equipotential lines for $F_{d-2}$ in Figure \ref{fig1}.
Clearly the $v$-coordinate
is periodic with period $2\pi$ for large $R$ and
not periodic for small $R$. We now find the critical $R$ value
where $v$ goes from being periodic to not being periodic.
It is clear from Figure \ref{fig1} that the critical curve that separates
$v$ being periodic or not, is the curve that goes through the point
$(r=0,z=\pi R_T)$. Computing the value of $F_{d-2}$ in this
point we find
\begin{equation}
F_{d-2} (0,\pi) = \frac{2}{\pi^{d-2}} \left( 1 - \frac{1}{2^{d-2}} \right)
\zeta ( d-2 ) \ .
\end{equation}
Then using the definition of $R$ \eqref{Rdef}  we obtain the
critical value $R_c$ of the $R$ coordinate as
\begin{equation}
R_c^{d-3} = \frac{1}{2} k_d \frac{(2\pi)^{d-2}}{2^{d-2}-1}
\frac{1}{\zeta(d-2)} \ .
\end{equation}

For $R < R_c$ we see that $v$ is not periodic and that the range
of $v$ is $[-\pi,\pi]$. For $R > R_c$ we have that $v$ is periodic
with period $2\pi$ and we can for instance choose
the range $[-\pi,\pi [$.
At the critical value $R=R_c$ we see that $v$ is periodic with
period $2\pi$, but the coordinate map has a singularity in $v=\pi$,
corresponding to the point $(r=0,z=\pi R_T)$.

\subsection{The flat metric}
\label{secflatmet}

Finally, using the coordinates \eqref{Rdef}, \eqref{pavr}, \eqref{pavz}
the flat metric on the $d$-dimensional cylinder $\R^{d-1} \times S^1$
can be written as
\begin{equation}
\label{newflat}
ds_d^2 = R_T^2 \left( A_{(0)} dR^2 + \frac{A_{(0)}}{(K_{(0)})^{d-2}} dv^2
+ K_{(0)} R^2 d\Omega_{d-2}^2 \right)
\end{equation}
where
\begin{equation}
\label{A0}
A_{(0)} = \frac{1}{R_T^2} \frac{(d-3)^2}{k_d^{\frac{2}{d-3}}}
\frac{F_{d-2}^{2 \frac{d-2}{d-3}}}{ (\partial_r F_{d-2})^2 +
(\partial_z F_{d-2})^2 }
\end{equation}
\begin{equation}
\label{K0}
K_{(0)} = \left(\frac{r}{R_T} \right)^2 \left(\frac{F_{d-2}}{k_d}
\right)^{\frac{2}{d-3}} \ .
\end{equation}
It is important to note here that since the functions $r(R,v)$, $F_{d-2}(R,v)$,
$\partial_r F_{d-2} (R,v)$ and $(\partial_z F_{d-2}(R,v))^2$
are periodic in $v$ with period $2\pi$, then
$A_{(0)}(R,v)$ and $K_{(0)}(R,v)$ are periodic in $v$
with period $2\pi$. Thus, even though $v$ is not a periodic coordinate
for $R < R_c$ we find that
the flat metric \eqref{newflat} is periodic in $v$ with
period $2\pi$ for all $R$.
Note also that the functions $A_{(0)}(R,v)$ and $K_{(0)}(R,v)$, and thereby
the flat metric \eqref{newflat},  are smooth
on the space given by $R \in [0,\infty [ $ and $v \in \, ]-\pi,\pi [$.

{}From the fact that  $A_{(0)}(R,v)$ and $K_{(0)}(R,v)$ are
even periodic functions in $v$
with period $2\pi$, follows that we can write the Fourier
expansions
\begin{equation}
\label{K0fourier}
K_{(0)}(R,v) = \sum_{n=0}^\infty \cos(nv) \, L_0^{(n)} (R)
\end{equation}
\begin{equation}
\label{A0fourier}
A_{(0)}(R,v) = \sum_{n=0}^\infty \cos(nv) \, B_0^{(n)} (R) \ .
\end{equation}
We emphasize that this expansion holds for all $(R,v)$.
The functions $L_0^{(n)} (R)$ and $B_0^{(n)} (R)$ are considered
in Appendix \ref{appcoord} for $n=1,2$, by working out the
$(R,v) \leftrightarrow (r,z)$ change of coordinates for large $R$
through second order.

The flat metric \eqref{newflat}
can be used to write the extremal dilatonic $p$-brane
solution \eqref{ordextrmet}-\eqref{theHarm} as
\begin{eqnarray}
\label{extrmet}
ds_{D}^2 &=& H^{-\frac{d-2}{D-2}}
\left[ - dt^2 + \sum_{i=1}^p (dx^i)^2 \right.
\nn \\ &&
\left. + H R_T^2
\left( A_{(0)} dR^2 + \frac{A_{(0)}}{(K_{(0)})^{d-2}} dv^2
+ K_{(0)} R^2 d\Omega_{d-2}^2 \right) \right]
\end{eqnarray}
\begin{equation}
e^{2\phi} = H^a
\spa
A_{01\cdots p} = 1 - H^{-1}
\end{equation}
with harmonic function
\begin{equation}
\label{extrend}
H = 1 + \frac{h_d}{R^{d-3}}
\spa
h_d \equiv k_d \frac{L^{d-2}}{R_T^{d-2}} \ .
\end{equation}

\section{Ansatz}
\label{secbigansatz}

In this section we find an ansatz that we believe can describe
non-extremal charged dilatonic branes with a transverse circle.
In order to do this, we first make precise what the conditions
on such an ansatz should be.

Part of these conditions is to demand that the solution
should interpolate between the usual black brane
with transverse space $\R^d$, which is a good description at
small mass, and the black brane smeared on the transverse circle,
which is a good description at large mass.
We are thus advocating the philosophy
that a small black hole on a cylinder can be continuously
deformed into a black string wrapped on the cylinder by
increasing the mass ad infinitum.

As mentioned before, one of the complications
in finding solutions of black holes on the cylinder
$\R^{d-1} \times S^1$ is that on its
covering space $\R^d$ the configuration is really a one-dimensional
array of black holes. Hence, due to the non-linear nature of
the interactions among the black hole, the geometry is expected
to be very complicated.
On the other hand, it is physically clear that such solutions should
exist, at least for small black holes.
One could naively think that they would be unstable since on
the covering space a slight perturbation of one of the black holes
would destroy the array. However, the black holes are constrained
to be at a fixed distance which is what removes that instability.

In the literature, only black holes on $\R^2 \times S^1$
have been considered previously \cite{Myers:1987rx,Bogojevic:1991hv,Lu:1997kg,Lavrinenko:1998rc}. This case
is very different from the generic case since the curvature
term of the $S^1$ symmetry of the $\R^2$ in the usual spherical
ansatz drops out of the Einstein equations.
It is therefore possible in this case to reduce the Einstein
equations to exactly solvable linear equations.

The class of black holes we are considering in this section
are the singly charged dilatonic black holes
that correspond to the $p$-branes of M-theory
and Type IIA/IIB String Theory compactified
in the longitudinal directions on a $p$-torus.
When discussing the solutions we discuss the $p$-brane
solutions rather than the compactified black hole solutions
(See Appendix \ref{appextrsol} for details on $p$-brane solutions of
String and M-theory and for the compactification on $T^p$).

We shall see that for this class of black holes
it is possible to make an ansatz for the solution that
reduces the EOMs (equations of motion) into three equations for one function
of two variables. This ansatz is constructed using the new coordinate
system found in Section \ref{secnewcoord}, and is tailored to fulfil the
appropriate boundary conditions.

\subsection{Conditions on a solution}
\label{secconditions}

In order to make an ansatz for the solution of dilatonic $p$-branes
with transverse space $\R^{d-1} \times S^1$
we need to determine the boundary conditions
that should be imposed.

As a preliminary, we define a general Schwarzschild radius
$R_0$ to be the maximal value of the $R$ coordinate on the horizon in
the $(R,v)$ coordinates.

We have three types of boundary conditions we want
to impose. First, we have the parts of the metric
that are independent of the charge of the solution.
For these, we impose the boundary conditions:

\begin{itemize}

\item[{\bf (i)}] The solution reduces
to an ordinary black $p$-brane with transverse space
$\R^{d}$ when $R_0 \leq R \ll 1$.

\item[{\bf (ii)}] The solution reduces to a black $p$-brane
smeared on the transverse circle when $R \geq R_0 \gg 1$.

\end{itemize}

Condition (i) tells us that for small $R_0$ we can
ignore the size of the transverse circle and regard it as non-compact
so that the solution should be the one corresponding to
transverse space $\R^d$.
In $(R,v)$ coordinates, this solution is
\begin{eqnarray}
\label{smallBHsol1}
ds_{D}^2 &=& H^{-\frac{d-2}{D-2}}
\left[ - f dt^2 + \sum_{i=1}^p (dx^i)^2 \right.
\nn \\ &&
\left. + H R_T^2 \left( f^{-1} A_{(0)}  dR^2
+ \frac{A_{(0)}}{(K_{(0)})^{d-2}}  dv^2
+ K_{(0)} d\Omega_{d-2}^2 \right) \right]
\end{eqnarray}
\begin{equation}
e^{2\phi} = H^{a}
\spa A_{01 \cdots p}= \coth \alpha \Big( 1 - H^{-1} \Big)
\end{equation}
\begin{equation}
\label{smallBHsol2}
f = 1 - \frac{R_0^{d-3}}{R^{d-3}}
\spa
H = 1 + \frac{R_0^{d-3} \sinh^2 \alpha}{R^{d-3}}
\end{equation}
with
\begin{equation}
\label{A0smallR}
A_{(0)} = \left(\frac{d-3}{d-2} \right)^2
\left(\frac{R_T}{k_d \rho} \right)^{\frac{2}{d-3}}
= \left(\frac{d-3}{d-2} \right)^2 (k_d R)^{- \frac{2}{d-2}}
\end{equation}
\begin{equation}
\label{K0smallR}
K_{(0)} = \frac{\rho^2}{R_T^2} \sin^2 \theta
\end{equation}
and  $\alpha$ is defined by
\begin{equation}
\label{alpdef1}
h_d = R_0^{d-3} \cosh \alpha \sinh \alpha \ .
\end{equation}
To obtain this result we used the non-extremal charged dilatonic
$p$-brane solution \eqref{nemet}-\eqref{nefH} and the limiting
coordinate transformations  $R(\rho)$ and $v (\theta)$ given in
\eqref{Randrho} and \eqref{vsmallR} respectively.
This solution is valid in $(R,v)$ coordinates for $R_0 \leq R \ll 1$.

Condition (ii) says that for large $R_0$ we can ignore the
coordinate in the circle direction, and thus the solution is
that of a $p$-brane smeared in one direction.
In $(R,v)$ coordinates, this solution is
\begin{equation}
\label{largeBHsol1}
ds_{D}^2 = H^{-\frac{d-2}{D-2}}
\left[ - f dt^2 + \sum_{i=1}^p (dx^i)^2 + H R_T^2
\left( f^{-1} dR^2 + dv^2
+ R^2 d\Omega_{d-2}^2 \right) \right]
\end{equation}
\begin{equation}
e^{2\phi} = H^{a}
\spa
A_{01 \cdots p}
= \coth \alpha \Big( 1 - H^{-1} \Big)
\end{equation}
\begin{equation}
\label{largeBHsol2}
f = 1 - \frac{R_0^{d-3}}{R^{d-3}}
\spa
H = 1 + \frac{R_0^{d-3} \sinh^2 \alpha}{R^{d-3}}
\end{equation}
with $\alpha$ defined again by \eqref{alpdef1}. This results follows
using the smeared non-extremal charged dilatonic $p$-brane solution
in \eqref{nemets}-\eqref{nefHs} and the limiting coordinate
transformations $R(r)$ and $v(z)$ given in \eqref{Rbigr} and \eqref{vbigR}
respectively.
This solution is valid in $(R,v)$ coordinates for $R \geq R_0 \gg 1$.

In addition to the two preceding conditions,
we want to ensure that the general black solution is the thermally
excited version of the extremal charged dilatonic $p$-brane
on $\R^{d-1} \times S^1$.
Hence, we impose the condition:

\begin{itemize}

\item[{\bf (iii)}] For $R_0/R \rightarrow 0$ the solution approaches the
solution \eqref{extrmet}-\eqref{extrend} corresponding to an extremal
dilatonic $p$-brane with transverse space $\R^{d-1} \times S^1$.

\end{itemize}

This condition has several important consequences.
First, we see that for $R_0 = 0$ the solution
should reduce to the extremal dilatonic $p$-brane
with transverse space $\R^{d-1} \times S^1$
given by \eqref{extrmet}-\eqref{extrend}.

Moreover, if we consider a small $R_0$, i.e. a small black hole on a
cylinder, the solution should approximately look like the extremal solution
\eqref{extrmet}-\eqref{extrend}. In Section \ref{secsmallBH} we take a closer look at the case of
small black holes on cylinders.

If we instead consider a fixed $R_0$, then condition (iii) has the
consequence that for $R/R_0 \rightarrow \infty$ the solution should approach
the extremal solution given by \eqref{extrmet}-\eqref{extrend}.
This ensures that the reference space of the black hole is
the extremal dilatonic $p$-brane on $\R^{d-1} \times S^1$
given by \eqref{extrmet}-\eqref{extrend}.

A corollary to this last remark is that for a given finite $R_0 > R_c$ the
solution cannot be translationally invariant along the circle
since for sufficiently large $R$ it has to be approximately equal
to the extremal solution \eqref{extrmet}-\eqref{extrend}
which is not translationally invariant along the circle.
We are thus forced to discard the usual smeared black $p$-brane solution
for finite $R_0 > R_c$ as the exact solution. The smeared black
$p$-brane solution is only exact for $R_0 \rightarrow \infty$.

The three preceding conditions are not enough. We need in addition
to specify a location of the horizon. We therefore assume the extra
condition:

\begin{itemize}

\item[{\bf (iv)}] The horizon is located at constant $R$.

\end{itemize}

The rationale behind this condition is that the equipotential surfaces
of the charge potential are defined by $R$ being constant
and we expect the horizon to be at an equipotential surface%
\footnote{This holds also for spinning brane solutions
(See for example \cite{Harmark:1999xt}).}.
Obviously, the horizon is then defined by the equation $R=R_0$.

In Section \ref{secsmallBH} we consider an additional condition on the
$g_{00}$ component of the metric for solutions
with $R_0 \ll 1$.

\subsection{The ansatz}
\label{secansatz}

We are now ready to specify our ansatz for the
dilatonic black $p$-brane with transverse space $\R^{d-1} \times S^1$.
In accordance with the conditions above we write
\begin{eqnarray}
\label{metans1}
ds_{D}^2 &=& H^{-\frac{d-2}{D-2}}
\left[ - f dt^2 + \sum_{i=1}^p (dx^i)^2 \right.
\nn \\ &&
\left. + H R_T^2
\left( f^{-1} A dR^2 + C dv^2
+ K R^2 d\Omega_{d-2}^2 \right) \right]
\end{eqnarray}
\begin{equation}
e^{2\phi} = H^{a}
\spa
A_{01 \cdots p}
= \coth \alpha \Big( 1 - H^{-1} \Big)
\end{equation}
along with the functions
\begin{equation}
\label{theH}
f = 1 - \frac{R_0^{d-3}}{R^{d-3}}
\spa
H = 1 + \frac{R_0^{d-3} \sinh^2 \alpha}{R^{d-3}}
\end{equation}
and
\begin{equation}
h_d = R_0^{d-3} \cosh \alpha \sinh \alpha \ .
\end{equation}
Here, we introduced the three
undetermined functions $A(R,v)$, $C(R,v)$ and $K(R,v)$.

In accordance with condition (iv) above,
the equation $R = R_0$ defines the horizon of the black hole.
The conditions (i)-(iii) above are satisfied provided
the functions
$A$, $C$ and $K$ reduce to their extremal
values if we consider the limit
$R_0/R \rightarrow 0$ or the limit $R_0 \gg R_c$.
Apart from the requirements coming from the conditions,
we have also imposed that the metric is diagonal.
Though it is not a priori obvious that this is possible,
we shall present strong evidence below that this in fact gives
consistent EOMs.

We can be even more restrictive in the ansatz for the metric.
If we consider the EOM for the field strength
\begin{equation}
\partial_\mu ( \sqrt{-g} e^{a\phi} F^{\mu \nu} ) = 0
\end{equation}
we get
\begin{equation}
\partial_R \left[ \sqrt{\frac{C K^{d-2}}{A}} R^{d-2} \partial_R H  \right]
= 0\ .
\end{equation}
Using \eqref{theH} we see that this requires
$C K^{d-2} / A$ to be independent of $R$. It then follows
from the boundary conditions above that $C = A K^{-(d-2)}$.
The ansatz for charged dilatonic black holes
on $\R^{d-1} \times S^1$ therefore becomes
\begin{eqnarray}
\label{metans2}
ds_{D}^2 &=& H^{-\frac{d-2}{D-2}}
\left[ - f dt^2 + \sum_{i=1}^p (dx^i)^2 \right.
\nn \\ &&
\left. + H R_T^2
\left( f^{-1} A dR^2 + \frac{A}{K^{d-2}} dv^2
+ K R^2 d\Omega_{d-2}^2 \right) \right]
\end{eqnarray}
\begin{equation}
e^{2\phi} = H^{a}
\spa
A_{01 \cdots p}
= \coth \alpha \Big( 1 - H^{-1} \Big)
\end{equation}
\begin{equation}
\label{fandH}
f = 1 - \frac{R_0^{d-3}}{R^{d-3}}
\spa
H = 1 + \frac{R_0^{d-3} \sinh^2 \alpha}{R^{d-3}}
\end{equation}
\begin{equation}
\label{endansa}
h_d = R_0^{d-3} \cosh \alpha \sinh \alpha
\end{equation}
with only two undetermined functions $A(R,v)$
and $K(R,v)$ at this point.
Below we find $A(R,v)$ in terms of $K(R,v)$ so
that ultimately the ansatz \eqref{metans2}-\eqref{endansa}
has only one undetermined function $K(R,v)$.

\section{Map of solution to neutral and near-extremal solutions}
\label{secmapsol}

In this section we use the fact that the equations for $A(R,v)$ and
$K(R,v)$ are independent of the charge parameter $h_d$ to map
the ansatz for non-extremal charged dilatonic branes with
a transverse circle to an ansatz for neutral black holes on
cylinders and to near-extremal charged dilatonic branes with
a transverse circle.

\subsection{Neutral black holes on a cylinder}
\label{neutralmap}

It can be checked that the EOMs for the ansatz \eqref{metans2}-\eqref{endansa}
are independent of the constant $h_d$ which is proportional to the charge.
Thus, the EOMs are the same for
a neutral non-dilatonic black hole on $\R^{d-1} \times S^1$
with metric%
\footnote{We have omitted the longitudinal directions of the
$p$-brane since these are trivial when the charge is zero.}
\begin{equation}
\label{neutBH}
ds_{d+1}^2 = - f dt^2 + R_T^2 \left[ f^{-1} A dR^2 + \frac{A}{K^{d-2}} dv^2
+ K R^2 d\Omega_{d-2}^2 \right]
\end{equation}
where
\begin{equation}
\label{moreneut}
f = 1 - \frac{R_0^{d-3}}{R^{d-3}} \ .
\end{equation}
The boundary conditions are then that $A \rightarrow A^{(0)}$
and $K \rightarrow K^{(0)}$ when%
\footnote{We note here that the condition for $R_0 \rightarrow 0$
is equivalent to $A \rightarrow A^{(0)}$
and $K \rightarrow K^{(0)}$ for $R/R_0 \rightarrow \infty$, i.e.
in the asymptotic region far away from the black hole.}
$R_0 \rightarrow 0$ or
$R_0 \rightarrow \infty$ for any $R$.
These boundary conditions are also natural for this neutral
black hole case, as they express the conditions that we want
a) the solution to reduce to a black hole solution
on $\R^d$ for $R_0 \rightarrow 0$; b)
the solution to reduce to a black string solution for
$R_0 \rightarrow \infty$; and finally c)
the solution to reduce to the flat space metric
on $\R^{d-1} \times S^1$ in $(R,v)$ coordinates when
$R_0 = 0$ or in the asymptotic region $R/R_0 \rightarrow \infty$,
so that our solution is asymptotically $\R^{d-1} \times S^1$ very
far away from the black hole.

Consequently, the problem of finding solutions of black dilatonic $p$-brane
with transverse space
$\R^{d-1} \times S^1$ is mapped to the problem of finding
neutral black holes on $\R^{d-1} \times S^1$.

\subsection{Near-extremal branes on transverse circle}
\label{secnemap}

The above stated fact that the black hole structure of
the solution of a $p$-brane on a transverse circle
is independent of the charge, means that we also can map the
non-extremal charged dilatonic $p$-brane solutions
to the corresponding near-extremal dilatonic $p$-brane solutions
with a transverse circle.

The general near-horizon limit of the non-extremal
$p$-brane ansatz \eqref{metans2}-\eqref{endansa} is
\begin{equation}
R_T \rightarrow 0 \spa
R \ \mbox{ fixed} \spa
R_0 \ \mbox{ fixed} \spa
\hat{h}_d \equiv R_T^2 h_d \  \mbox{ fixed} \ .
\end{equation}
The resulting near-extremal $p$-brane solution is then
\begin{eqnarray}
\label{NEmet}
R_T^{-\frac{2(d-2)}{D-2}} ds_{D}^2 &=& \hat{H}^{-\frac{d-2}{D-2}}
\left[ - f dt^2 + \sum_{i=1}^p (dx^i)^2 \right.
\nn \\ &&
\left. + \hat{H}
\left( f^{-1} A dR^2 + \frac{A}{K^{d-2}} dv^2
+ K R^2 d\Omega_{d-2}^2 \right) \right]
\end{eqnarray}
\begin{equation}
R_T^{2a} e^{2\phi} = \hat{H}^{a}
\spa
R_T^{-2} A_{01 \cdots p}
= - \hat{H}^{-1}
\end{equation}
\begin{equation}
\label{NEfct}
f = 1 - \frac{R_0^{d-3}}{R^{d-3}}
\spa
\hat{H} = \frac{\hat{h}_d}{R^{d-3}}
\end{equation}
where the functions $A(R,v)$ and $K(R,v)$ are the same as
in the ansatz \eqref{metans2}-\eqref{endansa}.
So, as promised, the ``black part'' of the near-extremal
solution is the same as that of the corresponding
non-extremal and neutral solutions.

\section{Newton limit of small black holes on cylinders}
\label{secsmallBH}

In this section we examine the limit of small black holes
on cylinders, i.e. the limit $R_0 \ll R_c$.
This is done for two purposes.
Firstly, we want to test the ansatz \eqref{metans2}-\eqref{endansa}
in this limit and verify that this case can be correctly
incorporated.
We find that the results are indeed consistent and that the
ansatz \eqref{metans2}-\eqref{endansa} works in this case.
Secondly, we shall see that the results of this section
have important consequences for the form of the
general solution.
We restrict ourselves to the neutral case with
ansatz \eqref{neutBH}-\eqref{moreneut} in this section
but all the results can trivially be extended to the charged
case due to the map discussed in Section \ref{neutralmap}.

\subsection{The Newton limit of Einsteins equation}
\label{secnewteins}

We consider Einsteins equations
in a $(d+1)$-dimensional space-time,
\begin{equation}
\label{einst}
R_{\mu \nu} - \frac{1}{2} g_{\mu \nu} R = 8 \pi G_{d+1} T_{\mu \nu}
\spa \mu,\nu = 0,1,...,d
\end{equation}
with a weak gravitational field
\begin{equation}
|g_{\mu \nu} - \eta_{\mu \nu} | \ll 1
\end{equation}
where $\eta_{\mu \nu} = \mbox{diag}(-1,1,...,1)$ is the Minkowski metric.
Other types of flat space coordinates will be considered below.
We also impose that the metric is static so that $\partial_t g_{\mu \nu} = 0$
and $g_{0i} = 0$, $i =1,...,d$ and hence  $R_{0i} = 0$.
We consider non-relativistic matter
\begin{equation}
T_{00} = \varrho
\spa
|T_{ij}| \ll \varrho \ \ , \ i,j = 1,...,d
\end{equation}
where $\varrho$ is the density of mass.
{}From the above equations we find to leading order
\begin{equation}
\label{R00}
R^0_{\ 0} = - \frac{d-2}{d-1} 8 \pi G_{d+1} \varrho
\spa
R^i_{\ j} = \delta^i_{j} \frac{1}{d-1} 8 \pi G_{d+1} \varrho \ .
\end{equation}
The Geodesic equation in a weak gravitational field
gives to leading order
\begin{equation}
\frac{\partial^2 x^i}{\partial t^2} = \frac{1}{2} \partial_i g_{00} \ .
\end{equation}
Comparing with Newtons Second law in a Newton gravitational potential $\Phi$
\begin{equation}
\frac{\partial^2 x^i}{\partial t^2} = - \partial_i \Phi
\end{equation}
we identify
\begin{equation}
\partial_i g_{00} = - 2 \partial_i \Phi \ .
\end{equation}
Since $g^{(0)}_{00} = - 1 $ we then see that we must have
\begin{equation}
\label{g00}
g_{00} = - ( 1 + 2\Phi )
\end{equation}
to leading order.

{}From the definition of the Ricci tensor we have
$R^0_{\ 0} = \frac{1}{2} \nabla^2 g_{00}$, where
$\nabla^2 = \partial^i \partial_i$.
Using \eqref{R00} and \eqref{g00}, we then find that Newtons equation
for the Newton gravitational potential is%
\footnote{This can of course be derived independently of the
Einstein equations.}
\begin{equation}
\label{newtpot}
\nabla^2 \Phi = 8 \pi G_{d+1} \varrho \frac{d-2}{d-1} \ .
\end{equation}
Thus, in terms of the gravitational potential we have
the equations
\begin{equation}
\label{newteinst}
R^0_{\ 0} = - \nabla^2 \Phi
\spa
R^i_{\ j} = \delta^i_{j} \frac{1}{d-2} \nabla^2 \Phi \ .
\end{equation}
We can instead consider other flat coordinates
so that the metric to leading order can be written
\begin{equation}
g_{\mu \nu} = g^{(0)}_{\mu \nu} + g^{(1)}_{\mu \nu} + \cdots
\end{equation}
where $g^{(0)}_{\mu \nu}$ is the flat space metric
in  the coordinates under consideration
(with $g^{(0)}_{00} = -1$ and $g^{(0)}_{0i} = 0$)
and $g^{(1)}_{\mu \nu}$ the leading correction expressed
in these new coordinates. In this more general case the equations
\eqref{newteinst} still hold for the metric $g_{\mu \nu}$
to leading order, provided the Laplacian is taken in its covariant
form
\begin{equation}
\nabla^2 \Phi = \frac{1}{\sqrt{-g^{(0)}}}
\partial_\mu \left( \sqrt{-g^{(0)}} \partial^\mu \Phi \right) \ .
\end{equation}
Clearly, also the relation \eqref{g00} holds for other choices of
flat coordinates.

\subsection{The Newton potential and the $g_{00}$ component}
\label{secg00}

In this section we describe how the standard connection \eqref{g00}
between the $g_{00}$ component of the metric and the Newton
potential works for small black holes using
the ansatz \eqref{neutBH}-\eqref{moreneut}.

If we consider a point mass of mass $M$ in flat space $\R^{d}$
we get from the equation for a Newtonian gravitational
potential \eqref{newtpot} that
\begin{equation}
\Phi = - \frac{8 \pi G_{d+1} M }{(d-1) \Omega_{d-1} r^{d-2}} \ .
\end{equation}
If we instead consider a point mass of mass $M$ on the cylinder
$\R^{d-1} \times S^1$ we obtain using
the superposition principle the potential
\begin{eqnarray}
\label{gravpot}
\Phi &=& - \frac{8\pi G_{d+1} M}{(d-1)\Omega_{d-1} } \sum_{n \in \Z}
\frac{1}{(r^2 + (z + 2 \pi n R_T )^2 )^{(d-2)/2}}
\nn \\
&=& - \frac{8\pi G_{d+1} M}{(d-1)\Omega_{d-1} R_T^{d-2} }
F_{d-2} \left( \frac{r}{R_T} , \frac{z}{R_T} \right) \ .
\end{eqnarray}
Using now the connection \eqref{g00} between $\Phi$  and $g_{00}$ and
in the Newtonian limit, we find that
\begin{equation}
\label{newtong00}
g_{00} = - 1 + \frac{16\pi G_{d+1} M}{(d-1)\Omega_{d-1} R_T^{d-2} }
F_{d-2} \left( \frac{r}{R_T} , \frac{z}{R_T} \right)
\end{equation}
for a point mass of mass $M$ on a cylinder $\R^{d-1} \times S^1$.

We can now compare with our ansatz \eqref{neutBH}-\eqref{moreneut}
which has
\begin{equation}
\label{ourg00}
g_{00} = - 1 + \frac{R_0^{d-3}}{R^{d-3}}
= -1 + R_0^{d-3} k_d^{-1}  F_{d-2} \left( \frac{r}{R_T} , \frac{z}{R_T} \right)
\ .
\end{equation}
We see that the ansatz precisely has the right form of $g_{00}$
to reproduce the gravitational potential \eqref{gravpot} and \eqref{newtong00}
for $R \gg R_0$. That $g_{00}$ should have this form for $R_0 \ll R_c$
is an additional condition for black hole solutions on cylinders
independent of the conditions of Section \ref{secconditions}.

Moreover, by comparing \eqref{newtong00} and \eqref{ourg00} we
can determine the mass
\begin{equation}
\label{smallBHmass}
M = \frac{(d-3)(d-1)}{d-2}
\frac{\Omega_{d-2} 2\pi R_T}{16 \pi G_{d+1}} (R_0 R_T)^{d-3} \ .
\end{equation}
This shows that the mass $M$ of a small black hole on the cylinder
$\R^{d-1} \times S^1$ is the same as the mass of a black
hole with the same horizon radius $R_0$ but in $\R^d$
($\R^d$ is obtained if we impose that \eqref{Randrho} holds exactly
for all $R$).
This means that the mass of the black hole is not affected by the
global structure of the space surrounding it, provided it is sufficiently
small. So a black hole obeys the locality principle in this respect.
This not a completely trivial result in the sense
that energy in General Relativity only can be defined globally.
However, this result is to be expected if the weak gravitational
region around a black hole should behave like that of Newtonian
gravity.

\subsection{The metric in the Newton limit}
\label{secnewton}

We now want to describe how, given a gravitational
potential $\Phi = \Phi(R)$, the leading order correction
to the metric is determined.
In \eqref{g00} this was given for the $g_{00}$ component.
However, this is highly gauge-dependent, so
we need to fix the gauge by writing an ansatz
for the corrections.
Since we want this ansatz to reduce to the weak gravitational
field limit of \eqref{neutBH}-\eqref{moreneut}
when $\Phi = - \frac{1}{2} \frac{R_0^{d-3}}{R^{d-3}}$
we write the ansatz
\begin{eqnarray}
\label{newtans}
ds_{d+1}^2 &=& - \Big(1+2\Phi \Big) dt^2
+ R_T^2 \left[ \Big( 1 - 2 u + 2g \Big) A_{(0)} dR^2
\right. \nn \\ && \left.
+ \Big( 1 + 2g - (d-2) 2h \Big) \frac{A_{(0)}}{K_{(0)}^{d-2}} dv^2
+ \Big( 1 + 2h \Big) K_{(0)} R^2 d\Omega_{d-2}^2 \right]
\end{eqnarray}
where $u$, $g$ and $h$ are undetermined functions.
The idea is now to find $u$, $g$ and $h$ as functions of $\Phi$ and
its derivatives so that the approximate Einstein equations
\eqref{newteinst} are satisfied.

Since the right hand side of \eqref{newteinst} has at most
two derivatives of $\Phi$ we expect that the metric can be
written in terms of $\Phi$ and $\Phi'$ only, since
e.g. a $\Phi''$ term in the metric would give $\Phi'''$ terms or
higher in the
Ricci tensor.
Thus, we should find $u$, $g$ and $h$ as function of
$R$, $v$, $\Phi$ and $\Phi'$. Clearly they have to be
linear combinations of $\Phi$ and $\Phi'$ since we only
consider leading corrections.
For $u$ this has the immediate consequence that
\begin{equation}
\label{ulamb}
u = (1-\lambda) \Phi - \lambda \frac{R}{d-3} \Phi'
\end{equation}
for some function $\lambda = \lambda(R,v)$.
This is because we want to impose that
$u = - \frac{1}{2} \frac{R_0^{d-3}}{R^{d-3}}$
whenever $\Phi = - \frac{1}{2} \frac{R_0^{d-3}}{R^{d-3}}$.

\smallsec{The metric for $R \gg R_c$}

For $R \gg R_c$ we have $A_{(0)} = K_{(0)} = 1$, so that the ansatz
\eqref{newtans} becomes
\begin{eqnarray}
ds_{d+1}^2 &=& - \Big(1+2\Phi \Big) dt^2
+ R_T^2 \left[ \Big( 1 - 2 u + 2g \Big) dR^2
\right. \nn \\ && \left.
+ \Big( 1 + 2g - (d-2) 2h \Big) dv^2
+ \Big(1 + 2h \Big) R^2 d\Omega_{d-2}^2 \right] \ .
\end{eqnarray}
{}From the Einstein equation $(d-2) R^v_{\ v} - \nabla^2 \Phi = 0$
we get
\begin{equation}
\nabla^2 \left[ (d-2) h - g - \frac{1}{d-2} \Phi \right] = 0 \ .
\end{equation}
Clearly, in order for this to hold for general potentials $\Phi$
we need
\begin{equation}
g = (d-2) h - \frac{1}{d-2} \Phi \ .
\end{equation}
We now write
\begin{equation}
h = h_1 \Phi - h_2 \frac{R}{d-3} \Phi'
\end{equation}
where $h_1$ and $h_2$ are constants. Using this
along with \eqref{ulamb} with $\lambda$ a constant
we find from the remaining Einstein equations
$(d-2) R^R_{\ R} = \nabla^2 \Phi$
and $(d-2) R^{\phi_1}_{\ \phi_1} = \nabla^2 \Phi$
that $h_1 = 1/((d-2)(d-3))$, $h_2 = 0$ and $\lambda = 1$.
Thus, we have
\begin{equation}
\label{res1}
u = - \frac{R}{d-3} \Phi'
\spa
h = \frac{1}{(d-2)(d-3)} \Phi
\spa
g = \frac{1}{(d-2)(d-3)} \Phi
\end{equation}
which uniquely determines the metric \eqref{newtans} in terms of the Newton
gravitational potential.
However, we have neglected above to explain why we can assume
$h_1$, $h_2$ and $\lambda$ to be constants. As we shall see below
at least $h_1$ is dependent on $R$. But, since we
neglect any correction to the metric of order $\exp(-R)$
our assumption that $h_1$, $h_2$ and $\lambda$ are constants
is really an assumption that they are constants up to
corrections of order $\exp(-R)$.
This is just another way of saying that we have obtained
the leading contribution to $h_1$, $h_2$ and $\lambda$ for $R \gg R_c$.

\smallsec{The metric for $R_0 \ll R \ll R_c$}

We now turn to the region $R_0 \ll R \ll R_c$.
Here it is convenient to work in the $(\rho,\theta)$ coordinates
in terms of which the ansatz \eqref{newtans} becomes
\begin{eqnarray}
ds_{d+1}^2 &=& - \Big(1+2\Phi \Big) dt^2
+ R_T^2 \left[ \Big( 1 - 2 u + 2g \Big) d\rho^2
\right. \nn \\ && \left.
+ \Big( 1 + 2g - (d-2) 2h \Big) \rho^2 d\theta^2
+ \Big( 1 + 2h \Big) \rho^2 \sin^2 \theta d\Omega_{d-2}^2 \right] \ .
\label{newtanss}
\end{eqnarray}
The $R_{\rho \theta}=0$ equation gives
$\partial_\rho g = (d-1) \partial_\rho h$, so that
$g = (d-1) h$. This means that the metric is spherically
symmetric.
All the Einstein equations are then solved if and only if
\begin{equation}
u = (d-2) h - \rho \partial_\rho h - \frac{\rho}{d-2} \partial_\rho \Phi \ .
\end{equation}
Since $u = - \frac{1}{2} \frac{\rho_0^{d-2}}{\rho^{d-2}}$
whenever $\Phi  = - \frac{1}{2} \frac{\rho_0^{d-2}}{\rho^{d-2}}$
we see that $h = 0$ since the above equation would otherwise
determine that $h$ is proportional to $\rho^{d-2}$.
Thus, we have
\begin{equation}
u = - \frac{\rho}{d-2} \partial_\rho \Phi
\spa
h = g = 0
\end{equation}
which determines the metric \eqref{newtanss} in this case.
In the $(R,v)$ coordinates these relations read
\begin{equation}
\label{res2}
u = - \frac{R}{d-3} \Phi'
\spa
h = g = 0 \ .
\end{equation}
%

\subsection{Consequences for small black hole on cylinder}
\label{secconsBH}

{}From \eqref{neutBH}-\eqref{moreneut} it follows that the ansatz for a
black hole on a cylinder $\R^{d-1} \times S^1$ is
\begin{equation}
ds_{d+1}^2 = - f dt^2 + R_T^2 \left[ f^{-1} A dR^2 + \frac{A}{K^{d-2}} dv^2
+ K R^2 d\Omega_{d-2}^2 \right]
\end{equation}
\begin{equation}
f = 1 - \frac{R_0^{d-3}}{R^{d-3}} \ .
\end{equation}
We first observe, as has already been remarked before, that this
metric fits into the general ansatz for the Newton limit
\eqref{newtans}
on a cylinder with
\begin{equation}
\label{phiak}
\Phi = u = - \frac{1}{2} \frac{R_0^{d-3}}{R^{d-3}}
\spa
A = A_{(0)} ( 1 + 2g )
\spa
K = K_{(0)} ( 1 + 2h ) \ .
\end{equation}
The results of Section \ref{secnewton} now give the two
limiting cases
\begin{eqnarray}
\label{thehg1}
  h = g = 0 &  & \mbox{for} \;\;R_0 \ll R \ll R_c \\
\label{thehg2}
2h = 2g = - \frac{1}{(d-2)(d-3)} \frac{R_0^{d-3}}{R^{d-3}}
 & & \mbox{for} \;\;R \gg R_c \ .
\end{eqnarray}

That $h = g = 0$ for $R_0 \ll R \ll R_c$ justifies our condition
(i) in
Section \ref{secconditions} that the solution we consider should
reduce to the solution with transverse space $\R^d$ when
$R_0 \leq R \ll R_c$. Essentially this means that the small black hole
is so small that in its close vicinity the cylindrical geometry
can be ignored. Again, this can be seen as a reflection
of the fact that the black hole
obeys the locality principle in the sense that
the black hole solution near the black
hole is not affected by the asymptotic structure
of the space-time.

That $h$ and $g$ are given by \eqref{thehg2} for $R \gg R_c$
means that the black hole on a cylinder has a potential
term in the metric at infinity. This term expresses the attraction
of the black hole to itself across the cylinder.
Moreover, in the following section we shall see that this term
is needed in order to compute the right value for the mass.
This is thus another consistency check on our ansatz
\eqref{metans2}-\eqref{endansa}.

\subsection{Consequences for measurements of mass}
\label{secconmass}

We have already seen how to compute the mass of the black hole
solution via the $g_{00}$ component in Section \ref{secg00}.
In this section we test our above results for the Newton limit
of the metric on the cylinder $\R^{d-1} \times S^1$ by
measuring the mass via the Hawking-Horowitz mass formula \cite{Hawking:1996fd}
\begin{equation}
\label{HHM}
M = - \frac{1}{8\pi G_{d+1}} \int \frac{d\Omega}{\sqrt{g_{\Omega}}}
\int^{2\pi}_0 dv N \sqrt{g_{d-1}} ( K-K_0 ) \Big|_{R=R_m} \ .
\end{equation}
Here we have evaluated the mass at $R = R_m \gg R_0$.
$N$ is the lapse function which is the extremal value of
$\sqrt{-g_{00}}$ which for the case at hand is equal to one.
$\sqrt{g_{d-1}}$ is the square root of the metric on the space
of constant $t$ and $R=R_m$.
$K$ is the extrinsic curvature given by
\begin{equation}
K = \frac{\left( \sqrt{g_{d-1}} \right)'}{\sqrt{g_d}}
\end{equation}
where $\sqrt{g_{d}}$ is the square root of the metric on the space
of $R=R_m$. $K_0$ is the extremal value of $K$.
In terms of the metric \eqref{newtans} the mass \eqref{HHM} becomes
\begin{equation}
\label{hhmass}
M = \frac{\Omega_{d-2}}{8\pi G_{d+1}}
R_T^{d-2} R_m^{d-2} \int^{2\pi}_0 dv \left[ - g'
- u \left( \frac{1}{2}\frac{A_{(0)}'}{A_{(0)}} + \frac{d-2}{R_m} \right)
 \right] \ .
\end{equation}
We now evaluate this mass for the small black hole on the cylinder.
{} From \eqref{phiak} we know we should set
$\Phi = u = - \frac{1}{2} \frac{R_0^{d-3}}{R^{d-3}}$.
We also calculate
\begin{equation}
\label{aovera}
\frac{1}{2\pi} \int^{2\pi}_0 dv \frac{1}{2}\frac{A_{(0)}'}{A_{(0)}}
= \left\{ \begin{array}{ccc} - \frac{1}{d-2} \frac{1}{R}
& \mbox{ for } & R_0 \ll R \ll R_c \\
0 & \mbox{ for } & R \gg R_c \end{array}   \right.
\end{equation}
where we used \eqref{A0smallR}, while using \eqref{thehg1}, \eqref{thehg2}
 we have that
\begin{equation}
\label{gprime}
g' = \left\{ \begin{array}{ccc}
0 & \mbox{ for } & R_0 \ll R \ll R_c \\
\frac{1}{2(d-2)} \frac{1}{R} \frac{R_0^{d-3}}{R^{d-3}}
& \mbox{ for } & R \gg R_c \ .
\end{array}   \right.
\end{equation}
Putting \eqref{aovera} and \eqref{gprime} into \eqref{hhmass} we
obtain for both $R_0 \ll R_m \ll R_c$ and $R_m \gg R_c$
the mass
\begin{equation}
M = \frac{(d-3)(d-1)}{d-2}
\frac{\Omega_{d-2} 2 \pi R_T}{16 \pi G_{d+1}} (R_0 R_T)^{d-3} \ .
\end{equation}
Since this is the same result as \eqref{smallBHmass},
we have successfully checked that our Newton limit results are consistent
with the mass measured at $R_0 \ll R \ll R_c$ or $R \gg R_c$.

One can also put the results \eqref{res1}
and \eqref{res2} of Section \ref{secnewton} into
\eqref{hhmass} and verify
that for both $R_0 \ll R_m \ll R_c$ and $R_m \gg R_c$ we get
the same result in terms of the potential $\Phi$.

\section{Finding solutions}
\label{secfindsol}

\subsection{Equations of motion}
\label{seceoms}

We now find the EOMs for the ansatz \eqref{metans2}-\eqref{endansa}.
As stated above, these can be found directly from the metric
\eqref{neutBH} of
a neutral non-dilatonic black hole on $\R^{d-1} \times S^1$.
The EOMs are then given by $R_{\mu \nu} = 0$.
We get four non-trivial EOMs, $R_{RR} = 0$, $R_{vv} = 0$ , $R_{Rv} = 0$
and $R_{\phi_1 \phi_1} =0$.
These four EOMs are
\begin{eqnarray}
\label{eqRRR}
0 &=& (R^{d-3} - R_0^{d-3}) \Big[ R a'' -(d-2)(R k' +1) a'  +\frac{1}{2}(d-2)
((d-1)Rk' +4)k') \Big]
\nn \\ &&
+ e^{(d-2)k} R^{d-2} ( (d-2) \dot{a} \dot{k} + \ddot{a})
\end{eqnarray}
\begin{eqnarray}
\label{eqRvv}
0 &=& (R^{d-3} - R_0^{d-3})R (a''  - (d-2) k'') + ((d-2) R^{d-3} - R_0^{d-3})
 (a' - (d-2) k')
\nn \\ &&
 + e^{(d-2)k} R^{d-2} \Big( \ddot{a} + (d-2) \ddot{k} +
 \frac{1}{2}(d-1)(d-2) \dot{k}^2 \Big)
\end{eqnarray}
\begin{eqnarray}
\label{eqRRv}
0 &=& (R^{d-3} - R_0^{d-3}) \Big( R [\dot{k} a' + (\dot{a}  - (d-1) \dot{k})k' -
2 \dot{k}'] - 2 \dot{k} \Big)
\nn \\ &&
+\frac{1}{(d-2)} \Big( 2(d-2) R^{d-3} - (d-1) R_0^{d-3} \Big) \dot{a}
\end{eqnarray}
\begin{eqnarray}
\label{eqRpp}
0 &=& (R^{d-3} - R_0^{d-3}) R k'' + ((d-2) R^{d-3} - R_0^{d-3}) k'\label{eom4}
\nn \\ &&
+ e^{(d-2)k} R^{d-2} ( \ddot{k} + (d-2) \dot{k}^2)+
2 (d-3) R^{d-4} ( 1 - e^{a-k})
\end{eqnarray}
here written in terms of the functions $a(R,v)$ and $k(R,v)$ defined by
\begin{equation}
A(R,v) = \exp( a(R,v) )
\spa
K(R,v) = \exp( k(R,v) ) \ .
\end{equation}
We now see from \eqref{eom4}, which comes from $R_{\phi_1 \phi_1} = 0$,
that we can find $A(R,v)$ in terms of $K(R,v)$ as
\begin{eqnarray}
\label{AfromK}
A &=& e^k + \frac{e^k}{2(d-3)R^{d-4}} \Big[
(R^{d-3} - R_0^{d-3}) R k'' + ((d-2) R^{d-3} - R_0^{d-3}) k'
\nn \\ &&
+ e^{(d-2)k} R^{d-2} ( \ddot{k} + (d-2) \dot{k}^2)
\Big] \ .
\end{eqnarray}
This means the Ansatz \eqref{metans2}-\eqref{endansa} only has
one unknown function $K(R,v)$.
We can then substitute \eqref{AfromK} in
\eqref{eqRRR}-\eqref{eqRRv} and thereby we obtain three equations
for $K(R,v)$ %
\footnote{We have not written these three equations here since they are quite
complicated.}.
Due to the complexity of the three equations we have not been able
to show that they pose a consistent integrable set of equations.
However, in Section \ref{secaneom} we show that they are consistent
to second order when making an expansion of $K(R,v)$ for large $R$.
Moreover, we consider the three equations on the horizon in Section
\ref{secprophor} and show consistency also in that case.

\subsection{General considerations}
\label{secgencons}

In the previous section we established that the solution is
determined by only one function $K(R,v)$
and that the EOMs gave three equations determining this function.
Here we comment on the boundary conditions we put on
$K(R,v)$ and we also provide an argument for existence
of solutions with $R_0 > R_c$ that are
non-translationally invariant along the $z$ direction.

The condition (iii) of Section \ref{secconditions} can now
be formulated as
\begin{equation}
\label{kcond}
K(R,v) \rightarrow K_{(0)}(R,v) \ \ \mbox{for} \ \
\frac{R_0}{R} \rightarrow 0 \ .
\end{equation}
This means that $K(R,v)$ should have an expansion in powers of
$R_0^{d-3}/R^{d-3}$ with $K_{(0)}(R,v)$ being the zeroth order
term.

In Section \ref{secflatmet} we established that the flat
metric in $(R,v)$ coordinates is periodic for all $v$,
even when $R \leq R_c$.
We therefore impose this on the full non-extremal solution.
Thus, $K(R,v)$ is required to be periodic in $v$ with period
$2\pi$ for all $R$ and $R_0$.
Moreover, since $K_{(0)}(R,v)$ is an even function with respect
to $v$ we also impose that on the general $K(R,v)$ function.

These two conditions on $K(R,v)$ follow from symmetry
arguments. That $K(R,v)$ should be even in $v$ originates from
the fact that we want the space-time to be symmetric around $v=0$.
Since the metric is a measure of distance this means
it has to be even under $v \rightarrow -v$, and hence $K(R,v)$ should be even.
The periodicity of the metric for $R < R_c$ is then the
statement that $K(R,v)$ is the same for $v=\pi$ and $v=-\pi$ and
this in fact follows from the fact that $K(R,v)$ is even.

That $K(R,v)$ is an even periodic function of $v$ with period $2\pi$
means we can make a Fourier expansion of $K(R,v)$ as
\begin{equation}
\label{FourK}
K(R,v) = \sum_{n=0}^\infty \cos(nv) \, L^{(n)}(R) \ .
\end{equation}
For $K_{(0)} (R,v)$ we have from Section \ref{secflatmet} that
\begin{equation}
K_{(0)}(R,v) = \sum_{n=0}^\infty
\cos(nv) \, L_0^{(n)} (R) \ .
\end{equation}
Thus we require from \eqref{kcond}
that $L^{(n)}(R) \rightarrow L_0^{(n)}(R)$
for $R_0/R \rightarrow 0$.

To further clarify the boundary conditions on $K(R,v)$
we consider the expansion of $L^{(n)}(R)$ for $R \gg 1$.
For $K_{(0)} (R,v)$ we have from Appendix \ref{appcoord} that
\begin{equation}
\label{bdnm}
L_0^{(n)}(R) = \sum_{m=0}^\infty e^{- (n + 2m) R} \,
\tilde{L}_0^{(n,m)} (R) \spa
\tilde{L}_0^{(0,0)} = 1
\end{equation}
for $R \gg 1$.
In analogy with this we define
the functions $\tilde{L}^{(n,m)} (R)$ by
\begin{equation}
\label{Lnm}
L^{(n)} (R) = \sum_{m=0}^\infty e^{- (n + 2m) R} \,
\tilde{L}^{(n,m)} (R) \ .
\end{equation}
That \eqref{Lnm} is a warranted expression will be clear from the analysis
of Section \ref{secaneom}.

Apart from the boundary condition \eqref{kcond} that we already
discussed we can now formulate a second crucial boundary condition,
namely that
\begin{equation}
\tilde{L}^{(0,0)} (R) = 1 - \chi (R_0) \frac{R_0^{d-3}}{R^{d-3}} \ .
+ \cdots
\end{equation}
This means that for $R \gg 1$
the leading correction to $K(R,v)$ with respect to
$K_{(0)}(R,v)$ is precisely that term, so we can also write
\begin{equation}
\label{leadK}
K(R,v) = 1 - \chi (R_0) \frac{R_0^{d-3}}{R^{d-3}} \ .
+ \cdots
\end{equation}
This is how we impose $K(R,v)$ to behave at $R \rightarrow \infty$.
We expect this to be the right type of behavior of $K(R,v)$ for
$R \rightarrow \infty$ from the fact we have a term
like that for $R_0 \ll R_c$, as explained in Section \ref{secconsBH}.

In some sense the function $\chi(R_0)$ contains all physical information
about the solution, for example the entire thermodynamics can be derived from
it as will be explained in Section \ref{secthermo}.
We do not at present know $\chi(R_0)$, though we know the limiting
values
\begin{equation}
\label{knowchi}
\chi(R_0) = \left\{ \begin{array}{ccc}
\frac{1}{(d-2)(d-3)} & \mbox{for} & R_0 \ll R_c \\
0 & \mbox{for} & R_0 \gg R_c \end{array} \right.
\end{equation}
where the $R_0 \ll R_c$ value was obtained in Section \ref{secconsBH}.
However, as will be clarified in Section \ref{secaneom}, our
expansion \eqref{Lnm} means that $K(R,v)$ is now
in principle completely determined for a given value of $\chi(R_0)$.

We note that from \eqref{leadK} and \eqref{AfromK} we find
\begin{equation}
\label{ARlarge}
A(R,v) = 1 - \chi (R_0) \frac{R_0^{d-3}}{R^{d-3}}
+ \cdots
\end{equation}
as the leading correction with respect to $A_{(0)}(R,v)$ for $R \gg 1$.
This is of course consistent with \eqref{phiak} and
\eqref{thehg1}, \eqref{thehg2}.

\smallsec{Existence of solutions with $R_0 > R_c$}

We are now in the position to argue for the existence of solutions
with $R_0 > R_c$. Such solutions would necessarily be
non-translationally invariant along $z$, but the horizon
for such a solution clearly connects to itself across the
cylinder.
These solutions, if they exist,
are thus a new class of solutions that are neither black holes
nor black strings.

The argument is simple. On general physical ground it is safe to
assume that solutions exist for which $R_0 < R_c$.
This is because these solutions just correspond to black
holes on a $\R^{d-1} \times S^1$ cylinder.
Indeed, we have seen in Section \ref{secsmallBH} that our ansatz
\eqref{metans2}-\eqref{endansa} seems to be able to describe
black holes on cylinders.
Thus, the question is now whether there are any fundamental
differences in solving the EOMs for $R_0 < R_c$ and $R_0 > R_c$.
The answer is no. The boundary conditions are the same,
$K(R,v)$ is periodic and even function of $v$ and the only
difference between small and large $R_0$ is in the value of $\chi(R_0)$.
This is just a boundary condition at infinity which
means the EOMs should have solutions for any value of $\chi(R_0)$,
as will be further supported in Section \ref{secaneom}.

Therefore, it seems that in our description there is no
fundamental difference in solving the EOMs for $R_0 < R_c$ and $R_0 > R_c$
and the fact that black holes on cylinders exist means that
solutions with $R_0 > R_c$ also should exist.

Alternatively, if the statement of existence
and uniqueness stated below in Section \ref{secaneom} is
true (we can only partly verify it) we
trivially have that these solutions exist.

\subsection{Analysis of equations of motion}
\label{secaneom}

In order to clarify the preceding section, we summarize here
the boundary conditions and make a statement about
existence and uniqueness of solutions that will be justified below.

As shown in Section \ref{seceoms} we have three equations for $K(R,v)$.
We also recall that $K(R,v)$ completely determines the solution
\eqref{metans2}-\eqref{endansa}.
Consider now any given value of $R_0$ and $\chi$ (we treat $R_0$ and $\chi$
as independent parameters in this section).
The full set of boundary conditions is then
\begin{itemize}
\item
$K(R,v)$ is an even periodic function in $v$ with period $2\pi$.
\item
The Fourier components $L^{(n)}(R)$ of $K(R,v)$ obey
\begin{equation}
\label{Lnbc}
\frac{L^{(n)} (R)}{L_0^{(n)}(R)} \rightarrow 1 \ \ \mbox{for}\ \
\frac{R_0}{R} \rightarrow 0
\end{equation}
\item
For $R$ much greater than both $1$ and $R_0$ we have
\begin{equation}
\label{Kchi}
K(R,v) = 1 - \chi \frac{R_0^{d-3}}{R^{d-3}}
+ \cdots
\end{equation}
\end{itemize}
Our general statement about existence and uniqueness is then
\begin{itemize}
\item
For any given value of $R_0$ and any given value of $\chi \geq 0$
the three equations for $K(R,v)$ have a solution
obeying the above three boundary conditions and this solution is unique.
\end{itemize}
The justification of this statement is the subject of the rest of this
section.

We start by considering the expression \eqref{Lnm}, which implies that
the equations of motion allow
a well-defined expansion in terms of powers of
the ``expansion parameter'' $e^{-R}$. First, using \eqref{FourK}, \eqref{Lnm}
in the expression \eqref{AfromK} for $A(R,v)$, implies that
this function can similarly be written as
\begin{equation}
\label{FourA}
A(R,v) = \sum_{n=0}^\infty \cos(nv) \, B^{(n)}(R)
\end{equation}
\begin{equation}
\label{Bnm}
B^{(n)}(R) =  \sum_{m=0}^\infty e^{- (n + 2m) R} \,
\tilde{B}^{(n,m)} (R) \ .
\end{equation}
To see that \eqref{FourK}, \eqref{Lnm} and \eqref{FourA}, \eqref{Bnm}
are consistent with the remaining three EOMs
\eqref{eqRRR}-\eqref{eqRRv}, can be established by induction.
Here, the two central ingredients are the specific forms of the EOMs,
i.e. the way the $R$ and $v$ derivatives occur, along with
standard multiplicative properties of $\cos (mv) $ and $\sin (nv)$.

For example, looking at \eqref{eqRRv}, one observes the
structures $\dot{z}$, $\dot{z}'$ and $\dot{z}z'$, where $z$ stands for
$a$ or $k$. Thus, the expansions given above generate terms like
$\sin(nv)$ and  $\sin (nv) \cos (mv) \simeq \sin( (n+m)v )  + \sin( (n-m)v) $.
The latter relation is responsible for the fact that
\eqref{Lnm}, \eqref{Bnm} have even shifts $2m$ in the exponential.
Then, using induction, it is not difficult to establish the consistency
of our proposed expansion. The other two  EOMs \eqref{eqRRR}, \eqref{eqRvv}
can be examined similarly.

Moreover, the expansions \eqref{FourK}, \eqref{Lnm} and \eqref{FourA},
\eqref{Bnm} are also strongly suggested by the fact that they hold in
particular for the extremal functions $K_{(0)}$ and $A_{(0)}$
(see Appendix \ref{appcoord}). Finally, we explicitly show below
the consistency of the truncation by considering the first terms
in the expansion.

Using \eqref{Lnm}
the first three terms in $K(R,v)$ take the form
\begin{eqnarray}
K(R,v) & = &  \tilde L^{(0,0)} (R)  + \cos (v) e^{-R} \tilde L^{(1,0)} (R)
\nn \\
 & & + e^{-2R}  [ \cos (2v) \tilde L^{(2,0)} (R) +\tilde L^{(0,1)} (R)  ]
+ {\cal{O}} (e^{-3R})   \\
 & \equiv  &
y(R) + \cos (v) b(R) +  \cos (2v) q(R) + p(R) + {\cal{O}} (e^{-3R})
\label{ybqp}
\end{eqnarray}
where, for simplicity of notation, we have defined the functions
$y(R)$, $b(R)$, $q(R)$ and $p(R)$ in the second line.
We now examine the EOMs that arise from substituting these
first terms in the expansion.
Here and in the following
we always implicitly assume that the solution \eqref{AfromK} for
$A(R,v)$  is substituted in \eqref{eqRRR}-\eqref{eqRRv}.
Moreover, it turns out that there exists a linear combination of the
two equations \eqref{eqRRR} and \eqref{eqRvv}
that gives rise to a simpler equation (with only up to 3rd order
$R$-derivatives of $K(R,v)$), which is given by the difference of
these two equations. To facilitate the discussion, we denote the resulting
three EOMs symbolically by ${\cal{E}}_{i=1,2,3} $ with
\begin{eqnarray}
{\cal{E}}_1: & & \;\;\; R_{Rv} (A(K)) = 0 \label{E1} \\
{\cal{E}}_2:  & & \;\;\; R_{RR} (A(K))  - R_{vv}  (A(K)) = 0 \label{E2}  \\
{\cal{E}}_3:  & & \;\;\; R_{RR} (A(K)) = 0 \label{E3}
\end{eqnarray}
where the argument $A(K)$ expresses the fact that we have substituted
\eqref{AfromK}, and we recall that the Ricci components
$R_{RR}$, $R_{vv}$, $R_{Rv}$ are the right hand sides in \eqref{eqRRR},
\eqref{eqRvv} and \eqref{eqRRv} respectively.
As some of the details become rather involved
we will describe the resulting structure below, leaving most of the
details to Appendix \ref{appdet}. This appendix also gives
the corresponding expressions of the expansion of $A(R,v)$ that
follow from substituting \eqref{ybqp} in \eqref{AfromK}.

Starting with the leading $v$-independent term $y(R) =\tilde L^{(0,0)} (R) $
in \eqref{ybqp},
it is immediately clear from \eqref{eqRRv} that ${\cal{E}}_1$ is satisfied, leaving us with the
two equations ${\cal{E}}_{2,3}$. We first analyze ${\cal{E}}_2$, which
to leading order gives a non-linear differential equation on $y(R)$
which is quartic in $y$ and its derivatives and contains up to three
derivatives
\begin{equation}
\label{ydifeq0}
\sum_{ 0 \leq k \leq l \leq m \leq n \leq 3} c_{klmn}
y_k y_l y_m y_n = 0 \spa
y_m \equiv R^m \frac{\partial^m y(R)}{\partial R^m} \ .
\end{equation}
The ten non-zero coefficients $c_{klmn}$ which are functions of $x=R_0/R$ and
$d$ are given in \eqref{c0001}-\eqref{c0013}.
This differential equation can be solved
perturbatively for $R \gg R_0$ by  substituting the power series expansion
\begin{equation}
\label{ysol0}
y(R) = \tilde L^{(0,0)} (R)
= 1 + \sum_{n=1}^\infty \alpha_n \left( \frac{R_0}{R}
\right)^{(d-3)n}
\spa \alpha_1 = -\chi  \ .
\end{equation}
It turns out that the resulting solution is
uniquely determined given $\chi$. In \eqref{ysolc1}-\eqref{ysolc5} we have
given the explicit expressions for $\alpha_n$, $1 \leq n \leq 5$ in terms of
$\chi$ for arbitrary $d$. To verify consistency, we still need to
consider the remaining EOM ${\cal{E}}_3$. Indeed, it turns out that
the solution \eqref{ysol0}
 of ${\cal{E}}_2$ also solves ${\cal{E}}_3$. This is the first important
non-trivial check of our system of EOMs.
In all, we see that up to this point in the expansion we have indeed
verified the form \eqref{Kchi} of $K(R,v)$ for $ R \gg 1,R_0$ and
the claim that the solution is unique given $ R_0$ and  $\chi$.

We continue by studying the first exponentially suppressed correction
to $y(R)$, i.e. the $b(R)$-term in \eqref{ybqp}. Starting again
with the EOM ${\cal{E}}_1$, we now find a second order homogeneous
differential equation on $b(R)$
\begin{equation}
\label{bdifeq0}
M_0 (R,y(R)) b(R) + M_1 (R,y(R)) R b'(R) + M_2(R,y(R)) R^2 b''(R) =0
\end{equation}
where the coefficients $M_{m=0,1,2}$ depend
on $y(R)$ and its derivatives. The explicit form of this differential
equation is given in \eqref{bdifeq}. Though the algebra is highly
non-trivial, we have explicitly checked that the two equations on $b(R)$
resulting from the other two EOMs, ${\cal{E}}_2$ and ${\cal{E}}_3$
(which are in fact 3rd and 4th order homogeneous differential equations
on $b(R)$ with $y(R)$-dependent coefficients)  are indeed satisfied
given the $y$-equation \eqref{ydifeq0} {\it and} the $b$-equation \eqref{bdifeq0}.
Again, this is a rather non-trivial check on the consistency of
our system.

Two remarks are in order here.
First, since $y(R)$ is uniquely determined given $(R_0,\chi)$, the
equation on $b(R)$ is uniquely determined given $(R_0,\chi)$. Second,
the two boundary conditions that need to be fixed in order to
integrate the second order system \eqref{bdifeq} are fixed as a consequence
of the boundary conditions \eqref{Lnbc}. Indeed, this condition (for $n=1$)
represents a boundary condition on $b(R)$ and, by differentiation of
\eqref{Lnbc} also its first derivative%
\footnote{Note that the boundary conditions on all higher derivatives
do not present further constraints, as the reference solution $L_0^{(n)}(R)$
solves the EOMs.}.
In conclusion, we have verified the claim at the beginning of this section
to (and including) first order in $e^{-R}$.

Our final explicit computation involves the second order corrections
$q(R)$ and $p(R)$ in \eqref{ybqp}.
We first discuss $q(R)$ in which case we obtain
from ${\cal{E}}_1$ a second order inhomogeneous differential equation
with $y(R)$-dependent coefficients (as in the $b(R)$ equation).
The inhomogeneous part is quadratic in $b(R)$ (and its derivatives). This
equation is given in \eqref{qdifeq}.
Just as for the $b(R)$ equation,
we have explicitly checked that the two equations on $q(R)$
resulting from the other two EOMs, ${\cal{E}}_2$ and ${\cal{E}}_3$
(which are again 3rd and 4th order) are indeed satisfied
given the $y$-equation \eqref{ydifeq0}, the $b$-equation \eqref{bdifeq0}
and the $q$-equation \eqref{qdifeq}.

Turning to $p(R)$, since this is the order $e^{-2R}$ correction to
the $v$-independent leading term $y(R)$, we have again that ${\cal{E}}_1$
is  immediately satisfied. Consequently, the first non-trivial equation
results from ${\cal{E}}_2$, which gives   a third order inhomogeneous
differential equation (of similar form as the one for $q(R)$).
This equation is given
in \eqref{pdifeq}. Again, we have verified the non-trivial fact that
the other fourth order inhomogeneous differential equation on $p(R)$ coming
from ${\cal{E}}_3$ is satisfied given
the $y$-equation \eqref{ydifeq0}, $b$-equation \eqref{bdifeq0}
and the $p$-equation \eqref{pdifeq}.

To summarize, our perturbative analysis above has explicitly shown
the validity of the expansion \eqref{Lnm} up to (and including)
second order (i.e. to order $e^{-2R}$).
We have checked explicitly that the EOMs
\eqref{eqRRR}-\eqref{eqRpp}, under the boundary conditions
given in the beginning of this section, are consistent to this order.
We note that some highly non-trivial cancellations in the EOMs were
necessary
in checking that all three EOMs ${\cal{E}}_1$-${\cal{E}}_3$
give equations for $y(R)$, $b(R)$, $p(R)$ and $p(R)$
that are mutually consistent.
We have thus provided calculational
support to our claim above concerning existence
and uniqueness of solutions for EOMs corresponding to the
ansatz \eqref{metans2}-\eqref{endansa}.


\section{Thermodynamics}
\label{secthermodynamics}

\subsection{Properties of the horizon}
\label{secprophor}

We now examine some of the properties of the $R=R_0$ hypersurface
in the solution given by the ansatz \eqref{metans2}-\eqref{endansa}.

It is clear that when $R_0 < R_c$ the horizon has topology
$S^{d-1}$, while for $R_0 > R_c$ the horizon overlaps with itself
on the cylinder and has therefore the topology $S^{d-2} \times S^1$.
For $R_0 = R_c$ the horizon precisely touches itself in one point.
If we think about the covering space with an array of black holes,
the picture is also clear: For $R_0<R_c$ the black holes are separated,
for $R_0=R_c$ their horizon touches in one point and for $R_0>R_c$
the horizons have merged.

If we consider the family of hypersurfaces $R = \mbox{\sl constant}$
then the normal vector is $\frac{\partial}{\partial R}$.
Since this vector has zero norm on the horizon (provided $A(R,v)$ is not
zero) it follows that the $R=R_0$ hypersurface is a null hypersurface.
Clearly, it is also a Killing horizon with respect to the Killing
vector $\frac{\partial}{\partial t}$ and also with respect to the
angular Killing vectors of the $S^{d-2}$ sphere.

Since we have a Killing horizon we can consider the surface gravity
$\kappa$.
Using $\frac{\partial}{\partial t}$ as the Killing vector, this is
computed to be
\begin{equation}
\label{surfgrav}
\kappa^2 = - \frac{1}{4} \lim_{R\rightarrow R_0}
\Big[ g^{tt} g^{RR} (\partial_R g_{tt})^2 \Big]
= \frac{(d-3)^2}{4 R_0^2 R_T^2 \cosh^2 \alpha \, A|_{R=R_0} } \ .
\end{equation}
Thus, in order for the surface gravity $\kappa$ to be constant on the
horizon, we need $A(R,v)$ to be independent of $v$ on the horizon,
i.e. we need $\dot{A} = 0$.
Fortunately, this follows from the equations of motion
\eqref{eqRRR}-\eqref{eqRpp}, as we now shall see.

However, we first need to discuss the behavior of
$A(R,v)$ and $K(R,v)$ on the horizon. From Section \ref{secconditions}
we see that $A(R,v)$ and $K(R,v)$ are non-singular on the horizon
in the $R_0 \ll R_c$ and $R_0 \gg R_c$ limits. We also know that the flat metric \eqref{newflat} is non-singular
on the space given by $R \in [0,\infty [ $ and $v \in \, ]-\pi,\pi [$.
We therefore assume that for all values of $R_0$ the functions $A(R,v)$ and $K(R,v)$ are non-singular on the horizon for
$v \in \, ] - \pi,\pi [$.

Under this assumption, the four equations of motion \eqref{eqRRR}-\eqref{eqRpp}
reduce for $R=R_0$ to the following two equations
\begin{equation}
\label{thehor1}
\dot{a} = 0
\end{equation}
\begin{equation}
\label{thehor2}
e^a = e^k + \frac{1}{2} R_0 e^k k' +  \frac{R_0^2}{2(d-3)} e^{(d-1)k}
\left( \ddot{k} + (d-2)  \dot{k}^2 \right)
\end{equation}
so that we indeed have $\dot{A} =0$, as desired.

Clearly we can combine the two equations \eqref{thehor1} and \eqref{thehor2}
into one equation for $k$.
We thus see that the three equations for
$K(R,v)$ reduce to one equation on the horizon and they are therefore
a consistent system of differential equations in this limit.

\subsection{Thermodynamics}
\label{secthermo}

In this section we consider the thermodynamics of non-extremal
charged dilatonic $p$-branes that follows from the ansatz \eqref{metans2}-\eqref{endansa}.
We also consider the near-extremal and neutral cases.

First we define  the function
\begin{equation}
\label{defgam}
\gamma (R_0) \equiv \frac{1}{\sqrt{ A|_{R=R_0} }}
\end{equation}
which  will play an essential role in the thermodynamics.
We note the limiting cases
\begin{equation}
\label{knowgamma}
\gamma(R_0) = \left\{ \begin{array}{ccc}
\frac{d-2}{d-3} ( k_d R_0)^{\frac{1}{d-2} }
& \mbox{for} & R_0 \ll R_c \\
1 & \mbox{for} & R_0 \gg R_c \end{array} \right.
\end{equation}
as found using \eqref{A0smallR} and \eqref{largeBHsol1} respectively.

We established in Section \ref{secprophor} that the surface gravity
$\kappa$ is constant on the horizon. This means we have a well-defined
temperature $T$ given as $2\pi T = \kappa$.
In particular, from \eqref{surfgrav} and \eqref{defgam}  we find
\begin{equation}
\label{thetemp}
T = \gamma(R_0) \frac{d-3}{4\pi R_0 R_T \cosh \alpha}  \ .
\end{equation}
We note that we obtain the same result by Wick rotating the metric
\eqref{metans2} and demanding absence of a conical singularity
near $R = R_0$.

Using the Bekenstein-Hawking formula for entropy as the horizon
area divided by $4G$ we get the entropy
\begin{equation}
\label{theentropy}
S = \frac{1}{\gamma(R_0)} \frac{V_p \Omega_{d-2} 2\pi R_T}{4G}
(R_0 R_T)^{d-2} \cosh \alpha \ .
\end{equation}
Note that from \eqref{thetemp} and \eqref{theentropy} one has
\begin{equation}
\label{theTSprod}
T S = \frac{V_p \Omega_{d-2} 2\pi R_T}{16\pi G} (d-3) (R_0 R_T)^{d-3}
\end{equation}
so that the product $TS$ is independent of the function $\gamma(R_0)$.

The chemical potential $\nu$ and charge $Q$ are given by
\begin{equation}
\label{themu}
\nu = \tanh \alpha
\end{equation}
\begin{equation}
\label{theQ}
Q = \frac{V_p \Omega_{d-2} 2\pi R_T}{16 \pi G}
(d-3) (R_0 R_T)^{d-3} \cosh \alpha \sinh \alpha \ .
\end{equation}
Finally, the mass is computed from the Hawking-Horowitz mass formula
\cite{Hawking:1996fd} which we considered in Section \ref{secconmass}.
Using as input the behavior \eqref{ARlarge} of $A(R,v)$ for $R \gg 1$
we find in this case
\begin{equation}
\label{theM}
M = \frac{V_p \Omega_{d-2} 2\pi R_T}{16 \pi G}
(d-3) (R_0 R_T)^{d-3} \left[ \frac{d-2}{d-3} - \chi(R_0)
+ \sinh^2 \alpha \right]
\end{equation}
where $\chi(R_0)$ is defined and discussed in Section \ref{secgencons}.

For $R_0 \ll R_c$ the thermodynamics \eqref{thetemp}-\eqref{theentropy},
\eqref{themu}-\eqref{theM} becomes the usual thermodynamics
of non-extremal charged dilatonic $p$-branes given in
Appendix \ref{appextrsol}%
\footnote{That \eqref{thetemp}-\eqref{theentropy},
\eqref{themu}-\eqref{theM} is the same as \eqref{rhoTS}-\eqref{rhoM}
can be seen by using the $R_0 \ll R_c$ expressions for
$\chi(R_0)$ and $\gamma(R_0)$ given in \eqref{knowchi} and \eqref{knowgamma}
along with the coordinate transformation from $\rho_0$ to $R_0$
given by \eqref{Randrho}.}.
For $R_0 \gg R_c$ it reduces to the usual thermodynamics
of non-extremal charged dilatonic $p$-branes smeared in one direction.

The first law of thermodynamics is
\begin{equation}
\label{firstlaw}
dM = T dS + \nu dQ
\end{equation}
Using the above formulas we see that \eqref{firstlaw} holds if
and only if $\gamma(R_0)$ and $\chi(R_0)$ are related by
\begin{equation}
\label{rell1}
R_0\frac{\gamma'}{\gamma}
= (d-3) \chi + R_0 \chi'
\end{equation}
where prime denotes a derivative with respect to $R_0$.
Using \eqref{knowchi} and \eqref{knowgamma} we see that this is
fulfilled for both $R_0 \ll R_c$ and $R_0 \gg R_c$
provided we take $\chi'(R_0) = 0$ in both limits.
The formula \eqref{rell1} relates the metric at the horizon
to the metric for $R \rightarrow \infty$.

\smallsec{Smooth interpolation}

As mentioned previously, we are in this paper advocating the
idea that there should be a smooth interpolation between the
small and large black hole solution on a cylinder.
For the thermodynamics, this means that the functions $\gamma(R_0)$
and $\chi(R_0)$ should interpolate smoothly between the two
limiting cases $R_0 \ll R_c$ and $R_0 \gg R_c$.
If we consider $\gamma(R_0)$ this seems very likely in
that $\gamma(R_0)$ is a concave function for $R_0 \ll R_c$
since the second derivative $\gamma''(R_0)$ is negative
at $R_0 =0$. This
fits nicely with the expectation that $\gamma(R_0)$ should
go from from $0$ to $1$ as $R_0$ increases
from $0$ to $\infty$.
Equivalently, the fact that $\chi'(R_0)$ should go to zero
for both $R_0 \rightarrow 0$ and $R_0 \rightarrow \infty$
fits nicely with the expectation that $\chi(R_0)$
monotonically decreases from $1/((d-2)(d-3))$ to $0$
as $R_0$ increases from $0$ to $\infty$.

We expect that $\chi(R_0)$ and $\gamma(R_0)$, and thereby
the thermodynamics, are smooth functions of $R_0$
because if we consider the metric for
$R \geq 0$ and $v \in \, ]-\pi,\pi [$ there is no difference
between $R \leq R_c$ and $R > R_c$. We have the same
periodicity properties with respect to $v$ and the horizon can
smoothly go from $R < R_c$ to $R > R_c$.
Note that we can consistently ignore the set of measure
zero defined by $v=\pi$. This
is because the boundary conditions of Section \ref{secaneom}
are not affected by this, so that the EOMs can be solved
consistently without including the set $v=\pi$.

However, when we consider the solution in the physical $(r,z)$ coordinates
on the cylinder, we have a naked singularity
at the event horizon when $R_0=R_c$.
This naked singularity precisely appears when the horizon
shifts topology. The singularity
is similar to a black hole evaporation effect since in
the slice $z=\pi$ it looks like an evaporating black hole.
Because of this
it seems likely that a small burst of energy will be released
which means that the thermodynamics will experience a first order
transition at this point. However, for macroscopic cylinders
this discontinuity in the thermodynamics at $R_0 = R_c$
will be negligible and it should be a good approximation to
consider the transition as smooth.

\smallsec{Thermodynamics for neutral case}

For the neutral black hole solution on the cylinder $\R^{d-1} \times S^1$
with solution of the form \eqref{neutBH}-\eqref{moreneut}
the thermodynamics is
\begin{equation}
\label{neutTS}
T = \gamma(R_0) \frac{d-3}{4\pi R_0 R_T}
\spa
S = \frac{1}{\gamma(R_0)} \frac{\Omega_{d-2} 2\pi R_T}{4G_{d+1}}
(R_0 R_T)^{d-2}
\end{equation}
\begin{equation}
\label{neutmass}
M = \frac{\Omega_{d-2} 2\pi R_T}{16 \pi G_{d+1}}
(d-3) (R_0 R_T)^{d-3} \left[ \frac{d-2}{d-3} - \chi(R_0) \right] \ .
\end{equation}
Note that the mass \eqref{neutmass}
correctly reduces to \eqref{smallBHmass} found in Section \ref{secg00}
for $R_0 \ll R_c$.

The thermodynamics \eqref{neutTS}-\eqref{neutmass} should correspond
to a Hawking radiating neutral black hole.
We expect therefore that the entropy $S$ and mass $M$ should be increasing
functions of $R_0$, since adding mass to the black hole should make
it bigger and increase its entropy.
We see from \eqref{neutTS} and \eqref{neutmass} that this requires
$R_0 \gamma' / \gamma < d-2$.
Moreover, since the black hole
radiates the heat capacity should be negative.
This means the temperature $T$ should be an decreasing function of $R_0$
since we already imposed that $S$ should be increasing.
We see from \eqref{neutTS} that this requires $R_0 \gamma' / \gamma < 1$.
Thus, the aforementioned three physical requirements on
the thermodynamics as function of $R_0$ hold, if and only if
\begin{equation}
\label{gamcond}
R_0 \frac{\gamma'}{\gamma} < 1  \ .
\end{equation}
If we consider the limiting behavior of $\gamma(R_0)$
given by \eqref{knowgamma} we see that $\gamma(R_0)$ would
have to behave in a very strange way to break \eqref{gamcond}.
Indeed, $R_0 \gamma'/\gamma$ is $1/(d-2)$ at $R_0 = 0$ and
goes to zero for $R_0 \rightarrow \infty$.

\smallsec{Thermodynamics for near-extremal case}

For the near-extremal dilatonic $p$-brane solutions of the form
\eqref{NEmet}-\eqref{NEfct} we have the thermodynamics
\begin{equation}
\label{neTS}
T = \frac{d-3}{4\pi \hat{h}_d^{1/2}} \gamma(R_0) R_0^{\frac{d-5}{2}}
\spa
S = \frac{V_p \Omega_{d-2} 2\pi \hat{h}_d^{1/2}}{4\hat{G}}
\frac{R_0^{\frac{d-1}{2}}}{\gamma (R_0)}
\end{equation}
\begin{equation}
\label{neE}
E = \frac{V_p \Omega_{d-2} 2\pi}{16 \pi \hat{G}} R_0^{d-3}
\left[ \frac{d-1}{2} - (d-3) \chi(R_0) \right]
\end{equation}
\begin{equation}
\label{neF}
F = - \frac{V_p \Omega_{d-2} 2\pi}{16 \pi \hat{G}} R_0^{d-3}
\left[ \frac{d-5}{2} + (d-3) \chi(R_0) \right]
\end{equation}
where $\hat{G} = G R_T^{-(d-2)}$ is the rescaled
Newtons constant in the $D$-dimensional space-time,
$E = M - Q$ is the energy above extremality and
$F = E - TS$ is the free energy.

As for the neutral black hole we expect the entropy $S$ and the energy
$E$ to be increasing functions of $R_0$.
{} From \eqref{neTS} and \eqref{neE} this requirement leads to
$R_0 \gamma' / \gamma <  (d-1)/2$.
We see that this is true provided \eqref{gamcond} is true.

We can also examine when the temperature $T$ is an increasing
function of $R_0$. From \eqref{neTS} we see that this is the
case when $R_0 \gamma' / \gamma > - (d-5)/2$.
We expect this to hold at least for $d \geq 6$, since
for $d \geq 6$ we know that $T(R_0)$ is increasing in the
$R_0 \ll R_c$ and $R_0 \gg R_c$ limits, and also since the
$R_0 \gamma' / \gamma > - (d-5)/2$ is a rather weak condition
on the behavior of $\gamma$ for $d \geq 6$.

We can clarify the physics of this by considering the heat capacity
$C = T dS/dT$ since we see that the heat capacity $C$ is positive
if and only if $T(R_0)$ is an increasing function,
provided of course that $S(R_0)$ is an increasing function.
So, since the heat capacity $C$ is positive
for $d \geq 6$ in the $R_0 \ll R_c$ and $R_0 \gg R_c$ limits,
we expect $C$ to be positive for all $R_0$ for $d \geq 6$.
For $d = 4$ we already know that
the heat capacity $C$ is negative for $R_0 \gg R_c$.
For $d=5$ we have that
\begin{equation}
T = \frac{1}{2\pi \hat{h}_d^{1/2}} \gamma (R_0) \ .
\end{equation}
So, here we encounter the very interesting situation that
the heat capacity for $R_0 \gg R_c$ depends
on the detailed behavior of $\gamma(R_0)$.
If $\gamma(R_0)$ crosses $1$ the heat capacity can be negative,
and if $\gamma(R_0)$ does not cross $1$
we expect the heat capacity to be positive for all $R_0$.
This will be considered further in Section \ref{secLST} when
we explain how this can be applied to study
the thermodynamics of Little String Theory.

\subsection{Physical subspace of solutions}
\label{secsolspace}

We now have the remedies to discuss a problem that can seem
surprising: We have too many solutions. In Section \ref{secaneom}
we justified the statement that any value of $R_0$ and $\chi \geq 0$
($R_0$ and $\chi$ treated as independent parameters)
corresponds to a solution.
However, clearly not all of these solution corresponds to
black holes on cylinders.
If we for instance consider $R_0 \ll R_c$ we know from the Newton
limit that $\chi = 1/((d-2)(d-3))$ from the physical requirement
that a small black hole should behave like a point particle
when observed from far away. However, we can pick any other value of
$\chi$ for $R_0 \ll R_c$ and it would still correspond to a solution.
So we need to find the appropriate physical subspace in this
two-dimensional space of solutions spanned by $R_0$ and $\chi$.

To do this, we first observe, as also pointed out in Section \ref{secaneom},
that given some values for $R_0$ and $\chi$ the function $K(R,v)$ is fully
determined. In particular, $\gamma$, defined as in \eqref{defgam} by
$\gamma = 1/\sqrt{A|_{R=R_0}} $, is determined.
So $\gamma = \gamma (R_0,\chi)$, i.e. $\gamma$ is a well-defined
function on the two-dimensional space of solutions.

The physical subspace of solutions should be such that to each value
of $R_0$ a corresponding value of $\chi$ can be found, i.e. so that
$\chi(R_0)$ is a well-defined function. One can imagine different
physical branches corresponding to different functions $\chi(R_0)$.
We claim here that the physical subspace of solutions for
black holes on cylinder is defined by
demanding that $\chi(R_0)$ is a continuous function that interpolates
as in \eqref{knowchi} and obeys the relation \eqref{rell1}, where
$\gamma(R_0) = \gamma(R_0,\chi(R_0))$.

It would be interesting to find other ways of characterizing the
physical subspace of solutions corresponding to black holes on
cylinders. One could for example imagine finding a requirement from
the Newtonian limit
as we did for small black holes in Section \ref{secsmallBH}.

\subsection{Further study of $\chi(R_0)$ and $\gamma(R_0)$}
\label{secgamchi}

We study in this section the behavior of the functions
$\chi(R_0)$ and $\gamma(R_0)$ as $R_0$ goes from $0$ to $\infty$.

\begin{figure}[ht]
\begin{center}
\includegraphics{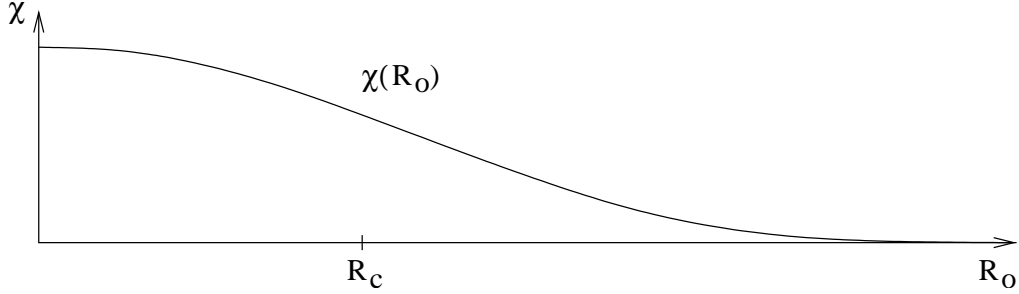}
\caption{\small Qualitative behavior of $\chi (R_0)$. }
\label{figchi}
\end{center}
\end{figure}

We first give a physical argument why $\chi(R_0)$ is a
monotonically decreasing function and positive for all $R_0$.
In Section \ref{secsmallBH} we found the Newton limit for the $R_0 \ll R_c$
case. We can generalize this so that we consider static
matter with both a mass density $T_{00} = \varrho$ and
a negative pressure
$T_{zz} = - n \varrho$. The negative pressure corresponds to the
binding energy of the black hole solution on the cylinder,
i.e. the fact that it takes a certain amount of energy to separate
two black holes from each other.
Repeating the analysis%
\footnote{The modified Einstein equations are $R^0_{\ 0} = - 8\pi G \varrho
\frac{d-2}{d-1} [ 1 - \frac{n}{d-2} ]$, $R^z_{\ z} = 8\pi G \varrho
\frac{1}{d-1} [ 1 - (d-2) n ]$ and $R^r_{\ r} = R^{\phi_1}_{\ \phi_1}
= 8\pi G \varrho \frac{1}{d-1} [ 1 + n ]$. Using this, along with
\eqref{newtpot}, $R^0_{\ 0} = \frac{1}{2} \nabla^2 g_{00}$
and $R^z_{\ z} = - \frac{1}{2} \nabla^2 g_{zz}$ we easily find
$g_{00}$ and $g_{zz}$ in terms of $\Phi$. The expression for $\chi$
is then obtained by comparing to our ansatz for $R$ much greater than both $R_c$ and $R_0$, which
identifies $\chi = 1/(d-3) \times (1-g_{zz})/(-1-g_{00})$.},
we get that
$A(R,v) \simeq 1 - \chi \frac{R_0^{d-3}}{R^{d-3}}$
and $K(R,v) \simeq 1 - \chi \frac{R_0^{d-3}}{R^{d-3}}$
at infinity with
\begin{equation}
\chi = \frac{1}{(d-2)(d-3)} \frac{1-(d-2)n}{1-\frac{1}{d-2}n} \ .
\end{equation}
So, we see that we can interpret $\chi$ directly in terms of the
binding energy relative to the mass of our object.
It is physically clear that as the radius of the cylinder
$R_T$ decreases, or equivalently, as
$R_0$ increases, the binding energy increases.
This requires that $\chi(R_0)$ is a monotonically decreasing
function of $R_0$ since $\chi$ is monotonically decreasing
as a function of $n$. Since $\chi(R_0) \rightarrow 0$ for
$R_0 \rightarrow \infty$ this means in particular that $\chi(R_0) \geq 0$.
We thus expect $\chi(R_0)$
to behave qualitatively as depicted in Figure \ref{figchi}.

\begin{figure}[ht]
\begin{center}
\includegraphics{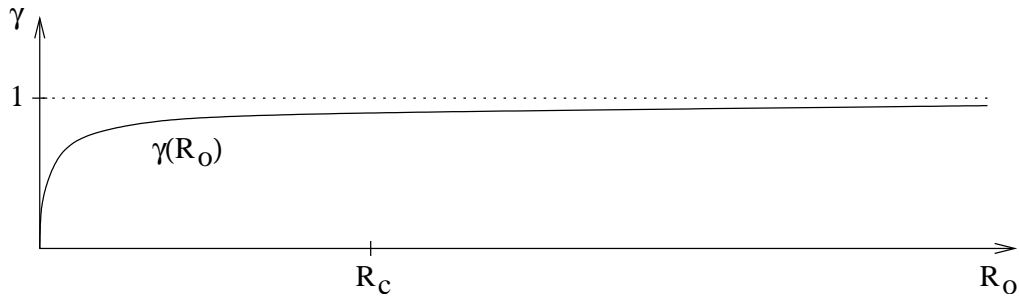}
\caption{\small Qualitative behavior of $\gamma (R_0)$: scenario I}
\label{figgam1}
\end{center}
\end{figure}

\begin{figure}[ht]
\begin{center}
\includegraphics{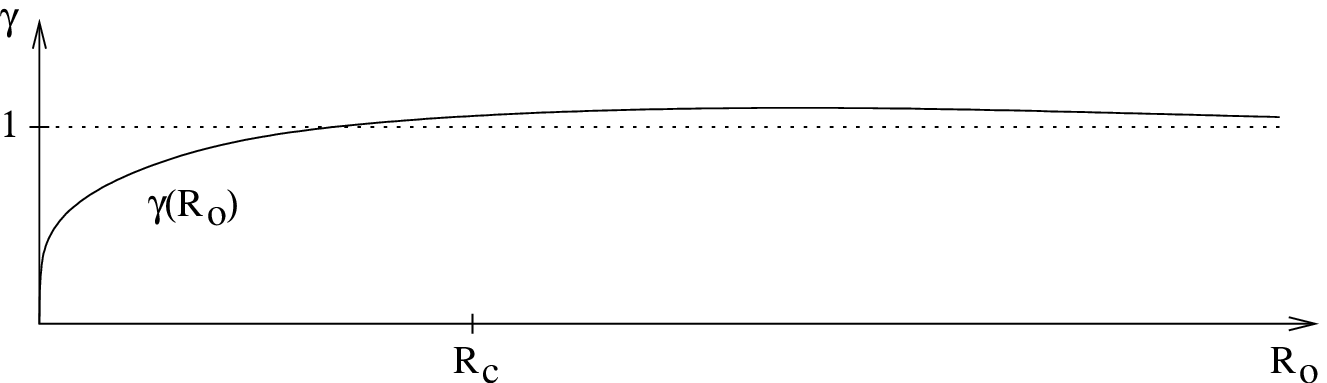}
\caption{\small Qualitative behavior of $\gamma (R_0)$: scenario II}
\label{figgam2}
\end{center}
\end{figure}

If we instead consider $\gamma(R_0)$ it seems that there are two
possibilities.
Either $\gamma(R_0)$ will never reach $1$ for any $R_0$ and
we expect $\gamma(R_0)$ to be a monotonically increasing function
so that it asymptotes to $1$ from below. This is depicted
in Figure \ref{figgam1}.
Or $\gamma(R_0)$ will reach $1$ at some finite $R_0$ and then have
a maximum and then asymptote to $1$ from above. This is depicted
in Figure \ref{figgam2}.

We can elaborate a bit more on the two scenarios for $\gamma(R_0)$.
First, we rewrite \eqref{rell1} as
\begin{equation}
\frac{\gamma'}{\gamma} = \frac{1}{R_0^{d-3}} \Big( R_0^{d-3} \chi \Big)'
\end{equation}
where prime denotes a derivate with respect to $R_0$.
We now see that since $\chi(R_0) \geq 0$ we have that
if $R_0^{d-3} \chi(R_0) \rightarrow 0$ for $R_0 \rightarrow \infty$
then we are in the case of Figure \ref{figgam2}
since clearly the derivative of $R_0^{d-3} \chi(R_0)$ should be negative
in that case.
Similarly, if $R_0^{d-3} \chi(R_0) \rightarrow \infty$
for $R_0 \rightarrow \infty$ then we are in the case of Figure \ref{figgam1}.

\subsection{Neutral solution has larger entropy than black string}
\label{secpunch}

In this section we argue that
our neutral solution on the cylinder $\R^{d-1} \times S^1$ given by
\eqref{neutBH}-\eqref{moreneut} has larger
entropy than a black string with the same mass.
We furthermore generalize the argument to non-extremal
and near-extremal charged dilatonic branes.

\smallsec{Neutral solution versus neutral black strings}

The neutral black string in $(d+1)$-dimensional space-time with
the string wrapped on an $S^1$ of radius $R_T$ has thermodynamics
\begin{equation}
\label{strTS}
T_{\rm str} = \frac{d-3}{4\pi r_0}
\spa
S_{\rm str} = \frac{\Omega_{d-2} 2\pi R_T}{4G_{d+1}} r_0^{d-2}
\end{equation}
\begin{equation}
\label{strmass}
M_{\rm str} = \frac{\Omega_{d-2} 2\pi R_T}{16 \pi G_{d+1}}
(d-2) r_0^{d-3}
\end{equation}
where $r_0$ is the horizon radius of the black string.
We note that these expressions are written
in terms of $r_0$ in order to emphasize that the
proper coordinate system for the black string solution is
the cylindrical $(r,z)$ coordinates.

We now want to compare the entropy of the black string solution
with the entropy of the non-translationally invariant
neutral solution given by \eqref{neutBH}-\eqref{moreneut}
for a given mass $M$.

{} From \eqref{neutmass} and \eqref{strmass} we see that we can
determine the horizon radius of our neutral solution $R_0(M)$
and the horizon radius of the black string $r_0(M)$ as functions of $M$.
Using the first law of thermodynamics
$dM = T_{\rm str} dS_{\rm str}$ and $dM = T dS$
for the two solutions we obtain
\begin{equation}
\label{ddMSS}
\frac{d}{dM} \left( \frac{S_{\rm str} (M)}{S(M)} \right)
= \frac{TS - T_{\rm str} S_{\rm str}}{T_{\rm str} T S^2}
\end{equation}
where $T$ and $T_{\rm str}$ are taken to be depending on $M$
through $R_0(M)$ and $r_0(M)$.

We now argue that \eqref{ddMSS} is positive.
We argued in Section \ref{secgamchi} that $\chi(R_0(M)) \geq 0$.
Comparing \eqref{neutmass} and \eqref{strmass}
we see that this gives that $R_0(M) R_T \geq r_0(M)$.
{} From \eqref{neutTS} and \eqref{strTS} we then have that
$TS \geq T_{\rm str} S_{\rm str}$.
It therefore follows that
\begin{equation}
\label{dms}
\frac{d}{dM} \left( \frac{S_{\rm str} (M)}{S(M)} \right) \geq 0
\end{equation}
and, moreover, \eqref{dms} is strictly positive if $\chi(R_0(M)) > 0$.

Consider now a given mass $M$. We may write
\begin{equation}
\frac{S_{\rm str} (M)}{S(M)}
= 1 -
\int_{R_0}^\infty dM' \frac{d}{dM'}
\left( \frac{S_{\rm str} (M')}{S(M')} \right) \ .
\end{equation}
Thus, since we know that $\chi(R_0')$ is non-zero
at least for some $R_0' \geq R_0(M)$ we see from \eqref{dms} that
\begin{equation}
S(M) > S_{\rm str} (M)
\end{equation}
showing that the entropy of the non-translationally invariant
solution given by \eqref{neutBH}-\eqref{moreneut}
is larger than that of the black string.

The reason for comparing the entropies at a given mass, i.e.
working in the microcanonical ensemble, is that we want
to check whether it is thermodynamically favorable
for a black string of a given mass to redistribute itself
into the non-translationally invariant solution in order to gain entropy.
What we have shown above is that
it is {\sl always} favorable for a black string to break
the translational invariance along the circle and redistribute
the mass according to the non-translationally invariant
solution we have described above.

In \cite{Gregory:1993vy} Gregory and Laflamme showed
that a neutral black string
wrapped on a circle is classically unstable if and only if
$r_0 \leq n_c R_T$
where $r_0$ is the horizon radius for the black string,
$R_T$ the radius of the circle and $n_c$ a number of order one.
This was explained physically by the fact that a black hole
of the same mass as the black string had larger entropy%
\footnote{See \cite{Gubser:2000mm,Reall:2001ag,Hubeny:2002xn}
for interesting recent work on classical
instabilities of near-extremal branes and the connection
to their thermodynamics.}.
We now see that even though no classical instability is present
for the black string solution for $r_0 > n_c R_T$ we still
have a quantum instability due to the fact that another
configuration with larger entropy but same quantum numbers exists.

The existence of new non-translationally invariant solution wrapping
a cylinder was also argued in \cite{Horowitz:2001cz} by Horowitz and Maeda.
They considered a black string wrapping $\R^4 \times S^1$
with $r_0 \ll n_c R_T$. In this case the black hole solution
on $\R^4 \times S^1$ has larger entropy than the black string,
which, as mentioned above, was already pointed out by Gregory and
Laflamme. They could then show using Raychaudhuri's equation
that, within reasonable assumptions, the black string solution cannot decay to the black hole
in finite affine parameter.
Since the black string is classically unstable in this case
this naturally leads to the conjecture that an intermediate
classical solution exists that is not translationally
invariant along the circle but which has larger entropy
than the black string%
\footnote{The existence of this solution was recently addressed
in \cite{Gubser:2001ac} using perturbative and numerical methods.}.
Physically this means that the Gregory-Laflamme instability
cannot, as thought previously, reveal a naked singularity when
the black string horizon bifurcates into a black hole horizon
not connected along the circle. So, the Horowitz-Maeda argument
removes this potential violation of the Cosmic Censorship Conjecture
of Penrose.
Horowitz and Maeda even generalized their argument to the statement
that a horizon cannot have any collapsing circles.
However, the new solutions that were argued to exist by Horowitz and Maeda
are apparently not related to our non-translationally
invariant neutral solution
since their arguments only apply when we have
a Gregory-Laflamme instability of the black string, which only
happens for $r_0 < n_c R_T$.

In conclusion we have found a new instability
of the neutral black string on the cylinder  $\R^{d-1} \times S^1$
for large masses $M > M_c$, where $M_c$ is given by $R_0(M_c)=R_c$.
This instability is not classical since Gregory and Laflamme
have shown that the black string is classically stable for $M > M_c$.
The instability occurs because we have a new solution
with the same mass and horizon topology but larger entropy.
The neutral black string therefore spontaneously breaks
the translation invariance along the circle and redistributes
its mass according the our solution given by
\eqref{neutBH}-\eqref{moreneut}. This transition occurs
without changing the topology of the horizon.

For small masses $M < M_c$ the black string is classically unstable,
as shown by Gregory and Laflamme.
Also in this case have we shown that the entropy of our solution
given by \eqref{neutBH}-\eqref{moreneut} has larger
entropy than the black string.
However, our solution cannot be reached by a classical evolution
as considered by Horowitz and Maeda.
According to Horowitz and Maeda the neutral black string
should in this case decay to
an intermediate non-translationally invariant solution with connected horizon
along the circle that does not seem immediately related
to our solution.

We should emphasize that we have not strictly
checked that $M = M_c$ marks the separation between the two regions where
the  neutral black string is classically stable and unstable.
If we call the $M_{\rm GL}$ the mass that marks the border between
classical stability and instability, we know that $M_{\rm GL}/M_c$
is some number of order one,
but whether it is exactly equal to $1$ is not clear.
To check this one would need to know the exact value for $\chi(R_0=R_c)$
and this we do not know at present.
It would be interesting to examine this further.

\smallsec{Non-extremal charged dilatonic branes}

We generalize here the above argument for the neutral case
to show that the entropy of our non-extremal
charged dilatonic $p$-brane
solution \eqref{metans2}-\eqref{endansa}
is larger than that of the non-extremal
charged dilatonic $p$-brane smeared on the transverse circle
for a given mass $M$ and charge $Q$.

The thermodynamics of the smeared solution is
\begin{equation}
\label{smeTS}
T_{\rm sme} = \frac{d-3}{4\pi r_0 \cosh \hat{\alpha}}
\spa
S_{\rm sme} = \frac{V_p \Omega_{d-2} 2\pi R_T}{4G}
r_0^{d-2} \cosh \hat{\alpha}
\end{equation}
\begin{equation}
\label{smemuQ}
\nu_{\rm sme} = \tanh \hat{\alpha}
\spa
Q_{\rm sme} = \frac{V_p \Omega_{d-2} 2\pi R_T}{16 \pi G}
(d-3) r_0^{d-3} \cosh \hat{\alpha} \sinh \hat{\alpha}
\end{equation}
\begin{equation}
\label{smeM}
M_{\rm sme} = \frac{V_p \Omega_{d-2} 2\pi R_T}{16 \pi G}
(d-3) r_0^{d-3} \left[ \frac{d-2}{d-3} + \sinh^2 \hat{\alpha} \right] \ .
\end{equation}
For given $M$ and $Q$ we
have from \eqref{theQ} and \eqref{theM}
the functions $R_0(M,Q)$ and $\alpha(M,Q)$
for the non-extremal charged dilatonic brane solution
\eqref{metans2}-\eqref{endansa}
and from \eqref{smemuQ} and \eqref{smeM}
we have the functions $r_0(M,Q)$ and $\hat{\alpha}(M,Q)$
for the non-extremal smeared charged dilatonic brane solution.

Writing the mass formulas \eqref{theM} and \eqref{smeM} as
\begin{eqnarray}
M &=& \frac{V_p \Omega_{d-2} 2\pi R_T}{16 \pi G}
(R_0 R_T)^{d-3} \left[ \frac{d-1}{2} - (d-3) \chi(R_0) \right]
\nn \\ &&
+ \sqrt{ \frac{1}{4} \left( \frac{V_p \Omega_{d-2} 2\pi R_T}{16 \pi G}
(d-3) (R_0 R_T)^{d-3} \right)^2 + Q^2 }
\end{eqnarray}
\begin{equation}
M = \frac{V_p \Omega_{d-2} 2\pi R_T}{16 \pi G}
r_0^{d-3} \frac{d-1}{2}
+ \sqrt{ \frac{1}{4} \left( \frac{V_p \Omega_{d-2} 2\pi R_T}{16 \pi G}
(d-3) r_0^{d-3} \right)^2 + Q^2 }
\end{equation}
we see that it follows from $\chi(R_0(M,Q)) \geq 0$ (see Section
\ref{secgamchi}) that $R_0(M,Q) R_T \geq r_0(M,Q)$.
Using \eqref{theTSprod} and \eqref{smeTS} this
gives that $TS \geq T_{\rm sme} S_{\rm sme}$.
Therefore, we find  using $dM = T dS + \nu dQ$
and $dM = T_{\rm sme} dS_{\rm sme} + \nu_{\rm sme} dQ$ that
\begin{equation}
\frac{\partial}{\partial M} \left( \frac{S_{\rm sme} (M,Q)}{S(M,Q)}
\right)
= \frac{TS - T_{\rm sme} S_{\rm sme}}{T_{\rm sme} T S^2} \geq 0
\end{equation}
where the expression is strictly positive if $\chi(R_0(M,Q)) > 0$.

Integrating in the same way as for the neutral case, we obtain
therefore for a given mass $M$ and charge $Q$ the inequality
\begin{equation}
S(M,Q) > S_{\rm sme}(M,Q) \ .
\end{equation}
So, indeed, we have that the entropy of our non-extremal brane solution
given by \eqref{metans2}-\eqref{endansa}
is larger than that of the non-extremal smeared brane solution
for a given mass $M$ and charge $Q$.
Thus, just as for the neutral black string we get
the result that the smeared charged dilatonic $p$-brane
can win entropy by spontaneously breaking the translational
invariance and redistribute its mass according to our new solution.

\smallsec{Near-extremal charged dilatonic branes}

The argument given above for the neutral case
trivially generalizes to the near-extremal
solution \eqref{NEmet}-\eqref{NEfct} with thermodynamics
\eqref{neTS}-\eqref{neF}.
Here, we find that
\begin{equation}
S(M) > S_{\rm sme}(M)
\end{equation}
for a given mass $M$,
where $S(M)$ is the entropy of our near-extremal charged dilatonic $p$-brane
solution \eqref{NEmet}-\eqref{NEfct} and
$S_{\rm sme}(M)$ is the entropy of the smeared near-extremal
charged dilatonic $p$-brane solution.

Thus, the near-extremal charged dilatonic $p$-brane smeared on a transverse
circle is thermodynamically unstable in a global sense.
It is interesting to consider the interplay between this statement
and the conjecture by Gubser and Mitra  \cite{Gubser:2000mm}, which was
further considered by Reall in \cite{Reall:2001ag}, that
a near-extremal black brane solution is classically stable if and only if
it is locally thermodynamically stable.
For near-extremal smeared branes with $d \geq 6$ the heat capacity
is positive for all energies.
This means that they are locally thermodynamically stable.
Thus they are classically stable for all energies
according to the Gubser-Mitra conjecture, contrary to usual non-extremal
branes which are classically unstable for low energies.
We see therefore that while local thermodynamic stability
for near-extremal branes is connected
to classical stability of the branes, global
thermodynamic stability is related to quantum stability of the branes.

\section{Black holes on cylinders and thermal Little String Theory}
\label{secLST}

We explain in this section how to use our results above to study
the thermodynamics of Little String Theory (LST).
We first review some of the known facts about supergravity duals of
LST in Sections \ref{secsusyLST} and \ref{secthLST},
and then we go on to study the effects of our new solutions
for the supergravity description of thermal LST in Section \ref{secnewsolLST}.

\subsection{Review of supersymmetric Little String Theory
from supergravity}
\label{secsusyLST}

To define $(2,0)$ LST of type $A_{N-1}$
\cite{Seiberg:1997zk,Berkooz:1997cq,Dijkgraaf:1997ku}
we consider $N$
coincident M5-branes in M-theory with a transverse circle
of radius $R_T$.
$(2,0)$ LST of type $A_{N-1}$ is then defined as the world-volume theory
on the M5-branes in the decoupling limit
$l_p \rightarrow 0$,
$l_p$ being the eleven-dimensional Planck length,
with $R_T/l_p^3$ kept fixed.

If we go to Type IIA String theory by S-dualizing on the
transverse circle we are considering
$N$ coincident NS5-branes. The string length $l_s$ is
given by $l_s^2 = l_p^3 / R_T$ and the string coupling
$g_s$ is given by $g_s^2 = R_T^3/l_p^3$ so the decoupling
limit is $g_s \rightarrow 0$ with $l_s$ kept fixed.

LST is a non-local theory without gravity in 5+1 dimensions with 16
supercharges, which flows in the infrared to respectively 5+1 dimensional SYM
or the (2,0) SCFT, depending on whether we are considering the (1,1) or
(2,0) LST. Stringy properties include little string
degrees of freedom, Hagedorn behavior and T-duality.

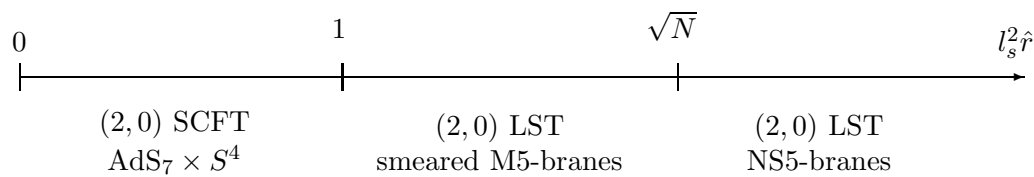
\begin{figure}[h]
\begin{picture}(390,75)(0,0)
\put(5,40){\vector(1,0){380}}
\put(375,50){$l_s^2 \hat{r}$}
\put(35,5){\shortstack{$(2,0)$ SCFT \\ $\ads_7 \times S^4$}}
\put(140,5){\shortstack{$(2,0)$ LST \\ smeared M5-branes}}
\put(280,5){\shortstack{$(2,0)$ LST \\ NS5-branes}}
\put(2,50){0}
\put(5,35){\line(0,1){10}}
\put(123,55){1}
\put(127,35){\line(0,1){10}}
\put(242,55){$\sqrt{N}$}
\put(254,35){\line(0,1){10}}
\end{picture}
\caption{Phase diagram of supersymmetric $(2,0)$ LST. \label{figLST} }
\end{figure}

The supergravity dual of supersymmetric $(2,0)$ LST of type $A_{N-1}$
\cite{Maldacena:1997cg,Aharony:1998ub}
is then the near-horizon limit
\begin{equation}
l_p \rightarrow 0 \spa
\hat{r} = \frac{r}{l_p^3} \spa
\hat{z} = \frac{z}{l_p^3} \spa
\hat{R}_T = \frac{R_T}{l_p^3}
\end{equation}
of the solution \eqref{m5circmet}-\eqref{m5circharm} of extremal M5-branes on
a transverse circle.
This gives the near-horizon metric
\begin{equation}
\label{LSTmet}
l_p^{-2} ds^2 = \hat{H}^{-1/3} \left[ - dt^2 + \sum_{i=1}^5 (dx^i)^2
+ \hat{H} \Big( d\hat{z}^2 + d\hat{r}^2
+ \hat{r}^2 d\Omega_3^2 \Big) \right]
\end{equation}
with
\begin{equation}
\label{LSTharm}
\hat{H} = \sum_{n=-\infty}^{\infty}
\frac{\pi N}{(\hat{r}^2 + (\hat{z} + 2\pi n \hat{R}_T)^2)^{3/2}} \ .
\end{equation}
We now consider the phase diagram of $(2,0)$ LST
as obtained from the supergravity dual in terms of the
rescaled radius $\hat{r}$, which should be thought of as
the energy scale of the theory.
We show in the following that by requiring $N \gg 1$ we
have small curvatures for all $\hat{r}$.
The phase diagram is depicted in figure \ref{figLST}.
Note that from the S-duality transformation to Type IIA theory
we get that \( \hat{R}_T = l_s^{-2} \).

Starting with $\hat{r}=0$ we have $\ads_7 \times S^4$ and $(2,0)$ LST
reduces to $(2,0)$ SCFT.
The metric \eqref{LSTmet}-\eqref{LSTharm} reduces to
\begin{equation}
\label{ads7}
l_p^{-2} ds^2 = \hat{H}^{-1/3} \left[ - dt^2 + \sum_{i=1}^5 (dx^i)^2
+ \hat{H} \Big( d\hat{\rho}^2 + \hat{\rho}^2 d\Omega_4^2 \Big) \right]
\end{equation}
\begin{equation}
\hat{H} = \frac{\pi N}{\hat{\rho}^3}
\end{equation}
where $\hat{\rho} = \rho / l_p^3$ where $\rho^2 = r^2 + z^2$.
Here we should rather use $\rho$ as
the energy scale coordinate than $\hat{r}$.
The curvature of the metric \eqref{ads7} is of order
$N^{-2/3}$ is units of $l_p^{-2}$ is clearly curvatures are
small for $N \gg 1$.  If we call the curvatures in units of $l_p^{-2}$
of the $S^1$ and $S^3$ by $C_{S^1}$ and $C_{S^3}$ respectively we
have $C_{S^1} \sim \hat{H}^{-2/3} \hat{R}_T^{-2}$
and $C_{S^3} \sim \hat{H}^{-2/3} \hat{r}^{-2}$.
Approaching $\hat{r} \sim \hat{R}_T$ we have $C_{S^1} \sim C_{S^3}$.
So, for both curvatures to be small we need
\begin{equation}
\sum_{n=-\infty}^{\infty}
\left[ 1 + \frac{\hat{R}_T^2}{\hat{r}^2} \left(
\frac{\hat{z}}{\hat{R}_T} + 2 \pi n \right)^2 \right]^{-\frac{3}{2}}
\gg \frac{1}{N} \ .
\end{equation}
The left-hand side is minimal for $\hat{z}/\hat{R}_T = \pi$ so we
require
\begin{equation}
\sum_{n=-\infty}^{\infty}
\left[ 1 + \frac{\hat{R}_T^2}{\hat{r}^2} \left(
2 \pi n + \pi \right)^2 \right]^{-\frac{3}{2}}
\gg \frac{1}{N} \ .
\end{equation}
For $\hat{r} \sim \hat{R}_T$ the left-hand side is of order one
so we clearly still have small curvatures when $N \gg 1$ around this
point.
For $\hat{r} \gg \hat{R}_T$ the condition reduces to
$\hat{r} / \hat{R}_T \gg 1/N$ which trivially is satisfied.

For $\hat{r} / \hat{R}_T \gg 1/N$ the M5-brane solution is effectively
a 6-brane solution consisting of smeared M5-branes
with metric \eqref{LSTmet} and harmonic function
\begin{equation}
\hat{H} = \frac{N l_s^2}{\hat{r}^2} \ .
\end{equation}
Increasing $\hat{r}$ we reach $\sqrt{N} l_s^{-2}$ where
the effective string coupling $g_s e^{\phi}$ is of order one, and we
have the near-horizon limit of $N$ coincident NS5-branes in
Type IIA String theory. Also in this case the curvatures are small
provided $N \gg 1$.

The upshot of this review of the supergravity dual of
supersymmetric $(2,0)$ LST
is that we have small curvatures of supergravity at all stages
in the phase diagram, even at the transition points.
So the supergravity description is valid for all $\hat{r}$.

\subsection{Review of thermal Little String Theory from supergravity}
\label{secthLST}

We briefly review in this section what is known about the
dual supergravity description of thermal $(2,0)$ LST.

\begin{figure}[h]
\begin{picture}(390,75)(0,0)
\put(5,40){\vector(1,0){380}}
\put(380,50){$\mu$}
\put(35,5){\shortstack{$(2,0)$ SCFT \\ black M5}}
\put(140,5){\shortstack{$(2,0)$ LST \\ smeared black M5}}
\put(280,5){\shortstack{$(2,0)$ LST \\ black NS5}}
\put(2,50){0}
\put(5,35){\line(0,1){10}}
\put(123,55){$l_s^{-6}$}
\put(127,35){\line(0,1){10}}
\put(240,55){$N \, l_s^{-6}$}
\put(254,35){\line(0,1){10}}
\end{picture}
\caption{Phase diagram of thermal $(2,0)$ LST
in terms of the thermodynamic energy density $\mu$. \label{figthermLST} }
\end{figure}
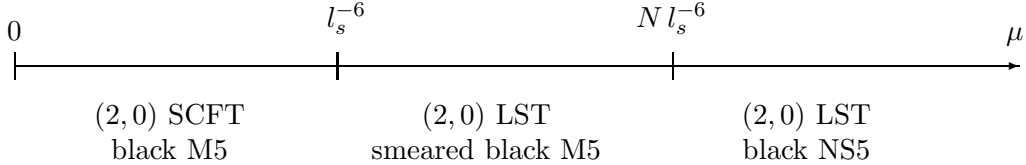

The supergravity description of thermal $(2,0)$ LST has the phase diagram
depicted in figure \ref{figthermLST}, as we shall review below.
This is similar to
the phase diagram for supersymmetric $(2,0)$ LST, though with
the crucial difference that in this case the supergravity
dual is not known when the Schwarzschild radius is of order the
size of the circle.
So, the transition between thermal $(2,0)$ SCFT and
thermal $(2,0)$ LST has not been described in terms of supergravity.

For small energy densities $\mu \ll l_s^{-6}$, we have thermal $(2,0)$ SCFT.
The supergravity dual is that of near-extremal M5-branes.
This is obtained from the solution
\eqref{nemet}-\eqref{nefH} (with $D=11$, $d=5$) of
$N$ coincident non-extremal M5-branes in the near-horizon
limit
\begin{equation}
l_p \rightarrow 0 \spa
\hat \rho = \frac{\rho}{l_p^3} \spa \hat \rho_0 = \frac{\rho_0}{l_p^3} \ .
\end{equation}
We have the metric
\begin{equation}
\label{nhm5met}
l_p^{-2} ds^2 = \hat H^{-1/3} \left[ - \hat f dt^2 + \sum_{i=1}^5 (dx^i)^2
+ \hat H \Big(\hat  f^{-1} d\hat \rho^2 + \hat \rho^2 d\Omega_4^2 \Big) \right]
\end{equation}
with potential
\begin{equation}
C_{012345} =  - \hat H^{-1}
\end{equation}
and harmonic functions
\begin{equation}
\hat f = 1 -\frac{\hat \rho_0^3}{\hat \rho^3} \spa
\hat H = \frac{\pi N}{\hat \rho^3} \ .
\end{equation}
The thermodynamics is
\begin{equation}
T = \frac{3}{4 \pi^{3/2}} \frac{\sqrt{\hat{\rho}_0}}{\sqrt{N}}
\spa
\mathfrak{s} = \frac{1}{24 \pi^{9/2}} \sqrt{N} \hat{\rho}_0^{5/2}
\end{equation}
\begin{equation}
\mu = \frac{5}{192\pi^6} \hat{\rho}_0^3
\spa
\mathfrak{f} = - \frac{1}{192 \pi^6} \hat{\rho}_0^3
\end{equation}
giving
\begin{equation}
\mathfrak{s} = \frac{2^7 \pi^3}{3^6} N^3 T^5
\spa
\mathfrak{f} = - \frac{2^6 \pi^3}{3^7} N^3 T^6
\end{equation}
with $\mathfrak{s}$ the entropy density and $\mathfrak{f}$ the free energy density.

For energies $l_s^{-6} \ll \mu \ll N l_s^{-6}$ we
have $(2,0)$ LST near the Hagedorn temperature. The supergravity
dual is $N$ coincident near-extremal M5-branes smeared on
the transverse circle. This is obtained from the solution
\eqref{nemets}-\eqref{nefHs} (with $D=11$, $d=5$)  of
$N$ coincident non-extremal M5-branes smeared on a transverse
circle in the near-horizon
limit
\begin{equation}
l_p \rightarrow 0
\spa
\hat r = \frac{r}{l_p^3}
\spa
\hat z = \frac{z}{l_p^3}
\spa
\hat{r}_0 = \frac{r_0}{l_p^3}
\spa
\hat{R}_T = \frac{R_T}{l_p^3}
\end{equation}
We have the metric
\begin{equation}
\label{nhblackM5cmet}
l_p^{-2} ds^2 = \hat{H}^{-1/3} \left[ - \hat f dt^2 + \sum_{i=1}^5 (dx^i)^2
+ \hat{H} \Big( d\hat{z}^2
+ \hat{f}^{-1} d\hat{r}^2 + \hat{r}^2 d\Omega_3^2 \Big) \right]
\end{equation}
with potential
\begin{equation}
\label{nhblackM5gp}
C_{012345} =  - \hat H^{-1}
\end{equation}
and harmonic functions
\begin{equation}
\label{nhHnm5circ}
\hat{f} = 1 - \frac{\hat r_0^2}{\hat r^2}
\spa \hat{H} = \frac{N l_s^2}{\hat r^2} \ .
\end{equation}
The thermodynamics is
\begin{equation}
\label{tsm6}
T = T_{\rm hg}
\spa
\mathfrak{s} = \beta_{\rm hg} \mu
\end{equation}
\begin{equation}
\label{mufm6}
\mu = \frac{\hat r_0^2}{(2\pi)^5 l_s^2}
\spa
\mathfrak{f} = 0
\end{equation}
with the definitions
\begin{equation}
T_{\rm hg} \equiv \frac{1}{2\pi \sqrt{N} l_s}
\spa
\beta_{\rm hg} \equiv \frac{1}{T_{\rm hg}} = 2\pi \sqrt{N} l_s \ .
\end{equation}

Raising the energy even further to $\mu \gg N l_s^{-6}$ bring us to
near-extremal type IIA NS5-branes.
The metric is
\begin{equation}
\label{nhblackNS5cmet}
l_s^{-2} ds^2
= \hat H^{-1/4} \left[ - \hat f dt^2 + \sum_{i=1}^5 (dx^i)^2
+ \hat H \Big( \hat f^{-1} d\hat{r}^2 + \hat{r}^2 d\Omega_3^2 \Big) \right]
\end{equation}
and dilaton
\begin{equation}
g_s^2 e^{2\phi} = \hat H
\end{equation}
with $\hat H$ and $\hat f$ as in \eqref{nhHnm5circ} and the gauge potential
the same as in \eqref{nhblackM5gp}.
The thermodynamics is given by \eqref{tsm6}-\eqref{mufm6}, since it is
invariant under the type IIA S-duality.

The thermodynamics \eqref{tsm6}-\eqref{mufm6}
of the near-extremal smeared M5-brane and the NS5-brane
is interpreted as the leading
order Hagedorn thermodynamics of a string theory
with Hagedorn temperature $T_{\rm hg}$
\cite{Maldacena:1996ya,Maldacena:1997cg}.
The next-to-leading terms in the entropy density
$\mathfrak{s}(\mu)$ gives information about the thermodynamics
near the Hagedorn temperature \cite{Harmark:2000hw,Berkooz:2000mz}.
Among other things, we can compute the density of levels
$\rho(\mu)$ and
if we compute $\mathfrak{s}(T)$ we can compute the
sign of the heat capacity and check whether the
thermodynamics is stable.

For high energies $\mu \gg N l_s^{-6}$ it was shown in
\cite{Harmark:2000hw,Berkooz:2000mz} that the one-loop correction to the
NS5-brane background gave the next-to-leading term in $\mathfrak{s}(\mu)$.
Under the assumption that the thermodynamics was stable,
this computation provided insight into the thermodynamics
of LST near the Hagedorn temperature.

However, in \cite{Kutasov:2000jp} a more precise one-loop computation
gave the result that the one-loop corrected thermodynamics
has a temperature $T > T_{\rm hg}$ and a negative specific
heat. So, this seems to indicate that the near-extremal NS5-brane at best
only is dual to LST above the Hagedorn temperature.
Because of the negative heat capacity,
the near-extremal NS5-brane solution has been suggested to be
classically unstable due to a Gregory-Laflamme like instability
\cite{Gubser:2000mm,Rangamani:2001ir,Buchel:2001dg}%
\footnote{The classical instability suggested in
\cite{Gubser:2000mm,Rangamani:2001ir} for near-extremal
NS5-branes in Type IIA is conjectured to be an instability which
spontaneously breaks the translational invariance on the 5-dimensional
world-volume of the NS5-brane, and the NS5-brane is expected to
settle in a new state without
translational invariance along the world-volume.}.
Therefore, a consequence of the computation of \cite{Kutasov:2000jp} is that
we do not know the thermodynamics of LST below the Hagedorn temperature.
That we do not know the thermodynamics below the Hagedorn temperature
means that we do not know the thermodynamics when LST is a string theory,
and we can therefore not expect to get information about the stringy degrees
of freedom of LST
from the thermodynamics of the near-extremal NS5-brane.

Another unknown aspect of thermal LST is
the transition between thermal $(2,0)$ SCFT and $(2,0)$ LST.
It is not known what type of phase transition this is.
Clearly the degrees of freedom in $(2,0)$ SCFT are different
from that of $(2,0)$ LST. The $(2,0)$ SCFT has tensionless selfdual
strings that are charged under the selfdual three-form field strength
living on the M5-brane. The $(2,0)$ LST has fundamental closed strings
(here called ``little strings'').
In the $(2,0)$ LST the tensionless strings of $(2,0)$ emerge as
D-string like solitonic objects that open ``little'' strings
presumably can end on.
Thus, the phase transition between thermal $(2,0)$ SCFT and $(2,0)$ LST
is non-trivial since the degrees of freedom are changing.

\subsection{The new solutions and thermal Little String Theory}
\label{secnewsolLST}

We now explain why our study of black holes on cylinders
is crucial in the supergravity description of
thermal $(2,0)$ LST.

We explained above that thermal $(2,0)$ LST has a
phase at energies $l_s^{-6} \ll \mu \ll N l_s^{-6}$
which is well-described by near-extremal M5-branes
smeared on a transverse circle in M-theory.
Since we have shown in Section \ref{secpunch} that
this near-extremal smeared brane in fact
has lower entropy than our non-translationally invariant
near-extremal M5-brane with a transverse circle,
this means that the smeared M5-brane does {\sl not} give
the correct description of the physics in classical supergravity.
It is our non-translationally invariant
near-extremal M5-brane with a transverse circle that gives
the correct description.

{} From the general ansatz \eqref{NEmet}-\eqref{NEfct} for near-extremal
branes we see that the solution for
the non-translationally invariant
near-extremal M5-brane with a transverse circle is
\begin{equation}
\label{newM5met}
l_p^{-2} ds^2 = \hat{H}^{-1/3} \left[
- \hat{f} dt^2 + \sum_{i=1}^5 (dx^i)^2
+ l_s^{-4} \hat{H} \left( A dR^2 + \frac{A}{K^3} dv^2 + K R^2 d\Omega_3^2
\right) \right]
\end{equation}
\begin{equation}
C_{012345} = - \hat{H}^{-1}
\end{equation}
with functions
\begin{equation}
\label{newM5fct}
\hat{f} = 1 - \frac{R_0^2}{R^2}
\spa
\hat{H} = \frac{N l_s^6}{R^2}
\end{equation}
where $A(R,v)$ and $K(R,v)$ only depend on $R$, $v$ and $R_0$,
as stated in Section \ref{secmapsol}.
We remark that $k_5 = 1/\pi$ and $R_c^2 = (2\pi)^2 / (7\zeta(3))$ for $d=5$.

For lower energies we have seen in the supersymmetric case
in Section \ref{secsusyLST}
that there is no point in the supergravity description
where we have large curvatures.
We see that this also should hold for \eqref{newM5met}-\eqref{newM5fct}.
For $R \sim R_0 \sim R_c$ we assume that $A(R,v)$ and $K(R,v)$ are
non-zero and of order one. Then it is clear that the curvatures
are suppressed by $1/N^{2/3}$.
This has two important consequences.
First, it means that the corrections to the thermodynamics
from using the solution \eqref{newM5met}-\eqref{newM5fct},
instead of the smeared one, dominate over any correction to the
thermodynamics coming from one-loop corrections to the supergravity.
Second, it means that the non-translationally invariant
near-extremal M5-brane with a transverse circle
\eqref{newM5met}-\eqref{newM5fct} describes
accurately the physics for all energies $\mu \ll N l_s^{-6}$
when $N \gg 1$.

Therefore, the non-translationally invariant
near-extremal M5-brane \eqref{newM5met}-\eqref{newM5fct}
not only gives the leading corrections to the thermodynamics
\eqref{tsm6}-\eqref{mufm6}
for energies $l_s^{-6} \ll \mu \ll N l_s^{-6}$, it
describes the thermodynamics
for all energies $\mu \ll N l_s^{-6}$. This thus includes the transition
between $(2,0)$ SCFT and $(2,0)$ LST.

{} From \eqref{neTS}-\eqref{neF} we get the thermodynamics of
\eqref{newM5met}-\eqref{newM5fct} as
\begin{equation}
\label{thlst1}
T = \gamma(R_0) T_{\rm hg}
\spa
\mathfrak{s} = \frac{\sqrt{N}}{(2\pi)^4 l_s^5} \frac{R_0^2}{\gamma(R_0)}
\end{equation}
\begin{equation}
\label{thlst2}
\mu = \frac{1}{(2\pi)^5 l_s^6} R_0^2 [ 1 - \chi(R_0) ]
\spa
\mathfrak{f} = - \frac{1}{(2\pi)^5 l_s^6} R_0^2 \chi(R_0) \ .
\end{equation}
{} From \eqref{knowchi} and \eqref{knowgamma} we know that
for $R_0 \gg R_c$ we have
\begin{equation}
\chi(R_0) \simeq 0 \spa
\gamma(R_0) \simeq 1
\end{equation}
and for $R_0 \ll R_c$ we have
\begin{equation}
\chi(R_0) \simeq \frac{1}{3}
\spa
\gamma(R_0) \simeq \frac{3}{2} \left( \frac{R_0}{\pi} \right)^{1/3} \ .
\end{equation}
We now analyze the thermodynamics \eqref{thlst1}-\eqref{thlst2}
from the point of view that it should be dual to LST.
As mentioned in Section \ref{secthermo}, the heat capacity
$\mathfrak{c} = T d\mathfrak{s}/dT$ is negative when $dT/dR_0$ and thereby $\gamma'(R_0)$
is negative. So the sign of $\mathfrak{c}$ is highly sensitive to the
behavior of $\gamma(R_0)$ for $R_0 > R_c$.
In Section \ref{secgamchi} we argued for the possibility of
two scenarios for $\gamma(R_0)$ depicted in the Figures
\ref{figgam1} and \ref{figgam2}.
Either $\gamma(R_0)$ increases monotonically for all $R_0$ and
thus always stays below $1$, as depicted in Figure \ref{figgam1},
or $\gamma(R_0)$ increases to a maximum above $1$ and then decreases
toward $1$, as depicted in Figure \ref{figgam2}.

In the first scenario corresponding to Figure \ref{figgam1}
the heat capacity $\mathfrak{c}$ is always positive. Thus, the
energies $\mu \ll N l_s^{-6}$ correspond to LST for temperatures
below the Hagedorn temperature $T_{\rm hg}$, and LST has
positive heat capacity for $T < T_{\rm hg}$.
If this is the case then the first correction to $\chi(R_0)$
or $\gamma(R_0)$ should give information about the density of
states of the $(2,0)$ LST in a regime where it should behave as
a string theory (this we cannot expect for $T > T_{\rm hg}$).
This can give insight into the microscopic behavior of LST.
Since we saw in Section \ref{secgamchi} that this first scenario
can be ruled out if $R_0^2 \chi(R_0) \rightarrow \infty$
for $R_0 \rightarrow \infty$ we expect that $\chi(R_0) \sim R_0^{-a}$
for $R_0 \gg R_c$ with $0 < a \leq 2$.
{} From this we get that $\gamma(R_0) = 1 - (b R_0)^{-a}$,
with $b$ a positive number.
Using this one can straightforwardly determine the leading
behavior of the thermodynamics for temperatures near the Hagedorn
temperature, e.g. the entropy will behave as
$\mathfrak{s}(T) \sim (T_{\rm hg}-T)^{-2/a}$.

In the second scenario corresponding to Figure \ref{figgam2}
the heat capacity starts out positive for low temperatures
and energies, but as the energy increases the temperature crosses the
Hagedorn temperature and continues to increase.
Then the temperature reaches a maximum, the heat capacity becomes negative
and the temperature goes toward the Hagedorn temperature from above,
as the energy increase toward $\mu \sim N l_s^{-6}$.
This scenario has the consequence that we cannot obtain information
about the LST thermodynamics near (but below) the Hagedorn temperature
unless we find the complete functions $\chi(R_0)$ and $\gamma(R_0)$.
It is not enough to find corrections for $R_0 \gg R_c$.
We note that in this scenario the Hagedorn temperature is
reached for energies $\mu \sim l_s^{-6}$.

It is very interesting to try and go beyond energies
$\mu \sim N l_s^{-6}$. For energies $\mu \gg N l_s^{-6}$
we stated in Section \ref{secthLST}
that we are in weakly coupled type IIA string theory
and the smeared near-extremal M5-branes on a transverse circle
become near-extremal NS5-branes.
However, since we believe M-theory is the fundamental theory
our non-translationally invariant near-extremal
M5-brane with a transverse circle \eqref{newM5met}-\eqref{newM5fct}
should be the correct solution for all energies, also for
$\mu \gg N l_s^{-6}$, which corresponds to $R_0 \gg \sqrt{N}$.

Since the $z$-direction is not an isometry of the
non-translationally invariant near-extremal
M5-brane \eqref{newM5met}-\eqref{newM5fct} we cannot S-dualize this
to a Type IIA solution, even though the string coupling is small.
However, this does not have to be a problem if we can show that
any effect due to the variation of the solution along $z$
is insignificant in the physics we want to study using
our solution.
Otherwise we would be led to the perhaps rather spectacular conclusion
that the physics of the
near-extremal NS5-brane cannot be understood in 10 dimensions
although the string coupling is small.

We now consider whether the leading order thermodynamics
for energies $\mu \gg N l_s^{-6}$ is sensitive to these
corrections due to our non-translationally invariant solution
\eqref{newM5met}-\eqref{newM5fct}.
Write for $R_0 \gg R_c$ the function $\gamma(R_0)$ as
\begin{equation}
\gamma(R_0) \simeq 1 + \gamma_1(R_0)
\end{equation}
meaning that $\gamma_1(R_0)$ is the leading correction to $\gamma(R_0)$
for $R_0 \gg R_c$.
This is the correction due to our non-translationally invariant solution
\eqref{newM5met}-\eqref{newM5fct} relative to the usual smeared solution.
Note that $\gamma_1(R_0)$ only depends on $R_0$ and not on $N$.

{} From the analysis of \cite{Harmark:2000hw,Berkooz:2000mz,Kutasov:2000jp}
we know that the leading correction to the near-extremal NS5-brane
is the one-loop term. The total correction
to the temperature from both effects is therefore
\begin{equation}
\label{tcor}
\frac{T}{T_{\rm hg}} = 1 + \gamma_1(R_0) + \frac{\delta}{N^2 R_0^2}
\ \ \ \mbox{for} \ \ R_0 \gg \sqrt{N}
\end{equation}
where $\delta$ is a constant which is computed in \cite{Kutasov:2000jp}.
The leading correction to the temperature in \eqref{tcor}
determines the leading behavior of the thermodynamics, including
the dependence of the entropy as function of temperature and the sign of
the heat capacity.

{} From \eqref{tcor} we now see that if $\gamma_1(R_0) = - (bR_0)^{-a}$
with $0 < a \leq 2$ (this corresponds to the first scenario depicted
in Figure \ref{figgam1})
the correction due to our non-translationally
invariant solution {\sl always} dominates
the one-loop correction to the solution.
Therefore, the heat capacity would be positive for all $R_0$,
even when $R_0 \gg \sqrt{N}$, and the essential physics of the near-extremal
NS5-brane cannot be computed in 10 dimensions no matter how weak
the string coupling is.
If this scenario is the correct one, then the Hagedorn temperature
of $(2,0)$ LST is a limiting temperature, in the sense that no matter
how high energies we are considering we are always below the
Hagedorn temperature.

On the other hand, if we are in the second scenario depicted
in Figure \ref{figgam2} where $R_0^2\gamma_1(R_0) \rightarrow \infty$
for $R_0 \rightarrow \infty$ then the one-loop term
will start to dominate over the $\gamma_1(R_0)$ term for sufficiently
large $R_0$.

\section{Discussion and Conclusions}
\label{secconc}

We have left out several important issues and questions in the main
text. We discuss some of these below.
After that we draw our conclusions with a summary of some of our results
and open directions.

\smallsec{The space of physical solutions}

We have in the above only discussed the specific branch of solutions
of our ansatz \eqref{metans2}-\eqref{endansa} defined by the
boundary conditions
listed in Section \ref{secaneom}. But from the discussion of the
thermodynamics in Section \ref{secthermo} it seems possible
that we have a larger space of physical solutions.
Specifically, we can see that we can not only describe our black hole solution
with the neutral ansatz \eqref{neutBH}-\eqref{moreneut}
but also the black string solution, by putting $A(R,v) = K(R,v) =1$.
Also, the thermodynamics \eqref{neutTS}-\eqref{neutmass}
only depends on the form
of the ansatz, the $\chi$ boundary conditions and the EOMs.
So the question is, if $R_0$ and $\chi$ are specified, does there
exist one and only one physical solution?
If we speculate that a physical solution indeed exists for any
$(R_0,\chi)$ it seems reasonable to expect that it is also
unique since the mass and binding energy
(discussed in Section \ref{secgamchi}) are completely
specified by $(R_0,\chi)$ and it seems reasonable that only one
object should exist with same mass and binding energy.

To formulate this in a more precise way, we introduce the parameter
$\lambda$ that is constant when we move on a curve in $(R_0,\chi)$
space given by a certain physical solution. So, we define
$\lambda(R_0,\chi)$ to be a smooth function and
we define $\lambda=0$ to correspond to $\chi=0$ which is
the black string solution and $\lambda=1$ to correspond to
$\chi=\chi(R_0)$ which is our black hole branch.

For a given $\lambda$ we then impose the boundary condtions
\begin{itemize}
\item
$K(R,v)$ is an even periodic function in $v$ with period $2\pi$.
\item
The Fourier components $L^{(n)}(R)$ of $K(R,v)$ obey
\begin{equation}
\label{Lnbc2}
\frac{L^{(n)} (R)}{L_0^{(n)}(R,\lambda)} \rightarrow 1 \ \ \mbox{for}\ \
\frac{R_0}{R} \rightarrow 0
\end{equation}
\item
For $R$ much greater than both $1$ and $R_0$ we have
\begin{equation}
\label{Kchi2}
K(R,v) = 1 - \chi \frac{R_0^{d-3}}{R^{d-3}}
+ \cdots
\end{equation}
\end{itemize}
Here $L_0^{(n)}(R,\lambda)$ obeys $L_0^{(n)}(R,1) = L_0^{(n)}(R)$
and $L_0^{(n)}(R,0) = \delta_{n,0}$.
So, we can write $K(R_0,\lambda,R,v)$ to display the full functional
dependence of this function.
Thus, we see that our ansatz \eqref{neutBH}-\eqref{moreneut} could in principle
describe a much larger space of physical solutions than just
the black hole and black string branches.

With the above assumptions it is also clear that $\gamma$, which
is defined as $1/\sqrt{A(R,v)}$ on the horizon, is purely a function
of $R_0$ and $\chi$, so we have the complete thermodynamics
with mass $M(R_0,\chi)$, entropy $S(R_0,\chi)$ and temperature
$T(R_0,\chi)$, if we consider the neutral case.
Clearly, we can reparameterize our solution space using
$(M,\lambda)$ as specifying the points. For example, this would
give  $S(M,\lambda)$, i.e. the entropy as function
of $M$ for a given branch specified by $\lambda$.

In the $(M,\lambda)$ parameterization of the space of solutions
we clearly have that $\chi(M,\lambda)$ is an increasing function
of $\lambda$ for a given $M$. Moreover, from the mass formula
we see that increasing $\chi$ forces $R_0$ to increase for a fixed
mass, so also $R_0(M,\lambda)$ is an increasing function
of $\lambda$ for a given $M$.
Since $R_0$ increases with $\lambda$ this means
that $S(M,\lambda)$ is an increasing function
of $\lambda$ for a given $M$, by the arguments of Section \ref{secpunch}.

The question stated above, formulated in term of $\lambda$,
is thus whether there exist solutions for other values of
$\lambda$ than for $\lambda=0,1$.
Clearly we can rule out other solutions with $\lambda \neq 1$
with horizon of topology $S^{d-1}$, because if we had
more solutions with that horizon topology for low masses it would
violate the ``no-hair theorem'' for static black holes in
$(d+1)$-dimensional Minkowski space.
On the other hand, it seems possible that solutions with
$0 < \lambda < 1$ could exist, provided they have a horizon of
topology $S^{d-2} \times S^1$.
In fact, for low masses these could correspond to
the new solutions that Horowitz and Maeda conjectured to exist
\cite{Horowitz:2001cz}.
Also, if we considered the solutions in the non-extremal
charged $p$-brane case, they could feasibly be obtained
by considering other charge configurations.
The $\lambda=0$ case corresponds to a completely smeared
charge on the circle, whereas the $\lambda=1$ case corresponds
to the charge localized on a point of the circle.
A $0 < \lambda < 1$ solution could thus be less localized
than the $\lambda=1$ solution but not completely delocalized.
On the other hand, solutions with $\lambda > 1$
would not seem physically sensible, from the charged
$p$-brane point of view, since they would correspond
to a charge becoming even more localized than a point,
which of course is impossible.
Therefore, while $0 < \lambda < 1$ solutions might possibly exist, it seems
doubtful that $\lambda > 1$ solutions exist. If this is true then
clearly the $\lambda = 1$ solutions would for any given mass correspond
to the ones with the highest entropy.

\smallsec{Heavy static and neutral black objects on cylinders have hair}

``Hair'' on black objects can be defined as free parameters
of the black object which are not subject to a Gauss law.
Mass, charge and angular momentum are all conserved
quantities subject to a Gauss law, so a ``no-hair theorem''
typically states that the black object can be uniquely characterized
by its mass, charge and angular momentum.

We discuss here General Relativity on cylinders, i.e. with
the action taken to be the Einstein-Hilbert action.
For Minkowski space, the so-called ``no-hair theorem'' states
in this case that all static solutions are spherically
symmetric and equivalent to the Schwarzschild black hole solution
with the mass as the only parameter.
See for example \cite{Bekenstein:1996pn} for a review of ``no-hair theorems''.

We defined above the $(M,\lambda)$ parameterization of the two-dimensional
space of solutions.
{} From this viewpoint we see that
we have a family of black objects, all with same mass,
but with the entropy being a function of {\sl another} parameter
$\lambda$.
Therefore, since we have shown that $S=S(M,\lambda)$
increases with $\lambda$, it follows that the parameter $\lambda$
is precisely a free parameter characterizing the black object
that is {\sl not} subject to a Gauss law, since in the case
described here only the mass is subject to a Gauss law.
The $\lambda$ parameter is therefore a ``hair'' on our
black object%
\footnote{Note that from the above discussion we imagine $\lambda$
to take values in either the continuous interval $[0,1]$ or
the discrete set $\{ 0 , 1 \}$. Obviously, even in the discrete
case we can call $\lambda$ a ``hair'', since we just need more than
one solution.}.

However, for small masses $\lambda = 0$ and $\lambda = 1$ correspond
to two different horizon topologies.
Thus in that case it is a triviality to have more than one solution
with the same mass, since the higher dimensional solution is just
obtained by dimensional oxidation.
To get rid of this possibility we therefore
say that we include the topology of the horizon as part of the
characteristics of the black object, so $\lambda$ is only allowed
to vary over solutions with the same horizon topology.
For large masses we thus still have that $\lambda$ is a ``hair''.
We can therefore conclude that, even for General Relativity given
by the Einstein-Hilbert action, one cannot have a ``no-hair theorem''
on cylinders%
\footnote{In the interesting recent work \cite{Emparan:2001wn}
it was found that the vacuum Einstein equations
in five dimensions admit two different solutions with same mass $M$
and angular momentum $J$, one of which is a black hole while
the other is a rotating ``black ring''. This suggest that one cannot
have a ``no-hair theorem'' in asymptotic five-dimensional Minkowski space.}.

We also comment that ``no-hair theorems'' typically are understood
intuitively as a kind of delocalization effect that seems to be
present in General Relativity%
\footnote{See \cite{Surya:1998dx,Marolf:1999uq}
for work in this spirit in String Theory.}.
Consequently. if we imagine starting with a non-spherical configuration
with a large energy in asymptotic Minkowski space, this should
settle down to a spherical Schwarzschild black hole.
By the same token, one could have imagined that a black object
on the cylinder with horizon topology $S^{d-2} \times S^1$
should delocalize and smear itself out, eventually becoming
the black string.
What we have shown in this paper
is that it works the other way around,
so that General Relativity in this case actually prefers the most
{\sl localized} solution.

\smallsec{Black hole microstates}

For General Relativity on cylinders
we have found that the entropy is higher for our new neutral
solution than that of neutral black strings
in the microcanonical ensemble, i.e. for constant mass.
For $M > M_c$ this is true despite the fact that the
black string solution is classically stable.
So the instability of the black string for $M > M_c$
is not a dynamical instability.
To understand the nature of this instability we have
to think about the microstates of our system.
Using the $(M,\lambda)$ parameterization described above, we can
write the total number of microstates of the system
available for a given $\lambda$ as $\Omega ( M , \lambda )$%
\footnote{If $\lambda$ is a continuous variable one should
rather talk about the number of states between $\lambda$
and $\lambda + d\lambda$.}.
The entropy is then $S( M , \lambda ) = \log \Omega ( M , \lambda )$.
Thus, we have shown that $\Omega(M,1) > \Omega(M,0)$, which means
that our system has more microstates available for $\lambda=1$.
If one thus start out constraining the system to have $\lambda=0$
and then relaxes this constraint, the system will settle
in $\lambda=1$ instead, since this maximizes $\Omega(M,\lambda)$.
We can thus call the instability of the neutral black string
a quantum instability since it occurs because the system has
more microstates (i.e. quantum states)
available for $\lambda=1$. Because the black string solution
is classically stable it follows that purely classical General Relativity
is not sensitive to this instability;
It is only through the semi-classical
computation of the Bekenstein-Hawking entropy that we learn of
this quantum instability.

\smallsec{Conclusions}

To summarize, we have found in this paper a precise way
to describe solutions for black holes on cylinders, or
non-extremal and near-extremal charged dilatonic branes
with a transverse circle. This was done by defining a new
coordinates system that interpolates between spherical and
cylindrical coordinates. Using this coordinate system
we were able to write down an ansatz for the solutions
which is completely specified by one function only.

We have examined this ansatz for small black holes by using
the Newton approximation of General Relativity.
Moreover, we have examined the thermodynamics of the solution.
This led to the conclusion that the neutral black string
will spontaneously break its translational invariance for large
masses $M > M_c$
since our non-translationally invariant solution has higher
entropy.

We furthermore explained how our new ansatz can be used to study
the thermodynamics of $(2,0)$ Little String Theory.
In particular, we have discussed two possible scenarios, which
are quite different in their physical implications.

There are many directions that would be interesting to pursue further.
It would be very interesting to compute actual corrections
to our function $K(R,v)$ from either the small or large $R_0$
limit to further test that our construction makes sense and to
find corrections to the thermodynamics.
One could perhaps even hope to find a way to compute the thermodynamics
for all $R_0$.

Learning about corrections to the thermodynamics of near-extremal
branes with transverse circles could also be interesting
via its appliation to the dual non-gravita{-}tional theories
of these branes, along the lines that we sketched for Little String Theory.
This might give further insight into the microstates of
our solutions since these should be dual to microstates
of the dual non-gravitational theory.

A particular example of what
one could study is 3+1 dimensional $\mathcal{N}=4$ $U(N)$
super Yang-Mills on $\R^2 \times S^1$ for non-zero temperature.
This should be dual to $N$ near-extremal D2-branes on a transverse circle as
described by the ansatz \eqref{NEmet}-\eqref{NEfct}
with $p=2$ and $d=7$, even for high energies.

It would also be useful to obtain a  better understanding
of whether there exist
solutions for $0 < \lambda < 1$.
This would perhaps enable one to find the non-translationally
invariant solutions with horizon topology $S^{d-2} \times S^1$
that Horowitz and Maeda conjectured to exist for small masses $M < M_c$.

Obviously, it would be very interesting to generalize our
construction to other more complicated geometries.
Then one could determine whether the spontaneous breaking
of translational invariance in General Relativity
persists for all non-contract{-}ible
geometries, or if there are other topological characteristics
that come into play.


\smallsec{Addendum}

During the final stages of completing this work, we received
the paper \cite{Horowitz:2002ym} by Horowitz and Maeda in which they argue
that certain classes of charged non-extremal $p$-branes have
stable non-translationally invariant solution with horizons of topology
$S^{d-2} \times S^1$. Contrary to this paper,
this is done without any explicit
construction of solutions but instead using
specially constructed initial data that is argued to evolve
into these types of solutions.
This overlaps with the  conclusions of this paper in the case of the
near-extremal D3, M2 and M5-brane.


\section*{Acknowledgments}

A special thanks to Jan de Boer and Robbert Dijkgraaf
for many useful discussions and comments.
Thanks to Michael Gutperle, Finn Larsen, Jakob L. Nielsen, Igor Pesando,
Mukund Rangamani, Jan Pieter van der Schaar
and Nick Toumbas for useful discussions.
TH thanks the Spinoza Institute in Utrecht, the Institute
for theoretical physics at Amsterdam University, the Department
of Physics at University of Michigan
and the Niels Bohr Institute in Copenhagen
for kind hospitality while part
of this work was carried out.

\begin{appendix}

\section{Charged dilatonic $p$-brane solutions}
\label{appextrsol}

The class of charged dilatonic $p$-brane solutions we are interested
in here are solutions of the supergravity action
\begin{equation}
I_D = \frac{1}{16\pi G_D} \int d^D x \sqrt{-g} \left( R
- \frac{1}{2} \partial_\mu \phi \partial^\mu \phi
- \frac{1}{2(p+2)!} e^{a\phi} (F_{(p+2)})^2 \right)
\end{equation}
where $D$ is the space-time dimension.
We are only considering electric branes, since magnetic branes
can be described as electric by electromagnetic duality, at least
for our purposes.
The M2 and M5-branes have $D=11$ and $a=0$.
The D-branes have $D=10$ and $a = (3-p)/2$.
The F-string has $D=10$ and $a = -1$ and the NS5-brane has
$D=10$ and $a=1$.

\smallsec{Extremal charged dilatonic $p$-brane solutions}

The extremal $p$-brane solution in $D$-dimensional space-time has
the metric
\begin{equation}
ds^2_D = H^{-\frac{d-2}{D-2}} \left[ - dt^2 + \sum_{i=1}^p (dx^i)^p
+ H \Big( d\rho^2 + \rho^2 d\Omega_{d-1}^2 \Big) \right]
\end{equation}
the dilaton
\begin{equation}
e^{2\phi} = H^a
\end{equation}
and $(p+1)$-form potential
\begin{equation}
A_{01 \cdots p} = 1 - H^{-1}
\end{equation}
with the harmonic function
\begin{equation}
H = 1 + \frac{L^{d-2}}{\rho^{d-2}} \ .
\end{equation}
Here $d=D-p-1$ is the number of transverse dimensions.
The constant $L^{d-2}$ is related to the tension $\mu_p$ of the extremal brane
as $N \mu_p = L^{d-2} (d-2) \Omega_{d-1} / (16 \pi G)$, $N$
being the number of coincident branes.
The tension $\mu_p$ is $1 / ( (2\pi)^2 l_p^3 )$ for the M2-brane,
$1 / ( (2\pi)^5 l_p^6 )$ for the M5-brane,
$1 / (2\pi l_s^2)$ for the F-string, $1 / ((2\pi)^5 g_s^2 l_s^6)$
for the NS5-brane and $1 / ((2\pi)^p g_s l_s^{p+1})$ for the
D$p$-branes.

\smallsec{Compactification on $T^p$}

We can compactify the $p$-brane solution on a $p$-dimensional torus $T^p$
in the longitudinal directions. This gives extremal charged dilatonic black
hole solutions in $d+1$ space-time dimensions.
The relation between the $D$-dimensional Einstein frame metric $ds_D^2$
and the $(d+1)$-dimensional Einstein frame metric $ds_{d+1}^2$ is
\begin{equation}
ds_D^2 = e^{-\frac{2p}{d-1}\eta} ds_{d+1}^2 + e^{2\eta} \sum_{i=1}^p (dx^i)^2
\end{equation}
where $\eta$ is a new scalar (parameterizing fluctuations of the
volume of the $p$-torus).
The action in $d+1$ dimensions is
\begin{eqnarray}
I_{d+1} &=& \frac{1}{16\pi G_{d+1}} \int d^{d+1} x \sqrt{-g} \left( R
- \frac{1}{2} \partial_\mu \phi \partial^\mu \phi
- \frac{(D-2)(D-d-1)}{d-1}\partial_\mu \eta \partial^\mu \eta \right.
\nn \\ &&
\left.
- \frac{1}{4} e^{- 2\frac{d-2}{d-1}(D-d-1) \eta } e^{a \phi} (F_{(2)})^2 \right)
\end{eqnarray}
were $F_{(2)} = dA_{(1)}$.
The metric is
\begin{equation}
ds^2_{d+1} = H^{-\frac{d-2}{d-1}}
\left[ - dt^2 + H \Big( d\rho^2 + \rho^2 d\Omega_{d-1}^2 \Big) \right]
\end{equation}
the scalars are
\begin{equation}
e^{2 \phi} = H^{a} \spa
e^{2\eta} = H^{-\frac{d-2}{D-2}}
\end{equation}
and one-form potential is
\begin{equation}
A_{0} = 1 - H^{-1} \ .
\end{equation}

\smallsec{Non-extremal charged dilatonic $p$-branes}

The non-extremal solutions of charged dilatonic $p$-branes are given
by
\begin{equation}
\label{nemet}
ds^2_D = H^{-\frac{d-2}{D-2}} \left[ - f dt^2 + \sum_{i=1}^p (dx^i)^p
+ H \Big( f^{-1} d\rho^2 + \rho^2 d\Omega_{d-1}^2 \Big) \right]
\end{equation}
\begin{equation}
\label{nedilA}
e^{2\phi} = H^a
\spa
A_{01 \cdots p} = \coth \alpha ( 1 - H^{-1} )
\end{equation}
with the functions
\begin{equation}
\label{nefH}
f = 1 - \frac{\rho_0^{d-2}}{\rho^{d-2}}
\spa
H = 1 + \frac{\rho_0^{d-2} \sinh^2 \alpha}{\rho^{d-2}}
\end{equation}
and with $\alpha$ defined by
\begin{equation}
L^{d-2} = \rho_0^{d-2} \cosh \alpha \sinh \alpha \ .
\end{equation}

The thermodynamics is
\begin{equation}
\label{rhoTS}
T = \frac{d-2}{4\pi \rho_0 \cosh \alpha}
\spa
S = \frac{V_p \Omega_{d-1}}{4G} \rho_0^{d-1} \cosh \alpha
\end{equation}
\begin{equation}
\label{rhomuQ}
\nu = \tanh \alpha
\spa
Q = \frac{V_p \Omega_{d-1}}{16 \pi G} (d-2) \rho_0^{d-2} \cosh \alpha
\sinh \alpha
\end{equation}
\begin{equation}
\label{rhoM}
M = \frac{V_p \Omega_{d-1}}{16 \pi G} \rho_0^{d-2} \Big[ d-1 +
(d-2) \sinh^2 \alpha \Big]  \ .
\end{equation}

\smallsec{Non-extremal charged dilatonic $p$-branes
smeared on transverse circle}

The non-extremal solutions of charged dilatonic $p$-branes
smeared on a transverse circle are given by
\begin{equation}
\label{nemets}
ds^2_D = H^{-\frac{d-2}{D-2}} \left[ - f dt^2 + \sum_{i=1}^p (dx^i)^2
+ H \Big( f^{-1} dr^2 + dz^2  + r^2 d\Omega_{d-2}^2 \Big) \right]
\end{equation}
accompanied by the fields \eqref{nedilA} with
\begin{equation}
\label{nefHs}
f = 1 - \frac{r_0^{d-3}}{r^{d-3}}
\spa
H = 1 + \frac{r_0^{d-3} \sinh^2 \alpha}{r^{d-3}} \ .
\end{equation}
The thermodynamics is given in Section \ref{secpunch}
in \eqref{smeTS}-\eqref{smeM}, where we write $\hat{\alpha}$
instead of $\alpha$ as above.

\section{The functions $F_{2s}$ and $G_{2s}$}
\label{appfctF}

In this appendix we derive some properties of the function
\begin{equation}
\label{Fs}
F_{2s} (a,b) = \sum_{n\in \Z} \Big( a^2 + (2\pi n + b)^2 \Big)^{-s}
\end{equation}
which plays a central role in the harmonic function of branes
with a transverse circle.

For $a^2 +b^2 \ll 1$ the $n=0$ term in \eqref{Fs} dominates and one has
\begin{equation}
F_{2s} (a,b) \simeq (a^2 + b^2)^{-s} \ .
\end{equation}
To derive the asymptotics when $a \gg 1$ we use Poisson resummation
\begin{equation}
\label{pois}
\sum_{n} e^{-\pi (n +\alpha)^t A (n+\alpha) + 2 \pi i n \beta }
= \frac{1}{\sqrt{\det A}} \sum_{ m} e^{-\pi
(m +\beta)^t A^{-1} (m+\beta) - 2 \pi i (m +\beta) \alpha }
\end{equation}
and we recall the the integral representation of the Bessel function
\begin{equation}
\label{bessel}
\int_0^{\infty} \frac{dx}{x^{1+s}} e^{-B/x-Cx}
=2 \left|\frac{C}{B }\right|^{s/2} K_{s}(2\sqrt{|BC|}) \ .
\end{equation}
This is an even function in $s$, and admits the asymptotic expansion
at large $x$
\begin{equation}
\label{asbessel}
K_{s}(x)= \sqrt{\frac{\pi}{2x}} e^{-x} \left(1
+ \sum_{k=1}^{\infty} \frac{1}{(2x)^k}
\frac{\Gamma\left(s+k+\frac{1}{2}\right)}
{k!\Gamma\left(s-k+\frac{1}{2}\right)}
\right)\ .
\end{equation}
which truncates when $s$ is half-integer. In further detail, $K_s$ is
the modified Bessel function of the second kind, which solves
the second order differential equation
\begin{equation}
\label{BesselDE}
x^2 K_s''(x) + x K_s'(x) - (x^2 +s^2) K_s(x) = 0 \ .
\end{equation}

Using the integral representation of the $\Gamma$ function,
\begin{equation}
\label{Gam}
\Gamma (s) = \int_0^{\infty} \frac{dt}{t^{1+s}} e^{-1/t}
\end{equation}
we write
\begin{equation}
F_{2s} (a,b) = \frac{\pi^s}{\Gamma(s)} \sum_{n \in \Z} \int_0^{\infty}
\frac{dt}{t^{1+s}} \exp\left(-\frac{\pi}{t} [ a^2 + (2 \pi n + b)^2 ] \right) \ .
\end{equation}
After Poisson resummation this becomes
\begin{equation}
F_{2s} (a,b) = \frac{\pi^s}{\Gamma(s)} \frac{1}{2 \pi}\sum_{m \in \Z} \int_0^{\infty}
\frac{dt}{t^{1+s-1/2}} e^{-\pi a^2/t} e^{ - \pi t (m/ 2 \pi)^2} e^{- i m  b} \ .
\end{equation}
Using again \eqref{Gam}, the $m=0$ term in this expression easily gives the leading behavior
\begin{equation}
F_{2s} (a,b) \simeq \frac{1}{2 \sqrt{\pi} } \frac{\Gamma(s-1/2)}{\Gamma(s)}
\frac{1}{a^{2s-1}}
\end{equation}
for $a \gg 1$.
More generally, we find using \eqref{bessel} for the $m \neq 0$ part,
\begin{eqnarray}
\label{F2sexp}
F_{2s}(a,b) &=& \frac{1}{a^{2s-1}} \left[
\frac{1}{2 \sqrt{\pi}} \frac{\Gamma(s-1/2)}{\Gamma(s)}
 +  2 \sqrt{\frac{2}{\pi}} \frac{1}{2^s \Gamma (s)}
 \sum_{m=1}^{\infty}(ma)^{s-\frac{1}{2}} K_{s-1/2} (ma) \cos ( mb) \right]
\nn \\
&=& \frac{\hat k_s}{a^{2s-1}}
\left[ 1 + \sum_{m=1}^\infty \hat f_s( ma) \cos (mb) \right]
\end{eqnarray}
where we have defined
\begin{equation}
\label{fsdef}
\hat f_s (y) \equiv \frac{\sqrt{2}}{2^{s-2}} \frac{1}{\Gamma(s-1/2)} y^{s-1/2}
K_{s-1/2} ( y) \spa \hat k_s \equiv \frac{1}{2 \sqrt{\pi}}
\frac{\Gamma(s-1/2)}{\Gamma (s)}
\end{equation}
It then follows from \eqref{BesselDE} that the function
$\hat f_s (y)$ satisfies
\begin{equation}
\label{fsDE}
\hat f_s''(y) - \frac{2 (s-1)}{y} \hat f_s' (y) - \hat f_s(y) =0 \ .
\end{equation}

To the function $F_{2s}(a,b)$ we may associate the function
$G_{2s}(a,b)$, defined by
\begin{equation}
\label{intsys}
\frac{\partial G_{2s}(a,b)}{\partial a} = a^{2s}
\frac{\partial F_{2s}(a,b)}{\partial b}
\spa
\frac{\partial G_{2s}(a,b)}{\partial b} =- a^{2s}
\frac{\partial F_{2s}(a,b)}{\partial a}
\end{equation}
which is integrable as a consequence of the harmonic condition
\begin{equation}
\left[ \frac{\partial^2}{\partial a^2} + \frac{2s}{a} \frac{\partial}{\partial a}
+\frac{\partial^2}{\partial b^2} \right] F_{2s} (a,b) = 0 \ .
\end{equation}
Substituting the expansion \eqref{F2sexp} we find by integration
\begin{equation}
G_{2s}(a,b) = (2s-1)\hat k_s \left[ b
+ \sum_{m=1}^\infty \left( \frac{\hat f_s(ma)}{m} - \frac{a}{2s-1} \hat
f_s'(ma) \right) \sin (mb)  \right] .
\end{equation}
This satisfies the second relation in \eqref{intsys} by direct
differentiation, while for the first relation the differential equation
\eqref{fsDE} is used as well.

Note that since the expansion
of the Bessel function $K_{s-1/2}$ (cf. \eqref{asbessel})
terminates for integer $s$,  it follows from \eqref{F2sexp}
that the function $F_{2s}$ can be written more explicit for integer
$s$. For example, in the simplest case $s=1$ one finds after some
algebra
\begin{equation}
2F_2 (a,b) = \frac{1}{a} \frac{\sinh a}{\cosh a - \cos b}
\end{equation}
\begin{equation}
2G_2 (a,b)=b +\frac{ a \sin b }{\cosh a - \cos b} + 2 \arctan \left(
\frac{ \sin b }{e^a - \cos b} \right) \ .
\end{equation}
Since the relation between $s$ in this appendix and $d$ in the
text is $s=(d-2)/2$, these functions are relevant for the
coordinate transformation in the case $d=4$.

\section{Coordinate change in large $R$ limit}
\label{appcoord}

In this appendix, we work out the non-trivial change of coordinates
$(r,v) \rightarrow (R,v)$ in the large $R$ limit.
Starting point are the relations
\begin{equation}
R = \left( \frac{k_d}{F_{d-2}} \right)^{\frac{1}{d-3}}
= \frac{r}{R_T} \left[ 1 + \sum_{n=1}^\infty
f_d \Big( n \frac{r}{R_T} \Big)
\cos \Big( n \frac{z}{R_T} \Big) \right]^{-\frac{1}{d-3}}
\end{equation}
\begin{equation}
v = \frac{z}{R_T} + \sum_{n=1}^\infty \sin\left( n \frac{z}{R_T} \right)
\left[ \frac{1}{n} f_d \left(n\frac{r}{R_T} \right) - \frac{1}{d-3}
\frac{r}{R_T} f_d'\left(n \frac{r}{R_T} \right)
  \right]
\end{equation}
which are read off from \eqref{theFF}, \eqref{Rdef} and \eqref{vexpress}.
Since $f_d(y) \sim y^{\frac{d-4}{2}} e^{-y}$, the $n$-expansion in
the two expressions above can be regarded as the correction terms in
a large $r$ (and hence large $R$) expansion, with increasing exponentially
suppressed terms $e^{-nr}$. Alternately, we can regard the series as
a Fourier expansion in terms of $\cos(nz)$ for $R$ and $\sin (nz)$ for $v$.
These two ways of expanding coincide in fact for the function $F_{d-2}$.

For the functions $r(R,v)$, $z(R,v)$ these two expansion are correlated in
a more complicated matter, but it is not difficult to establish that
\begin{equation}
\label{rinv}
\frac{r}{R_T} =R  \sum_{n=0}^\infty g_d^{(n)} (R) \cos  n v
\end{equation}
\begin{equation}
\label{zinv}
\frac{z}{R_T} =v  - \sum_{n=1}^\infty h_d^{(n)} (R) \sin  n v
\end{equation}
where
\begin{equation}
g_d^{(n)}(R) = \sum_{m=0}^\infty e^{- (n + 2m) R} \tilde g_d^{(n,m)} (R) \spa
\tilde g_d^{(0,0)} = 1
\end{equation}
\begin{equation}
h_d^{(n)}(R) = \sum_{m=0}^\infty e^{- (n + 2m) R} \tilde h_d^{(n,m)} (R)
\ .
\end{equation}
 Here, we explicitly give the inverse relations \eqref{rinv}, \eqref{zinv}
thru second order ($e^{-2R}$) in
  terms of the known functions $f_d$ in \eqref{fddef}. After some
 algebra one obtains
\begin{equation}
g_d^{(0)}(R) \simeq 1 + \frac{d-2}{4 (d-3)^2} f_d (R) ^2
\end{equation}
\begin{equation}
g_d^{(1)} (R)\simeq  \frac{1}{d-3} f_d(R)
\end{equation}
\begin{equation}
g_d^{(2)} (R) \simeq \frac{1}{d-3} \left[
f_d (2R) - \frac{3d-10}{4(d-3)} f_d(R)^2 + \frac{1}{d-3} R f_d(R) f_d'(R) \right]
\end{equation}
\begin{equation}
h_d^{(1)} \simeq f_d(R) - \frac{1}{d-3} R f_d'(R)
\end{equation}
\begin{eqnarray}
h_d^{(2)} &\simeq  & \frac{1}{2} \left[
- f_d(R)^2 + \frac{3d-11}{(d-3)^2} R f_d(R) f_d'(R) -
\frac{R^2}{(d-3)^2} [(f_d'(R))^2 + f_d(R) f_d''(R)] \right. \nn \\
 & & \left. + f(2R)
-\frac{2}{d-3} R f_d'(2R) \right] \ .
\end{eqnarray}

\smallsec{Fourier expansion of $K_{(0)}$ and $A_{(0)}$}

The above results can be used to find the large $R$ behavior
of the Fourier components of $K_{(0)}$ and $A_{(0)}$
in the Fourier expansions \eqref{K0fourier} and \eqref{A0fourier}.
Using \eqref{theFF} and \eqref{K0} one obtains the expression
\begin{equation}
K_{(0)}(r,z) = \left[ 1 + \sum_{n=1}^\infty f_d \Big( n \frac{r}{R_T} \Big)
\cos \Big( n \frac{z}{R_T} \Big) \right]^{\frac{2}{d-3}}
\end{equation}
and then it follows from the results
above that in terms of $(R,v)$ coordinates one has
\begin{equation}
K_{(0)}(R,v) = \sum_{n=0}^\infty
L_0^{(n)} (R) \cos nv
\end{equation}
\begin{equation}
L_0^{(n)}(R) = \sum_{m=0}^\infty e^{- (n + 2m) R} \tilde L_0^{(n,m)} (R)
\spa
\tilde L_0^{(0,0)} = 1 \ .
\end{equation}
Through second order we have that
\begin{equation}
L_0^{(0)}(R) \simeq 1 + \frac{d-1}{2(d-3)^2} f_d(R)^2
\end{equation}
\begin{equation}
\label{L01}
 L_0^{(1)}(R) \simeq  \frac{2}{d-3} f_d(R)
\end{equation}
\begin{equation}
 L_0^{(2)}(R) \simeq \frac{2}{d-3} \left[ f_d(2R) + R f_d(R) f_d'(R)
- \frac{3d-11}{4(d-3)^2} f_d(R)^2 \right]
\end{equation}

Similarly, for $A^{(0)}$ it follows from \eqref{theFF} and \eqref{A0}
that
\begin{equation}
A_{(0)}(R,v) = \sum_{n=0}^\infty
B_0^{(n)} (R) \cos nv
\end{equation}
\begin{equation}
B_0^{(n)}(R) = \sum_{m=0}^\infty e^{- (n + 2m) R} \tilde B_0^{(n,m)} (R)
\spa
\tilde B_0^{(0,0)} = 1
\end{equation}
where through second order
\begin{eqnarray}
B_0^{(0)}(R) & \simeq & 1 + \frac{1}{2(d-3)} \left[
\Big(d-1- R^2 \frac{d+1}{d-3} \Big) f_d(R)^2 \right. \nn \\
& & \left.  \phantom{\frac{d}{d}} + 2 (d-1) R f_d(R) f_d'(R)
 + R^2 f_d'(R)^2 \right]
\end{eqnarray}
\begin{equation}
 B_0^{(1)}(R) \simeq \frac{2}{d-3}
\left[ \Big(1 -\frac{R^2}{d-3} \Big) f_d(R)  + R f_d'(R)  \right]
\end{equation}
\begin{eqnarray}
 B_0^{(2)}(R) & \simeq & \frac{1}{2(d-3)} \left[
 4 \Big( 1 - R^2 \frac{10}{d-3} \Big)  f_d(2R)
 + 4 \frac{3d-10}{d-3} R f_d'(2R) \right. \nn \\
 & &
+ \frac{1}{(d-3)^2} \Big( (3d-11) + R^2 \frac{4d^3 -28d^2+53d-9}{d-3} \Big)
 f_d(R)^2 \nn \\
& &
 + 2 \frac{1}{d-3} \Big ( \frac{d-1}{d-3}(2d^2-14d+25) - 8 R^2  \Big)
  R f_d(R) f_d'(R) \nn \\
& &
\left. +\frac{1}{(d-3)^3} (8d^3 -72d^2 +211d -199) R^2 f_d'(R)^2
\right] \ .
\end{eqnarray}

\section{Details on the expansion of the equations of motion}
\label{appdet}

In this appendix we give some of the details relevant to the
expansion
\begin{equation}
\label{Kexp}
K(R,v) = y(R) + \cos (v) b(R) + \cos (2v) q(R) + p(R) + {\cal{O}} (e^{-3R})
\end{equation}
in the equations of motion \eqref{eqRRR}-\eqref{eqRpp}. For completeness
we recall the definitions
\begin{equation}
y(R) \equiv \tilde L^{(0,0)}(R) \spa b(R) \equiv
e^{-R} \tilde L^{(1,0)} (R)
\end{equation}
\begin{equation}
q(R) \equiv e^{-2R} \tilde L^{(2,0)}(R) \spa
p(R) \equiv e^{-2R} \tilde L^{(0,1)}(R)
\end{equation}
in terms of the expansion coefficients in \eqref{FourK}.

The non-linear differential equation on $y(R)$ that follows
from ${\cal{E}}_2$ in \eqref{E2} is
\begin{equation}
\label{ydifeq}
\sum_{ 0 \leq k \leq l \leq m \leq n \leq 3} c_{klmn}
y_k y_l y_m y_n = 0
\end{equation}
where we have defined
\begin{equation}
\label{ymdef}
y_m \equiv R^m \frac{\partial^m y(R)}{\partial R^m} \ .
\end{equation}
In \eqref{ydifeq} the ten non-zero coefficients are
\begin{eqnarray}
\label{c0001}
c_{0001} &=& 2(d-4) [ 2 (d-2)^2 - (5d-11) x + (d-1) x^2] \\
c_{0011} &=& 2(d-2)(2d^2 -11d +21) - 2 (d^3-8d^2 + 31 d -36)x  +
 6 (d-1)x^2 \;\; \\
c_{0111} &=& (d-1) ( 1 -x) [ (d^2-8d + 12) + d x] \\
c_{1111}&=& - (d-1)(d-2) (1-x)^2 \\
c_{0002}&=&  -12(d-2) - 2 (3d^2 - 25 d +36) x+ 2 (d-1)(d-6)x^2 \\
c_{0012}&=&  2 (1-x) [ (d^2-4) - (d^2 -2d +2)x] \\
c_{0022} &=& 2 (d-2) ( 1-x)^2 \\
c_{0112} &=& (d-1)(d-2) (1-x)^2 \\
c_{0003} &=& - 2(1-x) [ 2 (d-2) - (d-1)x] \\
\label{c0013}
c_{0013} &=& -2(d-2) (1-x)^2
\end{eqnarray}
with the definition
\begin{equation}
\label{xdef}
x \equiv (R_0/R)^{d-3} \ .
\end{equation}

The perturbative solution of \eqref{ydifeq} is given by the power series
\begin{equation}
\label{ysol}
y(R) = 1 + \sum_{n=1}^\infty \alpha_n x^n
\end{equation}
where the first five orders are given by
\begin{eqnarray}
\label{ysolc1}
\alpha_1 &=& - \chi \\
\alpha_2 &=& -\frac{1}{12(d-2)} \chi [ (d-2)(d-9) \chi + 6d -14) ] \\
\alpha_3 & =& \frac{1}{12(d-2)} \chi (\chi -1) [ (d - 2)(d-5)\chi + 4d-10] \\
\alpha_4 &= & \frac{1}{1440(d-2)^2} \chi [ (d-2)^2 (d^3-4d^2-63d+258)\chi^3
+8(d-2)(22d^2-157d+243) \chi^2 \nn \\
 & &
- 4(9d-22) (3d^2-32d+53) \chi
 - 72 (d-2) (5d-13) ] \\
\alpha_5 & =& -\frac{1}{1440(d-2)^2} \chi (\chi-2) [ (d-2)^2  (d-5)
(d^2+d-18) \chi^3  \nn \\
   & &
  +4(d-2)(19d^2-123d+184)\chi^2  \nn \\ & & +
(-48d^3 + 564d^2 -1836d + 1816)\chi -48(d-2)(3d-8) ]
\label{ysolc5}
\end{eqnarray}
and, more generally, all coefficients $\alpha_{n \geq 2}$ are seen to be
 uniquely determined in terms of $\alpha_1 = -\chi$.

The second order differential equation on $b(R)$ that follows from
\eqref{E1} takes the form
\begin{equation}
\label{bdifeq}
M_2(R,y(R))R^2 b''(R) + M_1 (R,y(R)) R b'(R) +M_0 (R,y(R)) b(R)   =0
\end{equation}
where in terms of the notation defined in \eqref{ymdef}, \eqref{xdef} the
functionals $M_m$ are given by
\begin{eqnarray}
\label{M2}
M_2 & \equiv & (1-x) y_0^2 [ (2d -4 - (d-1)x ) y_0 + (d-2)(1-x) y_1 ] \\
M_1 & \equiv &  -y_0^2 [ (2d^2+16-12d +(-3d^2+19d-26)x + (-d+1)x^2) y_0
\nn \\
 & & + (1-x)(d^2-4  - (3d-4)x ) y_1
+ 2(1-x)^2 (d-2)  y_2 ] \label{M1} \\
\label{M0}
M_0 & \equiv &
M_0^{(0)}
+ R^2 y_0^d [ (-2d + 4 + (d-1)x)y_0 - (1-x)(d-2) y_1]
\end{eqnarray}
\begin{equation}
\label{M00}
M_0^{(0)} \equiv  [ 2 x (d-3)^2 y_0^3 + (1-x)  (d-2) ( -2d^2+15d-28 + (d-2)x]
y_0^2 y_1
\end{equation}
$$
 - (1-x) (d-1) (d^2-6d+8 +x) y_0 y_1^2
+ (1-x) (d-2) ( d-2 + (d-4)x ) y_0^2 y_2
$$
$$
- (1-x)^2 (d-1) (d-2) y_0 y_1 y_2
 + (1-x)^2 (d-2) y_0^2 y_3
+ (1-x)^2 (d-1)(d-2) y_1^3 ] \ .
$$
As a check note that for $y(R)=1$ and $R_0=0$ this differential
equation correctly reduces to \eqref{fsDE} with $s=(d-2)/2$, the
solution of which is given by $\hat f_{s=(d-2)/2} (R) \equiv f_d(R) $
given in \eqref{fddef}. Indeed, we know from \eqref{L01} that in this
case $b(R) = e^{-R} \tilde L_0^{(1,0)}(R) \propto f_d(R)$.

The second order differential equation on $q(R)$ takes the form
\begin{equation}
\label{qdifeq}
N_2(R,y(R)) R^2q''(R) + N_1 (R,y(R)) Rq'(R) + N_0 (R,y(R)) q(R)   =
\sum_{0 \leq m\leq n \leq 3} w_{mn} b_m b_n
\end{equation}
\begin{equation}
\label{bmdef}
b_m \equiv R^m \frac{\partial^m b(R)}{\partial R^m}
\end{equation}
where the functionals entering the homogeneous parts are
\begin{equation}
N_2 \equiv \lambda M_2 \spa N_1 \equiv \lambda  M_1 \spa N_0 \equiv \lambda M_0^{(0)}
+ R^2 N_0^{(1)}
\end{equation}
\begin{equation}
\lambda \equiv 4 y_0 [2(d-3)y_0^2+((d-2-x) y_1+(1-x)y_2)y_0-(1-x)y_1^2]
\end{equation}
\begin{equation}
N_0^{(1)} \equiv 16 y_0^{d+1}\Big[
- 2(d-3)\Big(2(d-2) -(d-1)x \Big)y_0^3
\end{equation}
$$
- \Big( [2(d-2)(2d-5) - (d-2)(3d-5)x + (d-1)x^2] y_1
$$
$$
+ (1-x)[2(d-2) - (d-1)x ] y_2 \Big) y_0^2
$$
$$
 -(1-x)\Big(  [(d-2)(d-4)+x]y_1 + (1-x)(d-2) y_2 \Big) y_0
+ (d-2)(1-x)^2 y_1^3 \Big]
$$
with $M_{m=1,2}$ defined in \eqref{M2}, \eqref{M1} and
$M_0^{(0)}$ defined in \eqref{M00}.
The six non-zero functionals $w_{mn}$ in the inhomogeneous part of
\eqref{qdifeq} are
\begin{equation}
\label{w00}
w_{00} = w_{00}^{(0)} +R^2 y_0^{d-2} w_{00}^{(1)}
\spa w_{01} = w_{01}^{(0)} +R^2 y_0^{d-2} w_{01}^{(1)}
\end{equation}
\begin{equation}
\label{w02}
w_{02} \spa w_{03}  \spa w_{11}  \spa w_{12}  \spa
w_{22} \ .
\end{equation}
As the expressions become rather lengthy we refrain from giving
the exact forms of these $y$-dependent functionals $w_{mn}$,
though we note that $w_{00}^{(0,1)}$, $w_{01}^{(0,1)}$ in
\eqref{w00} and the remaining five $w_{mn}$ in \eqref{w02}
are of fifth order in $\{y_m; m=0 \ldots 3\}$ with coefficients
that depend on $x$ and $d$.

Finally, the third order differential equation on $p(R)$ takes the form
\begin{equation}
\label{pdifeq}
\sum_{m=0}^3 J_m(R,y(R)) p_m = \sum_{0 \leq m \leq n \leq 3} v_{mn}
b_m b_n \spa
p_m \equiv R^m \frac{\partial^m p(R)}{\partial R^m}
\end{equation}
where $b_m$ was defined in \eqref{bmdef}. Here, the functionals
$J_m $ are fifth order polynomials in $\{y_m ;m=0 \ldots 3\}$ with coefficients
that depend on $x$ and $d$. We also comment on the form of the
 seven non-zero functionals $v_{mn}$
\begin{equation}
v_{00} = v_{00}^{(0)} + R^2 y_0^{d-2} v_{00}^{(1)} + R^4 y_0^{2(d-2)}
v_{00}^{(2)}
\end{equation}
\begin{equation}
v_{mn} = v_{mn}^{(0)} + R^2 y_0^{d-2} v_{mn}^{(1)} \spa (mn)=(01),(02),(11)
\end{equation}
\begin{equation}
v_{03} \spa v_{12} \spa v_{13} \spa v_{22}
\end{equation}
where in this case the  functionals are of fourth order in
$\{y_m; m=0 \ldots 3\}$.

We finally give some of the corresponding results for $A(R,v)$
in \eqref{FourA}, \eqref{Bnm} obtained from substituting \eqref{Kexp}
in \eqref{AfromK}. For the leading $v$-independent part one finds
\begin{equation}
\tilde B^{(0,0)}(R) = \frac{1}{2(d-3)y_0} [
2(d-3) y_0^2 + (d-2 -x) y_0 y_1 + (1-x) y_0 y_2 -(1-x) y_1^2]
\end{equation}
where we recall that $y_m$ is defined in \eqref{ymdef}.
Substitution of the power series solution \eqref{ysol}-\eqref{ysolc5}
of $y(R)$ gives
\begin{equation}
 \tilde B^{(0,0)}(R) = 1 + \sum_{n=1}^\infty \beta_n x^n
\end{equation}
with the first five orders given by
\begin{eqnarray}
\beta_1 &=& -\chi \\
\beta_2 &=& -\frac{1}{12} \chi [ d(d-5)\chi + 4] \\
\beta_3 &=&\frac{1}{12} \chi (\chi -1) [ (d-1)(d-4))\chi + 2] \\
\beta_4 &=& \frac{1}{1440(d-2)} \chi [
(3d-11)(d-2)(2d^3-13d^2+9d+6) \chi^3 \nn \\
& & + 4(d-2)(39d^2-194d+171) \chi^2 \nn \\
 & &  + (-108d^3+756d^2-1404d+644) \chi
 - 144 (d-2)  ] \\
\beta_5 &=& -\frac{1}{1440(d-2)} \chi (\chi-2) [
(d-1)(d-2)(2d-7)(3d^2-17d+18) \chi^3 \nn \\
& & + 4(d-2)(14d^2-69d+67)\chi^2  \nn \\
& & + -4(d-1)(12d^2-72d+97) \chi -48(d-2) ] \ .
\end{eqnarray}
We also give the expression for the first correction term
that multiplies $\cos (v)$,
\begin{eqnarray}
e^{-R} \tilde B^{(1,0)}(R) & = & \frac{1}{2(d-3) y_0^2}
\left[ ( 2(d-3) y_0^2 +(1-x) y_1^2 -R^2 y_0^d) b(R)
\right. \nn \\
  & & \!\!\!\!\!\!\!\! \left. +((d-2-x) y_0^2 -2(1-x) y_0 y_1) R b(R)'
+ (1-x) y_0^2 R^2 b(R)'' \right] \;\;\;\;\;\;\;
\end{eqnarray}
where we recall that $b(R) = e^{-R} \tilde L^{(1,0)}(R)$.

\section{M5 and NS5-branes}
\label{appM5NS5}

The solution of $N$ coincident M5-branes with a transverse
circle of radius $R_T$ has the metric
\begin{equation}
\label{m5circmet}
ds^2 = H^{-1/3} \left[ - dt^2 + \sum_{i=1}^5 (dx^i)^2
+ H \Big( dz^2 + dr^2 + r^2 d\Omega_3^2 \Big) \right]
\end{equation}
and electric six-form potential
\begin{equation}
\label{m5circc}
C_{012345} = 1 - H^{-1}
\end{equation}
with
\begin{equation}
\label{m5circharm}
H = 1 + \sum_{n=-\infty}^{\infty}
\frac{\pi N l_p^3}{(r^2 + (z + 2\pi n R_T)^2)^{3/2}}
\end{equation}
where $z$ is the coordinate of the circle with radius $R_T$.
For $r \gg R_T$ we have
\begin{equation}
\label{Hm5circ}
H = 1 + \frac{N l_p^3}{R_T} \frac{1}{r^2} = 1 + \frac{N l_s^2}{r^2}
\end{equation}
where we used that \( l_p^3 = R_T l_s^2 \) and \( R_T = g_s l_s \)
from the IIA/M-theory S-duality. Using the standard
S-duality transformation on the solution
\eqref{m5circmet}-\eqref{m5circc}
we get the NS5-brane solution in Einstein frame
\begin{equation}
\label{ns5met}
ds^2 = H^{-1/4} \left[ - dt^2 + \sum_{i=1}^5 (dx^i)^2
+ H \Big(  dr^2 + r^2 d\Omega_3^2 \Big) \right]
\end{equation}
\begin{equation}
e^{2 \phi } = H  \spa {\cal{E}}_{0  12345} = 1 - H^{-1}
\end{equation}
where ${\cal{E}}$ is the 6-form potential dual to the
Kalb-Ramond two-form $B$ and the function
$H$ is given by \eqref{Hm5circ}.

\end{appendix}

\addcontentsline{toc}{section}{References}
\label{refs}

\providecommand{\href}[2]{#2}\begingroup\raggedright\endgroup


\end{document}